\documentclass[final,3p,times,authoryear]{elsarticle}

\usepackage{amsmath}
\usepackage{amsfonts}
\usepackage{amssymb}
\usepackage[titletoc,title]{appendix}
\usepackage{booktabs}
\usepackage[mmddyyyy]{datetime}
\usepackage{dutchcal}
\usepackage{enumitem}
\usepackage{float}
\usepackage{graphicx}
\usepackage{hyperref}
\usepackage[utf8]{inputenc}
\usepackage{lineno}
\usepackage{lscape}
\usepackage{longtable}
\usepackage[round]{natbib}
\usepackage[percent]{overpic}
\usepackage{pgfplots}
\usepackage{relsize}
\usepackage{siunitx}
\usepackage{tabularx}
\usepackage{subfigure}
\usepackage{tikz}
 \usepackage{url,textcomp}

\newcommand{\ra}[1]{\renewcommand{\arraystretch}{#1}}
\modulolinenumbers[5]
\numberwithin{equation}{section}
\renewcommand{\bigr}{\mathcal{R}}
\newcommand{\bigs}{\mathcal{S}}
\usetikzlibrary{calc}
\usetikzlibrary{plotmarks}
\usetikzlibrary{positioning}
\usetikzlibrary{fit}
\usetikzlibrary{decorations.pathmorphing}
\usetikzlibrary{backgrounds}
\makeatother
\modulolinenumbers[5]
\numberwithin{equation}{section}
\pgfplotsset{compat = newest}
\pgfplotsset{ legend style={font=\tiny} }
\definecolor{bgreen}{rgb}{0.0,0.5,0.0}
\definecolor{bblue}{rgb}{0.0,0.0,0.9}
\definecolor{bgold}{rgb}{0.7,0.5,0.0}
\definecolor{bred}{rgb}{0.9,0.0,0.0}

\begin{document}

\begin{frontmatter}



\title{Self-similar orbit-averaged Fokker-Planck equation for isotropic spherical dense clusters  (iii) application of pre-collapse solution to Galactic globular clusters}


\author[address1,address2]{Yuta Ito}
\ead{yito@gradcenter.cuny.edu}

\address[address1]{Department of Physics, CUNY Graduate Center, \tnoteref{footnote1}} 
\address[address2]{Department of Engineering and Physics,
	CUNY College of Staten Island\tnoteref{footnote2}}
\tnotetext[footnote1]{ 365 Fifth Avenue, New York, NY 10016, USA}
\tnotetext[footnote2]{2800 Victory Boulevard, Staten Island, NY 10314, USA}

\begin{abstract}
This is the third paper of a series of our works on the self-similar orbit-averaged Fokker-Planck (OAFP) equation. The first paper provided an accurate spectral solution of the equation for isotropic pre-collapse star clusters and the second detailed the physical feature of the model. Based on the works, the present work applies the solution to the observed structural profiles of Galactic globular clusters. For fitting to the profiles, the most fundamental (quasi-)stationary model, the King model, and the variants have shown successful results while they can not apply to core-collapsing and core-collapsed clusters at the late stage of the relaxation evolution. We propose an energy-truncated self-similar OAFP model that can apply to clusters at both the early and late stages of the evolution. This new model fits the structural profiles of at least half of Galactic globular clusters while it also applies to core-collapsed stars with resolved cores. As a main result, we provide the completion rate of core collapse against concentration for the clusters. Also, we show our new model can apply to the globular clusters even in a broad range of radii (0.01$\sim$10 arcminutes). However, since our model includes polytrope (elongated outer halo), the tidal radius of the model becomes unrealistically large for some clusters.  To avoid the issue, we also propose an approximated form of the new model. Lastly, we report that Milky Way globular clusters with low concentrations have the same spatial structures as stellar polytropes and discuss whether such polytropic cluster is a reasonable concept.

\end{abstract}


\end{frontmatter}

\section{Introduction}\label{sec:intro}

This is the third paper of a series of our works on the self-similar orbit-averaged Fokker-Planck (OAFP) equation for isotropic dense star clusters in pre-collapse phase. The first paper \citep{Ito_2020_1} showed an accurate Gauss-Chebyshev spectral solution of the equation and the second \citep{Ito_2020_2} detailed the physical feature of the model focusing on the negative heat capacity of the core of the model. Based on their results, the present paper proposes a phenomenological model that reasonably fits to the projected structural profiles of Galactic globular clusters that include not only normal (King model) clusters but also core-collapsed (or core-collapsing) clusters with resolved cores. In the rest of the present section, we review the applicability of the most fundamental fitting model (the King model \citep{King_1966}) to Galactic globular clusters (Section \ref{sec:app_King}) and also time-dependent OAFP model (Section \ref{sec:app_OAFP}) together with explaining the reason why the pre-collapse solution can even apply to post-collapsed clusters (Section \ref{sec:ssOAFP_model_PCC}).

\subsection{Applicability of King model to Galactic globular clusters}\label{sec:app_King}

The most fundamental fitting model for the structures of globular clusters is the King model \citep{King_1965} that can reasonably apply to globular clusters that are \emph{not} in the late stage of relaxation evolution. Fitting of the King model depends on the three numerical parameters; the central projected density $\Sigma_\text{c}$ (or surface brightness (SB)), core radius $r_\text{c}$ and dimensionless central potential $K(=\varphi(r=0)/\sigma_{c})$ where the central potential $\varphi(r=0)$ is normalized by the central velocity dispersion $\sigma_{c}$. Only with the three degrees of freedom, the King model well fits the surface brightness or projected density for approximately 80\% of globular clusters in Milky Way; the rest of 20\% is considered to be on the verge of core collapse or have undergone core collapse at least once \citep{Djorgovski_1986}. In this sense, the clusters that can be fitted by the King model is sometimes called 'normal' or 'King-model' (KM) clusters and those that can not is 'post-collapsed-core' or `post-core-collapse' (PCC)  clusters. The difference between the KM- and PCC- clusters are that (i) the projected structural profile of a typical KM cluster flattens in the core while a typical PCC cluster has a cusp with a power law approximately $r^{-1}$ and (ii) the concentrations of PCC clusters are high $c\gtrapprox 2.0$ while those of KM clusters are low $0.7\lessapprox c\lessapprox1.8$ \citep[See e.g.][]{Meylan_1997}.  The high concentration and cusp are considered as the signatures of post core collapse in the late stage of relaxation evolution.  

\subsection{Applicability of OAFP- and other models to PCC clusters}\label{sec:app_OAFP}

The structural profiles of PCC clusters may be fitted by time-dependent OAFP models for spherical clusters in post core-collapsed phase with realistic effects though, alternative models have been developed for homogeneous survey. It is not generally a easy task to self-consistently solve a time-dependent OAFP equation coupled to Poisson equation.  As application, a time-dependent OAFP model applies to a certain globular cluster as a case study to discuss its detail structure e.g. \citep{Murphy_2011} for NGC 7088, \citep{Drukier_1992} for NGC 6838 and \citep{Drukier_1995} for NGC 6397. On one hand, the King model is based only on Poisson equation easy to be solved for the potential of a spherical cluster and it is rather used in homogeneous survey to capture the common properties (the characteristic size of clusters and dynamical states) of as many globular clusters as possible by neglecting the detail information of each cluster. In fact, the concentration, core- and tidal- radii obtained from the King model have been the fundamental structural parameters in compilation works for globular-cluster studies \citep[e.g.][]{Peterson_1975,Trager_1995,Miocchi_2013,Merafina_2017} and in \cite[][(2010 edition)]{Harris_1996}'s catalog. To the best of our knowledge, up to date, there does not exist a single-mass isotropic model only based on Poisson equation that applies to both PCC- and KM- clusters due to their different core structures. For PCC clusters, a modified power-law profile \citep[e.g.][]{Lugger_1995,Ferraro_2003}  or non-parametric model \citep[e.g.][]{Noyola_2006} has been employed in place of the single-component King model, its variants \citep[e.g.][]{Woolley_1961, Wilson_1975} and generalized models \citep[e.g.][]{Gomez_Leyton_2014,Gieles_2015}\footnote{In addition to those generalization, some more recent models can readily integrate into our model, hence we do not discuss the detail relation of our model to those models. e.g. we do not consider the relation of our model to collisionless relaxation based on $f_\nu$ model \citep{de_Vita_2016} and the effect of escapers to discuss the elongated outer halos of some clusters \citep{Claydon_2019}.}. The present work focuses on a single-mass model for simplicity though, a multi-mass King model is known to be able to fit some PCC cluster \citep{King_1995} and there is no a strict argument that rules out such multi-mass King model from a proper model for PCC clusters \citep{Meylan_1997}. 

\subsection{Relation of the ss-OAFP model with PCC clusters}\label{sec:ssOAFP_model_PCC}

Due to its more proper description for the relaxation evolution of star clusters, we expect the ss-OAFP model can well model the structure of globular clusters in the center and inner halo at core-collapsing phase, compared to the King model. Not only this, but also we expect the structure of the ss-OAFP model is similar to those of PCC cluster under certain conditions, which motivated us to apply the ss-OAFP model to cluster structures. 

The ss-OAFP model for pre-collapse clusters may be able to fit even the projected structural profiles of PCC clusters due to the structural similarities between pre-collapse- and post-collapse- clusters. In principle, the ss-OAFP model in pre-collapse phase is the model that applies only to globular clusters at the moment of complete core collapse (with infinite central density) and approximately collapsing-core clusters in the late stage of relaxation evolution before the core completely collapses. Also, the ss-OAFP model itself is unrealistic in the sense that binaries halt the collapse in the core before the infinite density develops; the core can have a high density that is enough to form binaries from single stars. After the core collapse holds, time-dependent- and self-similar- conducting gaseous models predict that clusters successively repeat core expansion (due to the energy release from binaries) and core collapse (due to the two body relaxation with self-gravity) \citep{Sugimoto_1983, Bettwieser_1984,Goodman_1984, Goodman_1987}. This process is called 'gravothermal oscillation' in post core-collapse phase since it shows a nonlinear oscillation of core density with time. Time-dependent OAFP models \citep{Cohn_1989, Murphy_1990, Takahashi_1996} and $N$-body simulations \citep{Makino_1996,Breen_2012} also predict the oscillation.  \cite{Sugimoto_1983, Bettwieser_1984} found that the velocity dispersion seems approach the singular isothermal sphere in the outer halo while the core structure is alike the (non-singular) isothermal sphere (except at the moment of core collapse) against the result of the self-similar gaseous model with a central cusp \citep{Inagaki_1983}. The latter seems proper to describe PCC clusters reported in \citep[e.g.][]{Djorgovski_1986} though, formation of a cusp in the core is a conditional result. The self-similar gaseous model of \cite{Goodman_1984} showed that the core radius gets smaller with increasing $N$, which induces a cluster to be overstable or unstable and the cluster undergoes a gravothermal oscillation \emph{if} the cluster has enough stars in it ($N\gtrapprox7\times10^{3}$) as shown by \cite{Goodman_1987} based on the same model but with different functional forms of energy source or more efficient binary heating. The \cite{Goodman_1984}'s model results in forming a cusp in the core while \cite{Goodman_1987}'s model has a core like a non-singular isothermal sphere. In fact, to avoid unrealistically small- and large- core, efficient binary heating with primordial binaries is expected to occur, which still can form possibly resolved cores \citep{Goodman_1989}. In addition, the structural profile with a resolved core in the post-collapse phase of the gaseous model \citep{Goodman_1987} and OAFP model \citep{Takahashi_1996} are similar to the profiles for the corresponding pre-collapse core.  There is no way to differentiate the structural profiles in the two phases only from observational data \citep{Meylan_1997} unless one acquires accurate kinematic data to see the temperature inversion \footnote{A distinct difference between the OAFP models in core-collapsing and core-collapsed phases appears in the radial profile for velocity dispersion \citep{Sugimoto_1983}. When the core-collapsed core expands, the temperature (velocity dispersion) increases with radius near the center of cluster, which causes heat to flow inward toward the center and cools down the center. Kinematic survey, however, in general provides much larger uncertainty in velocity dispersion compared to structural data \citep{Meylan_1997}. Hence, one can not easily determine if a well-relaxed (or high-concentration) cluster is currently in pre-collapse- or post-collapse- phases.}. This infers that some PCC clusters with efficient binary heating may be modeled by the ss-OAFP model. This is a motivation to apply the (core-collapsing) ss-OAFP model to PCC clusters and the present purpose of our new model is, of course, to find structural parameters of PCC clusters with resolved core rather than establishing strict modeling of them.

The present paper proposes an energy-truncated ss-OAFP model that can fit the structural profiles of Galactic KM- and PCC- clusters \emph{with resolved cores} reported in \citep{Kron_1984, Djorgovski_1986, Trager_1995, Lugger_1995, Drukier_1993, Ferraro_2003, Noyola_2006, Miocchi_2013}. Since we did not have access to the data of \citep{Djorgovski_1986,Lugger_1995,Miocchi_2013}, we employed \emph{WebPlotDigitizer} \citep{Rohatgi_2019} to extract data points and their uncertainties for the projected density and surface brightness profiles depicted on figures of their works. 
The present paper is organized as follows. Section \ref{sec:finite_OAFP} introduces the energy truncated ss-OAFP model to be fitted to the projected structural profiles of Galactic globular clusters. Section \ref{sec:Fit_OAFP_PCC_example} explains the result of fitting the new model to PCC clusters.  Section \ref{sec:Tr_compl} shows the relationship between the completion rate of core collapse and concentration based on the results of the fitting of our model to KM- and PCC- clusters. Section \ref{sec:discuss_finite_OAFP} discusses application of our model with higher index $m$ to globular clusters in a broad range of radii, proposes the approximated form of the new model and suggests that low-concentration globular clusters may have structures described by stellar polytropes (or polytropic spheres) of index $m$. Section \ref{sec:conclusion_finite_OAFP} concludes the paper. For the sake of brevity, in Appendixes \ref{Appendix_OAFP_Fit}, \ref{Appendix_poly_Fit} and \ref{Appendix_OAFP_Fit_PCC}, we show the majority of the projected density profiles and surface brightness fitted by the energy-truncated ss-OAFP model.

\section{Energy-truncated ss-OAFP model}\label{sec:finite_OAFP}

The present section introduces an energy-truncated ss-OAFP model. First, Section \ref{sec:ssOAFP_model_ISO} shows the relation of the ss-OAFP model to the isothermal sphere with a motivation for truncating energy of the ss-OAFP model. Section \ref{sec:finite_OAFP_model} details the new model. The new model does not depend on dimensionless central potential $K$, hence Section \ref{sec:c_rk} explains how to regularize the concentration and core radius of the new model to compare the new model to the King model. Also, our model is composed of a polytrope of $m$ and the ss-OAFP model and we must choose the value of $m$ for our model, hence Section \ref{sec:choice_m} explains how we found an optimal value of $m$.

\subsection{Relation of the ss-OAFP model with the isothermal sphere}\label{sec:ssOAFP_model_ISO}

\begin{figure}[H]
	\centering
	\begin{tikzpicture}
	\begin{loglogaxis}[width=10cm,height=6cm, grid=major,xlabel=\Large{$\bigr$ },ylabel=\Large{$D$},xmin=1e-1,xmax=1e3,ymin=1e-5,ymax=3,legend pos=south west]
	\addplot [color = black ,mark=no,thick,densely dashed ] table[x index=0, y index=1]{Riso_Diso.txt};
	\addlegendentry{\normalsize{Isothermal sphere}} 
	\addplot [color = orange ,mark=no,thick, solid] table[x index=0, y index=1]{Rfp_Dfp.txt};
	\addlegendentry{\normalsize{ss-OAFP model}} 
	\end{loglogaxis}
	\end{tikzpicture}
	\caption{ Dimensionless densities $D(\bigr)$ of the isothermal sphere and ss-OAFP model.}
	\label{fig:D_iso_FP}
\end{figure}
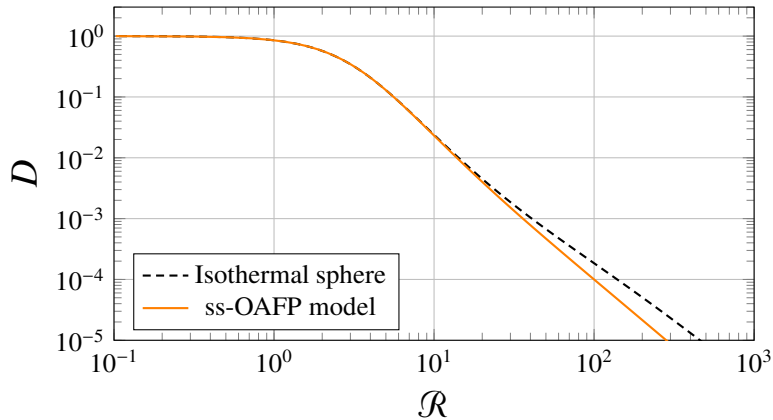

The ss-OAFP model can model KM clusters since the model has flat (quasi-)isothermal core that can be directly inferred from our numerical result of \citep{Ito_2020_2}.  There exists a similar structure in the cores between the ss-OAFP model and the isothermal sphere model as discussed in \citep{Ito_2020_2} for the local properties. The ss-OAFP model has almost the same morphology in density profile as that of the isothermal sphere as shown in Figure \ref{fig:D_iso_FP}. In the figure, the radius of the ss-OAFP model is rescaled by multiplying by $3.739$ so that the core size of the two models is approximately the same. This core structure infers that by properly truncating energy of the ss-OAFP model one may also obtain a model similar to the King model at small radii; we explain how to truncate the energy in Section \ref{sec:finite_OAFP_model}.

\subsection{Energy-truncated ss-OAFP model}\label{sec:finite_OAFP_model}

We energy-truncate the ss-OAFP model by considering the outer halo of our new model behaves like a polytrope of $m$, hence the model is to a \emph{phenomenological} model unlike the King model \citep{King_1966}.  The energy truncation of the King model is based on relatively-simple physical models and arguments using `test particle' method assuming particles (stars) other than the test particle (star) follow Maxwellian DF e.g. the King model may be a star cluster described by stationary FP model \citep{Spitzer_1958,Michie_1962,King_1965} and OAFP model \citep{Spitzer_1972} of isothermal sphere enclosed in a square well,  and the stellar DF proportional to $-E$ is an asymptotic stationary solution of the OAFP model with constant stellar flux at the fringe \citep{Spitzer_1972}\footnote{The stellar DF proportional $-E$ is the higher-energy limit of the lowered-Maxwellian DF.}. Our new model incorporates the effect of the escaping stars in a similar way to \citep{King_1966} by controlling high binding energy, but the mathematical operation for combining DF is opposite to \cite{King_1966}'s method. To obtain the King model (or lowered-Maxwellian DF), one must subtract DF for polytrope of $m=2.5$ from Maxwellian DF. On one hand, our energy-truncated ss-OAFP model adds DF for a polytrope of $m$ to the reference DF $F_\text{o}(E)$ for the ss-OAFP model (found in \citep{Ito_2020_1} as follows
\begin{align}
\tilde{F}(E)\equiv\frac{\rho_\text{c}}{4\sqrt{2}\pi \sigma_\text{c}^{3}}\frac{F_\text{o}(E)+\delta\, (-E)^{m-3/2}}{D_\text{o}(\varphi=-1)+\delta\, B(m-1/2,3/2)},\label{DF_finite_ss_OAFP}
\end{align}
where $\delta$ and $m$ are positive real numbers, $D_\text{o}(\varphi)$ is the reference density for the ss-OAFP model  (found in \citep{Ito_2020_1} and $B(a,b)$ is the beta function defined as $B(a,b)=2\int^{1}_{0}t^{2a-1}(1-t^{2})^{b-1}\,\text{d}t$ with $a>1/2$ and $b>1$.  The factor $1/(D_\text{o}(\varphi=-1)+\delta\, B(m-1/2,3/2))$ is inserted in the DF so that the density profile for the DF $\tilde{F}(E)$ has a certain central density $\rho_\text{c}$ as $R\to0$. This new DF behaves like the ss-OAFP model beyond order of $\delta$ in equation \eqref{DF_finite_ss_OAFP} while below $\delta$ it is approximately a polytropic sphere of index $m$. Yet, the value of $m$ must be further fixed\footnote{Since the index $m$ in our model is to be determined at first place, our model has the same parameter-dependence as the truncated $\gamma$ exponential (fractional-power) model proposed in \citep{Gomez_Leyton_2014} in the sense that the outer halo is controlled by a polytropic sphere.}  based on physical arguments  (with numerical experiments) and from observational data. In this work, we do only the latter process  (Section \ref{sec:choice_m}). In this sense, we consider our model a phenomenological model.

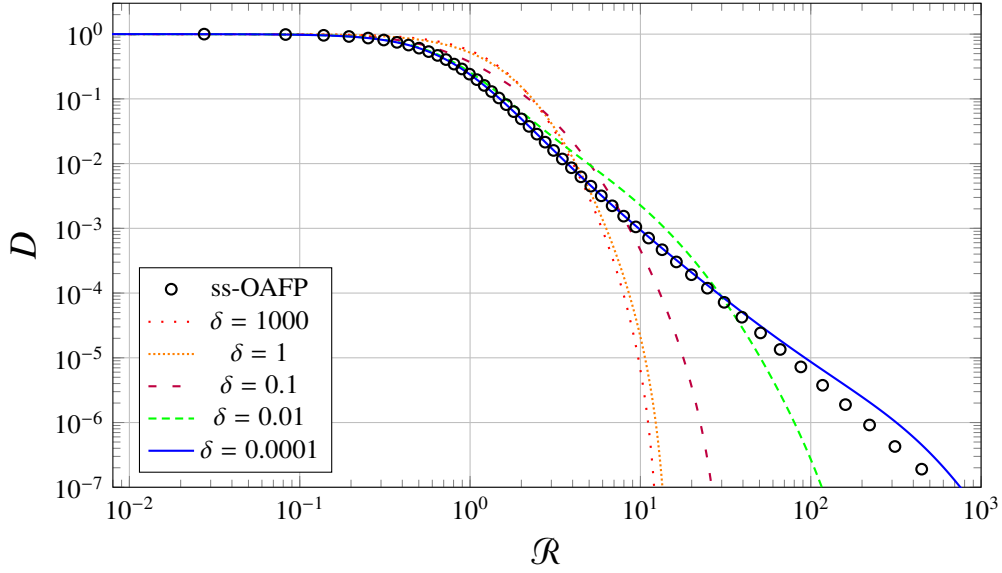
\begin{figure}[H]
	\centering
	\begin{tikzpicture}
	\begin{loglogaxis}[width=13cm,height=8cm, grid=major,xlabel=\Large{$\bigr$},ylabel=\Large{$D$},xmin=8e-3,xmax=1e3,ymin=1e-7,ymax=3,legend pos=south west]
	\addplot [only marks,color = black ,mark=o,thick, solid ] table[x index=0, y index=1]{DssOAFP.txt};
	\addlegendentry{\normalsize{ss-OAFP}} 
	\addplot [color = red ,mark=no,thick,loosely dotted] table[x index=0, y index=1]{DfssOAFP_del1000.txt};
	\addlegendentry{\normalsize{$\delta=1000$}} 
	\addplot [color = orange ,mark=no,thick,densely dotted] table[x index=0, y index=1]{DfssOAFP_del1.txt};
	\addlegendentry{\normalsize{$\delta=1$}}
	\addplot [color = purple ,mark=no,thick,loosely dashed] table[x index=0, y index=1]{DfssOAFP_del01.txt};
	\addlegendentry{\normalsize{$\delta=0.1$}}  
	\addplot [color = green ,mark=no,thick,densely dashed ] table[x index=0, y index=1]{DfssOAFP_del001.txt};
	\addlegendentry{\normalsize{$\delta=0.01$}}  
	\addplot [color = blue ,mark=no,thick,solid] table[x index=0, y index=1]{DfssOAFP_del00001.txt};
	\addlegendentry{\normalsize{$\delta=0.0001$}} 
	\end{loglogaxis}
	\end{tikzpicture}
	\caption{Dimensionless density $D(\bigr)$ of the energy-truncated ss-OAFP model for different $\delta$. The corresponding profile of the ss-OAFP model is also depicted.}
	\label{fig:density_fssOAFP}
\end{figure}

\begin{figure}[H]
	\centering
	\begin{tikzpicture}
	\begin{loglogaxis}[width=13cm,height=8cm, grid=major,xlabel=\Large{$\bigr$},ylabel=\Large{$\bar{\varphi}$},xmin=1e-3,xmax=1e6,ymin=3e-5,ymax=3,legend pos=south west]
	\addplot [only marks,color = black ,mark=o,thick, solid ] table[x index=0, y index=1]{PhissOAFP.txt};
	\addlegendentry{\normalsize{ss-OAFP}} 
	\addplot [color = red ,mark=no,thick,loosely dotted] table[x index=0, y index=1]{PhifssOAFP_del1000.txt};
	\addlegendentry{\normalsize{$\delta=1000$}} 
	\addplot [color = orange ,mark=no,thick,densely dotted] table[x index=0, y index=1]{PhifssOAFP_del1.txt};
	\addlegendentry{\normalsize{$\delta=1$}}
	\addplot [color = purple ,mark=no,thick,loosely dashed] table[x index=0, y index=1]{PhifssOAFP_del01.txt};
	\addlegendentry{\normalsize{$\delta=0.1$}}  
	\addplot [color = green ,mark=no,thick,densely dashed] table[x index=0, y index=1]{PhifssOAFP_del001.txt};
	\addlegendentry{\normalsize{$\delta=0.01$}}  
	\addplot [color = blue ,mark=no,thick,densely dashdotted] table[x index=0, y index=1]{PhifssOAFP_del0001.txt};
	\addlegendentry{\normalsize{$\delta=0.001$}} 
	\addplot [color = black ,mark=no,thick,solid] table[x index=0, y index=1]{PhifssOAFP_del000001.txt};
	\addlegendentry{\normalsize{$\delta=0.00001$}} 
	\end{loglogaxis}
	\end{tikzpicture}
	\caption{Dimensionless potential $\bar{\varphi}(\bigr)$ of  the energy-truncated ss-OAFP model. The corresponding potential of the ss-OAFP model is also depicted.}
	\label{fig:potential_fssOAFP}
\end{figure}
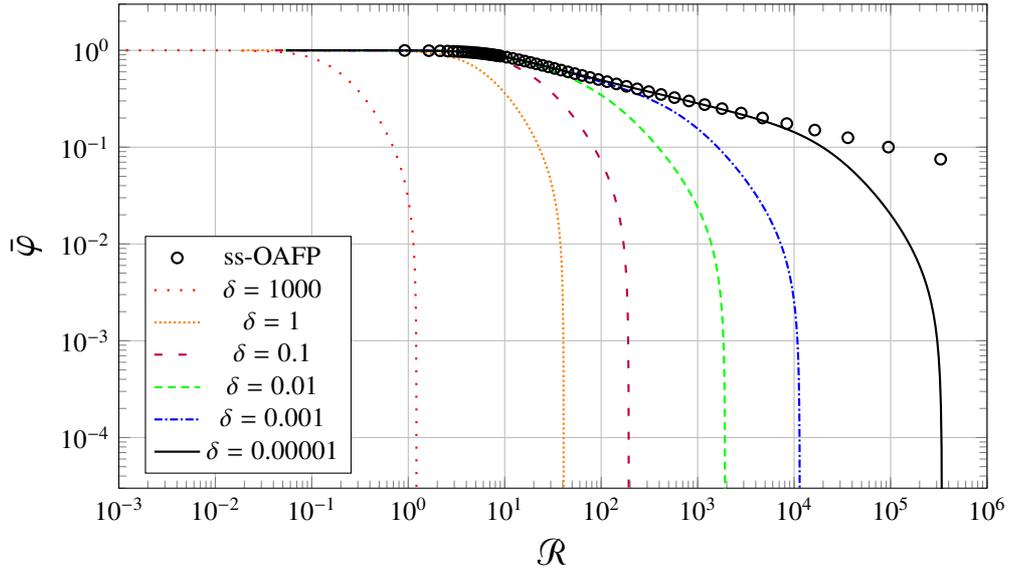

The rest of the present section shows numerically calculated density profile, m,f. potential and projected density profile for the energy-truncated ss-OAFP model. One can analytically derive the explicit form of the density profile for polytropes, hence the density for our new truncated model reads
\begin{align}
D(\varphi)=\rho_\text{c}\frac{D_\text{o}(\varphi)+\delta\, B(m-1/2,3/2)\left(-\varphi\right)^{m}}{D_\text{o}(\varphi=-1)+\delta\, B(m-1/2,3/2)}.
\end{align}
Poisson equation for potential $\varphi(r)$ of an isotropic spherical cluster reads
\begin{align}
\frac{\text{d}^{2} \varphi }{\text{d} r^{2} }+\frac{2}{r}\frac{\text{d} \varphi }{\text{d} r }=\rho\left[\varphi(r)\right]\equiv16\pi^{2}GD(\varphi).\label{Poisson_eqn}
\end{align}
where $G$ is the gravitational constant. The dimensionless form of the Poisson equation reads
\begin{align}
\frac{\text{d}^2\bar{\varphi}}{\text{d}\bar{r}^2}+\frac{2}{\bar{r}}\frac{\text{d}\bar{\varphi}}{\text{d}\bar{r}}=\bar{\rho}=\frac{D_\text{o}(\bar{\varphi})+\delta\, B(m-1/2,3/2)\left(-\bar{\varphi}\right)^{m}}{D_\text{o}(\bar{\varphi}=-1)+\delta\, B(m-1/2,3/2)},\label{Eq_Etrunc_Pois}
\end{align}
where the potential $\varphi(r)$, radius $r$ and density $\rho(r)$ are made in dimensionless form using equations 
\begin{subequations}
	\begin{align}
	&\bar{\varphi}(r)=-\frac{\varphi(r)}{\sigma^{2}_{c}},\label{Eq.bar_phi}\\
	&\bar{r}= r\sqrt{\frac{4\pi G\rho_{c}}{\sigma^{2}_{c}}},\label{Eq.bar_r}\\
	&\bar{\rho}(r)= \frac{\rho(r)}{\rho_{c}}.\label{Eq.bar_rho}
	\end{align}
\end{subequations}
where the variables with subscript $c$ corresponding to the time-dependent variables in self-similar analysis \citep{Ito_2020_1} and the variables at a certain time $t_c$ of \citep{Ito_2020_2}. The boundary conditions for Poisson equation \eqref{Eq_Etrunc_Pois} are
\begin{equation}
\bar{\varphi}(\bar{r}=0)=1, \qquad \frac{\text{d}\bar{\varphi}}{\text{d}\bar{r}}(\bar{r}=0)= 0. \label{BC_Etrunc_FP}
\end{equation}
Since $\bar{\varphi}$ is an independent variable for the ss-OAFP model, following the method of inverse mapping \citep{Ito_2018}, we solved the Poisson equation in its inverse form for $\bigr(\bar{\varphi})$ 
\begin{align}
\bigr\,\frac{\,\text{d}^2 \bigr}{\,\text{d}\bar{\varphi}^2}-2\left(\frac{\,\text{d}\bigr}{\,\text{d}\bar{\varphi}}\right)^{2}=\left(\frac{\,\text{d}\bigr}{\,\text{d}\bar{\varphi}}\right)^{3}\frac{D_\text{o}(\bar{\varphi})+\delta\, B(m-1/2,3/2)\left(-\bar{\varphi}\right)^{m}}{D_\text{o}(\bar{\varphi}=-1)+\delta\, B(m-1/2,3/2)}.
\end{align}
The numerical integration of the Poisson equaiton provided the density profile (Figure \ref{fig:density_fssOAFP}) and  m.f. potential (Figure  \ref{fig:potential_fssOAFP}) for an optimal index $m=3.9$. (The reason why $m=3.9$ is an optimal choice is explained in Section \ref{sec:choice_m}.) In the figures the value of $\delta$ spans $10^{-5}$ through $10^{3}$. For large $\delta>1$, the profiles show almost the same morphology since they behave like a polytrope of $m=3.9$. On one hand, the profiles approach the ss-OAFP model for small $\delta\,(\lessapprox10^{-2})$.

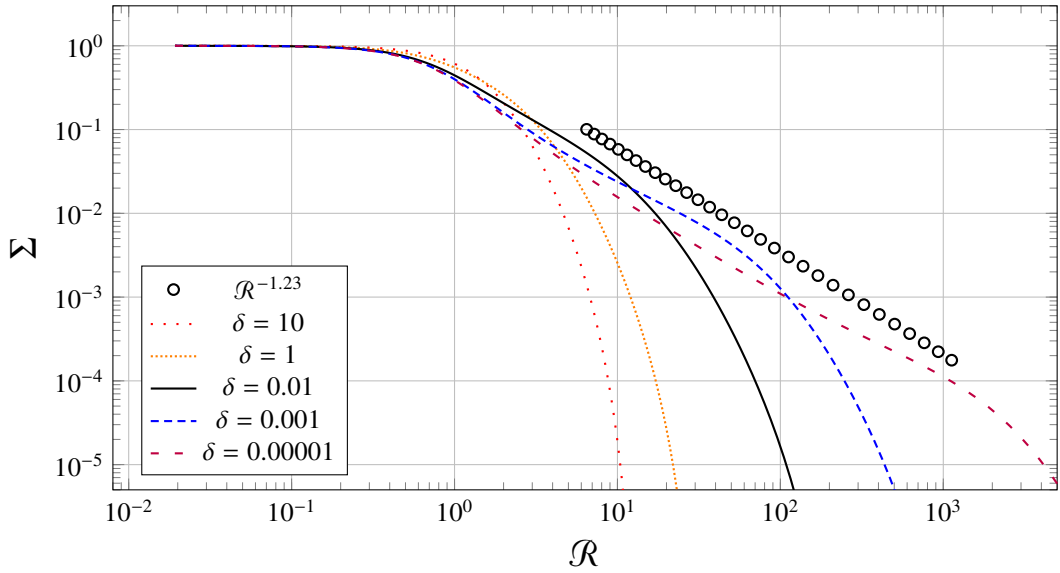
\begin{figure}[H]
	\centering
	\begin{tikzpicture}
	\begin{loglogaxis}[width=14cm,height=8cm, grid=major,xlabel=\Large{$\bigr$},ylabel=\Large{$\Sigma$},xmin=8e-3,xmax=5e3,ymin=5e-6,ymax=3,legend pos=south west]
	\addplot [only marks,color = black ,mark=o,thick, solid ] table[x index=0, y index=1]{R_sig_fssOAFPasymp.txt};
	\addlegendentry{\normalsize{$\bigr^{-1.23}$}} 
	\addplot [color = red ,mark=no,thick,loosely dotted] table[x index=0, y index=1]{R_sigFssOAFPdel_del10.txt};
	\addlegendentry{\normalsize{$\delta=10$}} 
	\addplot [color = orange ,mark=no,thick,densely dotted] table[x index=0, y index=1]{R_sigFssOAFPdel_del01.txt};
	\addlegendentry{\normalsize{$\delta=1$}}
	\addplot [color = black ,mark=no,thick, solid] table[x index=0, y index=1]{R_sigFssOAFPdel_del001.txt};
	\addlegendentry{\normalsize{$\delta=0.01$}}  
	\addplot [color = blue,mark=no,thick,densely dashed ] table[x index=0, y index=1]{R_sigFssOAFPdel_del0001.txt};
	\addlegendentry{\normalsize{$\delta=0.001$}}  
	\addplot [color = purple ,mark=no,thick,loosely dashed] table[x index=0, y index=1]{R_sigFssOAFPdel_del000001.txt};
	\addlegendentry{\normalsize{$\delta=0.00001$}} 
	\end{loglogaxis}
	\end{tikzpicture}
	\caption{Projected density $\Sigma$ of  the energy-truncated ss-OAFP model for different $\delta$. The power-law $R^{-1.23}$ corresponds to the asymptotic approximation of the ss-OAFP model as $R\to\infty$.}
	\label{fig:Projected_density_fssOAFP}
\end{figure}

For application of density profile to globular clusters, one needs to convert the density profile $D(\bar{\varphi})$ to the projected density profile using the following expression
\begin{align}
\Sigma(r)=2\int^{\infty}_{0}\frac{D(\varphi)}{\sqrt{1-(r/r')^{2}}}\text{d}r',
\end{align}
and the corresponding inverse form with dimensionless variables is 
\begin{align}
\Sigma(\bar{\varphi})=-2\int^{\bar{\varphi}}_{0}  \sqrt{\frac{1-\mu_{\bigr}(\bar{\varphi},\bar{\varphi}')}{1+\mu_{\bigr}(\bar{\varphi},\bar{\varphi}')}}\left[ -2\frac{\text{d}D}{\text{d}\bar{\varphi}'}\frac{\bigr(\bar{\varphi}')}{\bigs(\bar{\varphi}')}+D(\bar{\varphi}')\bigs(\bar{\varphi}')\frac{1+2\mu_{\bigr}(\bar{\varphi},\bar{\varphi}')}{1+\mu_{\bigr}(\bar{\varphi},\bar{\varphi}')} \right]\text{d}\bar{\varphi}',
\end{align}
where $\mu_{\bigr}(\bar{\varphi},\bar{\varphi}')\equiv\bigr(\bar{\varphi})/\bigr(\bar{\varphi})$ and $\bigs\equiv-\text{d}\bigr/\text{d}\bar{\varphi}(<0)$. Figure \ref{fig:Projected_density_fssOAFP} depicts the projected density pofiles for different $\delta$. As $\delta$ decreses the slope of $\bigr^{-1.23}$ in the inner halo develops more clearly (as expected from the asymptotic density profile for the ss-OAFP model; $D\propto \bigr^{-2.23}$). This power law occurrs at radii between $\bigr\sim10$ and $\bigr\sim100$ for $\delta=10^{-4}$ though, one also can find a similar power law for larger $\delta$. For $\delta=10^{-2}$ and $10^{-3}$, $\Sigma$ shows power-law-like structures $\bigr^{-1.0}\sim \bigr^{-1.1}$ at radii between $\bigr\sim1$ and $\bigr\sim10$. This property is a desirable feature to fit our model to the projected density profiles od PPC clusters whose projected desnity has similar power law profiles near the core \citep[e.g][]{Djorgovski_1986,Lugger_1995}. 

\subsection{Regularization for the concentration and King radius}\label{sec:c_rk}

\begin{figure}[H]
	\centering
	\begin{tikzpicture}
	\begin{semilogxaxis}[width=11cm,height=6cm, grid=major,xlabel=\Large{$\delta$},ylabel=\Large{$c$},xmin=1e-3,xmax=1e2,ymin=0.6,ymax=3.6,legend pos=north east]
	\addplot [color = black ,mark=no,thick, solid] table[x index=0, y index=1]{mudel_c_cPoly4.txt};
	\addlegendentry{\normalsize{Finite ss-OAFP}} 
	\addplot [color = orange ,mark=no,thick,densely dashed] table[x index=0, y index=2]{mudel_c_cPoly4.txt};
	\addlegendentry{\normalsize{Polytrope of $m=3.9$}} 
	\end{semilogxaxis}
	\end{tikzpicture}
	\caption{Concentration of  the energy-truncated ss-OAFP model. The horizontally dashed line represents the concentration for polytropic sphere of $m=3.9$.}
	\label{fig:mudel_c_cPoly4}
\end{figure}
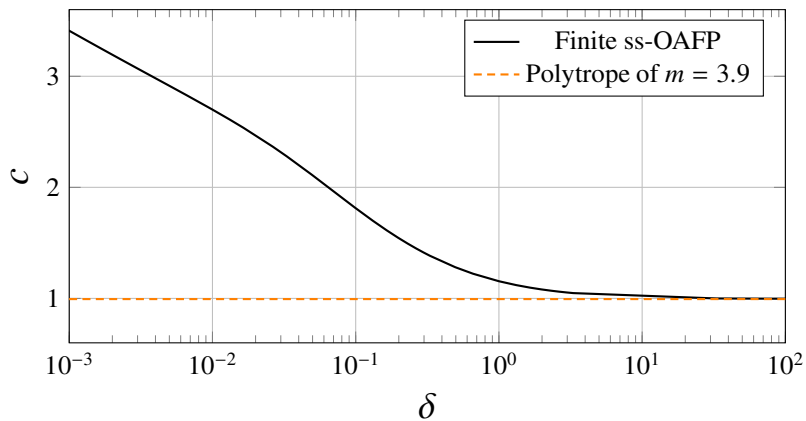

The energy-truncated ss-OAFP model is different from the King model in the sense of how concentration and core radius depend on the dimensionless central potential $K$, hence one must properly regularize the structural parameters for comparison. The King model \citep{King_1966} may be wirtten in the following form by regularizing the m.f. potential as $\bar{\bar{\varphi}} \equiv\bar{\varphi}/K$ 
\begin{align}
\frac{\text{d}^2\bar{\bar{\varphi}}}{\text{d}\bar{r}^2}+\frac{2}{\bar{r}}\frac{\text{d}\bar{\bar{\varphi}}}{\text{d}\bar{r}}-\frac{1}{K}\frac{I(\bar{\bar{\varphi}})}{I(1)}=0,
\end{align}
where the function $I(x)$ is
\begin{align}
I(x)=\exp(x)\text{erf}\left(\sqrt{x}\right)-\sqrt{\frac{4x}{\pi}}\left[1+\frac{2x}{3}\right],
\end{align}
where $\text{erf}(x)$ is the error function; $\text{erf}(x)=2\exp\left[{x^{2}}\right]\int^{x}_{0}\exp\left[{-x^{2}}\right]\text{d}x/\sqrt{\pi}$. Due to the $K$-dependence of the equation, as $K\to0$ the concentration is also $c\to0$. Of course, the minimum radius of the King model can be the tidal radius of polytrope of $m=2.5$, that is $5.355275459...$ \citep[e.g.][]{Boyd_2011} if one regularizes the radius as $\bar{r}=\sqrt{K}\bar{\bar{r}}$. On one hand, the ss-OAFP model dose not depend on $K$.  To find the same value of concentration (if necessary) or at least the same order as that of the King model, one must regularize the core radius and concentration\footnote{The tidal radius for the energy-truncated ss-OAFP model should not be regularized since it is still the radius at which the projected density reaches zero which can be found after the model is properly fitted to the projected structural profile of a globular cluster on graph.} as follows
\begin{subequations}
	\begin{align}
	&\bar{r}_\text{Kin}\equiv\frac{ r_\text{Kin}}{\sqrt{K^\text{(m)}}},\\
	&\bar{c}\equiv \log\left[ \frac{r_\text{tid}}{r_\text{King}}\sqrt{K^\text{(m)}}\right],
	\end{align}
\end{subequations}
where  $K^\text{(m)}$ is the dimensionless central potential that can be given when the the tidal radius divided by $\sqrt{K}$ of the King model is approximately the same as that of polytrope of $m$. Using this regularization, one can obtain the concentration of the energy-truncated ss-OAFP model. For example, if $m=3.9$ is chosen, then the tidal radius of the polytrope of $m=3.9$ is $13.4731$. In this case, $K$ of the King model must be chosen so that the tidal radius divided by $\sqrt{K}$ of the King model is close to 13.4731. This can be achieved when $K=4.82$ with the tidal radius 13.444 (based on our calculation). Hence, $K^\text{(m)}=4.82$ for $m=3.9$. (We show in Section \ref{sec:choice_m} that the concentrations calculated by this scaling are reasonably close to those of the King model.) Figure \ref{fig:mudel_c_cPoly4} depicts the concentration $\bar{c}$ for $m=3.9$. As $\delta$ increases, the concentration approaches a constant value that is corresponding to the concentration of polytrope of $m=3.9$. The present focus is $\delta<c_{4}^{*}(=0.3032)$ with which the energy-truncated ss-OAFP model approaches the ss-OAFP model and differentiates itself from the isothermal sphere and King model. The corresponding concentration is $\bar{c}>1.45$. On one hand, our model is expected to behave like the King model for $1<\bar{c}<1.45$ and like a polytrope of $m=3.9$ in the limit of $\bar{c}\to1$.

\subsection{An optimal choice for polytropic index: $m=3.9$}\label{sec:choice_m}

We determined the index $m$ to be 3.9 in the energy-truncated ss-OAFP model after having preliminarily applied the model to the projected surface densities of six KM clusters and a PCC cluster that we chose. For application of our model to KM clusters, we chose an optimal value for the index $m$ so that our model gives the same order of the tidal radius, King radius and exponential of concentration of six chosen KM clusters as those reported in the previous works. Initially, we expected $m=2.5$ could be an optimal choice for the energy-truncated ss-OAFP model following \citep{Spitzer_1972} though, it was not the case. The useful values were found in $3.5<m<4.4$ with which our model reasonably fits the projected surface densities for KM clusters reported in \citep{Kron_1984,Noyola_2006,Miocchi_2013} (See Appendix \ref{Appendix_OAFP_Fit}). Among the values of $m$, we chose 3.9 as an optimal value for the present work. This is since it provides the same order of magnitudes of structural parameters for six chosen clusters as that of the existing works based on the King model, as shown in Table \ref{table:core_tidal_radius}. The table provides the results for fitting of our model with $m=3.9$ to six Milky Way globular clusters reported in \citep{Kron_1984}. Only the six clusters are reported in all the compilation works \citep{Peterson_1975, Kron_1984, Chernoff_1989, Trager_1993, Miocchi_2013} that we chose to compare this time. \cite{Miocchi_2013} did not show the numeric values of projected density profiles but the six clusters in Table \ref{table:core_tidal_radius} have approximately the same maximum radius points of the projected density profiles on graphs reported in \citep{Kron_1984}. While many of the compilation works based on the King model are inhomogeneous surveys in the sense they depend on different instruments, photometry methods and statistical analyses, the structural parameters obtained from our model are reasonably close to the results of the compilation works. On one hand, If $m<3.8$ or $m>4.3$ is chosen for the fitting, the magnitudes of the structural parameters are less or greater by over a factor of ten from the compilation works' results. Interestingly, our structural parameters are close to those obtained from the King model rather than the Wilson model \citep{Wilson_1975}; the latter has a higher index $m$ than the former in the limit of $K\to0$. The Wilson model \citep{Wilson_1975} relies on the polytrope of $m=3.5$ in the limit $K\to0$ which provides greater values of the structural parameters since the polytrope of $m=3.5$ reaches further in radius compared to the polytrope of $m=2.5$ that the King model follows in the same limit \citep[e.g.][]{Chandra_1939}. The reason why our model does not overestimate the structural parameters with high $m(=3.9)$ would be that the density profile of the ss-OAFP model more rapidly decays compared to the isothermal sphere in the inner halo (Figure \ref{fig:D_iso_FP}). One can find in Appendix \ref{Appendix_OAFP_Fit} the energy-truncated ss-OAFP model with $m=3.9$ reasonably fitted to the projected structural profiles of KM clusters reported in \citep{Kron_1984,Noyola_2006,Miocchi_2013}.

\begin{table*}\centering
	\ra{1.3}
	\scalebox{0.8}{
		\begin{tabular}{@{}l|llr|llr|llr@{}}\toprule
			&\multicolumn{3}{l|}{NGC 1904}    & \multicolumn{3}{l|}{NGC 2419} &\multicolumn{3}{l}{NGC 6205} \\ 
			\midrule
			&$c$ &$r_\text{c}$&$r_\text{tid}$&$c$ &$r_\text{c}$&$r_\text{tid}$&$c$ &$r_\text{c}$&$r_\text{tid}$\\
			\midrule
			\textbf{Finite ss-OAFP}&    &          &              &    &            &              &    &     &    \\
			The present work based on&1.86 &0.191  &     13.9     &1.24 &  0.410    &    7.16      &1.54 &0.779 &27.0\\
			data of \citep{Kron_1984}&    &        &              &    &            &              &    &     &    \\
			\midrule
			\textbf{King model}  &    &            &              &    &            &              &    &     &    \\
			\citep{Miocchi_2013} &1.76&0.15        &   9.32       &1.51&  0.27      &    9         &1.32&0.825&18.5\\
			\citep[][(2010 edition)]{Harris_1996}  &1.70&0.16        &   8.0        &1.37&  0.32      &    7.5      &1.53&0.62 &21.0\\
			\citep{Trager_1993}  &1.72&0.159       &     8.35     &1.4 &  0.348     &    8.74       &1.49&0.875&27.0\\
			\citep{Chernoff_1989}&1.90&0.132       &     10.5     &1.6 &  0.373     &    14.8      &1.35&0.745&16.7\\		
			\citep{Kron_1984}&1.75&0.178        &     10.0        &1.00&  0.398      &   3.98        &1.25&0.83 &14.8\\		
			\citep{Peterson_1975}&1.60&0.27        &     10.7     &1.41&  0.42      &    10.7      &1.55&0.76 &26.9\\
			\midrule
			\textbf{Wilson model}&    &            &              &    &            &              &    &     &    \\
			\citep{Miocchi_2013} &2.14&0.18        &      28      &1.73&  0.32      &    20        &1.77&0.841&57\\
			\bottomrule
	\end{tabular}}
	
	\vspace{0.7cm}
	
	\scalebox{0.8}{
		\begin{tabular}{@{}l|llr|llr|llr@{}}
			\toprule
			&\multicolumn{3}{l|}{NGC 6229}    & \multicolumn{3}{l|}{NGC 6341} &\multicolumn{3}{l}{NGC 6864} \\ 
			\midrule
			&$c$ &$r_\text{c}$&$r_\text{tid}$&$c$ &$r_\text{c}$&$r_\text{tid}$&$c$ &$r_\text{c}$&$r_\text{tid}$\\
			\midrule
			\textbf{Finite ss-OAFP}&    &            &              &    &            &              &    &     &    \\
			The present work based on    &1.45 &0.178        &     5.00     &1.68 &  0.314      &    15.0      &1.83 &0.116 &7.85\\
			data of \citep{Kron_1984}&   &            &              &    &            &              &    &     &    \\
			\midrule        
			\textbf{King model}  &    &            &              &    &            &              &    &     &    \\          
			\citep{Miocchi_2013} &1.65&0.13        &   6.12       &1.74&  0.243     &    13.9      &1.79&0.082&5\\
			\citep[][(2010 edition)]{Harris_1996}  &1.50&0.12        &   3.79       &1.68&  0.26      &    12.4      &1.80&0.09 &6.19\\
			\citep{Trager_1993}  &1.61&0.13        &     5.39     &1.81&  0.235     &    15.2      &1.88&0.096&5.68\\
			\citep{Chernoff_1989}&1.40&0.167       &     4.19     &1.70&  0.132     &    6.64      &1.85&0.084&5.91\\
			\citep{Kron_1984}&1.25&0.173        &     3.08        &1.50&  0.308      &   9.75       &1.75&0.095 &5.34\\
			\citep{Peterson_1975}&1.41&0.22        &     5.62     &1.78&  0.275     &    16.6      &N/A &0.12 &$>3.2$\\
			\midrule
			\textbf{Wolley model}&    &            &              &    &            &              &    &     &    \\
			\citep{Miocchi_2013} &1.82&0.16        &      12.0    &2.17&  0.33      &    46        &2.38&0.095&25\\
			\bottomrule
		\end{tabular}
	}
	\caption{Concentration and core- and tidal- radii obtained from the energy-truncated ss-OAFP model. The structural parameters are compared to the previous compilation works based on the King- and Wilson- models.}
	\label{table:core_tidal_radius}
\end{table*}

\begin{figure}[H]
	\centering
	\begin{tikzpicture}
	\begin{axis}[width=7.5cm,height=8cm, grid=major,xmin=-2.5,xmax=1.8,ymin=-4.5,ymax=0.3,ylabel=\normalsize{$\log[\Sigma]$},legend style={cells={align=left}},legend entries={\normalsize{ss-OAFP}\\\normalsize{$(m=3.0)$},\normalsize{modified power-law}, \normalsize{NGC 6752}},yticklabel style={/pgf/number format/.cd,fixed zerofill,precision=1},legend pos=south west]
	\addlegendimage{no markers, orange};
	\addlegendimage{no markers, black,dashed};
	\addlegendimage{only marks, mark=ball, blue};
	\addplot [color = orange ,mark=no,thick,solid] table[x index=0, y index=1]{R_Sig_NGC6752_m3.txt};
	\addplot[color = blue, only marks, mark=ball, mark options={blue}, error bars,y dir=both, y explicit] table [x index=0, y index=1, y error index=2]{Rd_Sigd_Errbar_Err_NGC6752_m3.txt};
	\addplot [color = black ,mark=no,thick,densely dashed] table[x index=0, y index=1]{NGC6752_LogR_LogSig_guide.txt};
	\end{axis}
	\end{tikzpicture}\hspace{0.3cm}
	\begin{tikzpicture}
	\begin{axis}[width=7.5cm,height=8cm, grid=major,xmin=-2.5,xmax=1.8,ymin=-4.5,ymax=0.3,legend style={cells={align=left}},legend entries={\normalsize{ss-OAFP}\\\normalsize{$(m=3.5)$}, \normalsize{NGC 6752}},yticklabel style={/pgf/number format/.cd,fixed zerofill,precision=1},legend pos=south west]
	\addlegendimage{no markers, orange};
	\addlegendimage{only marks, mark=ball, blue};
	\addplot [color = orange ,mark=no,thick,solid] table[x index=0, y index=1]{R_Sig_NGC6752_m35.txt};
	\addplot[color = blue, only marks, mark=ball, mark options={blue}, error bars,y dir=both, y explicit] table [x index=0, y index=1, y error index=2]{Rd_Sigd_Errbar_Err_NGC6752_m35.txt};
	\end{axis}
	\end{tikzpicture}

	\vspace{0.6cm}
	
	\begin{tikzpicture}
	\begin{axis}[width=7.5cm,height=8cm, grid=major,xmin=-2.5,xmax=1.8,ymin=-4.5,ymax=0.3,ylabel=\normalsize{$\log[\Sigma]$},legend style={cells={align=left}},legend entries={\normalsize{ss-OAFP}\\\normalsize{$(m=3.9,\,\bar{c}=2.56)$}, \normalsize{NGC 6752}},yticklabel style={/pgf/number format/.cd,fixed zerofill,precision=1},legend pos=south west]
	\addlegendimage{no markers, orange};
	\addlegendimage{only marks, mark=ball, blue};
	\addplot [color = orange ,mark=no,thick,solid] table[x index=0, y index=1]{R_Sig_NGC6752_m39.txt};
	\addplot[color = blue, only marks, mark=ball, mark options={blue}, error bars,y dir=both, y explicit] table [x index=0, y index=1, y error index=2]{Rd_Sigd_Errbar_Err_NGC6752_m39.txt};
	\node[black] at (-0.9,-2.8) {(\normalsize{$r_\text{c}=0\textquotesingle.149, \, \delta=0.0044$})} ; 
	\end{axis}
	\end{tikzpicture}\hspace{0.3cm}
	\begin{tikzpicture}
	\begin{axis}[width=7.5cm,height=8cm, grid=major,xmin=-2.5,xmax=1.8,ymin=-4.5,ymax=0.3,legend style={cells={align=left}},legend entries={\normalsize{ss-OAFP}\\\normalsize{$(m=4.2, \,\bar{c}=2.46)$}, \normalsize{NGC 6752}},yticklabel style={/pgf/number format/.cd,fixed zerofill,precision=1},legend pos=south west]
	\addlegendimage{no markers, orange};
	\addlegendimage{only marks, mark=ball, blue};
	\addplot [color = orange ,mark=no,thick,solid] table[x index=0, y index=1]{R_Sig_NGC6752_m427.txt};
	\addplot[color = blue, only marks, mark=ball, mark options={blue}, error bars,y dir=both, y explicit] table [x index=0, y index=1, y error index=2]{Rd_Sigd_Errbar_Err_NGC6752_m427.txt};
	\node[black] at (-0.9,-2.8) {(\normalsize{$r_\text{c}=0\textquotesingle.188, \, \delta=0.0079$})} ; 
	\end{axis}
	\end{tikzpicture}
	
	\vspace{0.3cm}
	
	\begin{tikzpicture}
	\begin{axis}[width=7.5cm,height=4cm, grid=major,xmin=-2.5,xmax=1.8,ymin=-0.8,ymax=0.8,xlabel=\normalsize{$\log[R\text{(arcmin)}]$ },ylabel=\normalsize{$\Delta\log[\Sigma]$},legend style={cells={align=left}},legend entries={\normalsize{ $m=3.9$}},legend pos=south west]
	\addlegendimage{only marks, mark=ball, blue};
	\addplot[color = blue, only marks, mark=ball, mark options={blue}, error bars,y dir=both, y explicit] table [x index=0, y index=3, y error index=2]{Rd_Sigd_Errbar_Err_NGC6752_m39.txt};
	\draw[dashed] (-2.5,0)--(1.0,0);
	\end{axis}
	\end{tikzpicture}\hspace{0.3cm}
	\begin{tikzpicture}
	\begin{axis}[width=7.5cm,height=4cm, grid=major,xmin=-2.5,xmax=1.8,ymin=-0.8,ymax=0.8,xlabel=\normalsize{$\log[R\text{(arcmin)}]$ },legend style={cells={align=left}},legend entries={\normalsize{ $m=4.2$}},legend pos=south west]
	\addlegendimage{only marks, mark=ball, blue};
	\addplot[color = blue, only marks, mark=ball, mark options={blue}, error bars,y dir=both, y explicit] table [x index=0, y index=3, y error index=2]{Rd_Sigd_Errbar_Err_NGC6752_m42.txt};
	\draw[dashed] (-2.5,0)--(1.0,0);
	\end{axis}
	\end{tikzpicture}
	\caption{Fitting of energy-truncated ss-OAFP model to the projected density profile of NGC 6752 \citep{Ferraro_2003} for different $m$. The unit of $\Sigma$ is number per square of arcminutes and $\Sigma$ is normalized so that the density is unity at the smallest radius of data points. In the legends, (c) means PCC cluster as judged so in \citep{Ferraro_2003}. In the left top panel, double-power law profile is depicted as done in \citep{Ferraro_2003}. In the two bottom panels, $\Delta\log[\Sigma]$ for $m=3.9$ and $m=4.2$ depicts the corresponding deviation of $\Sigma$ from our model.}
	\label{fig:NGC6752}
\end{figure}
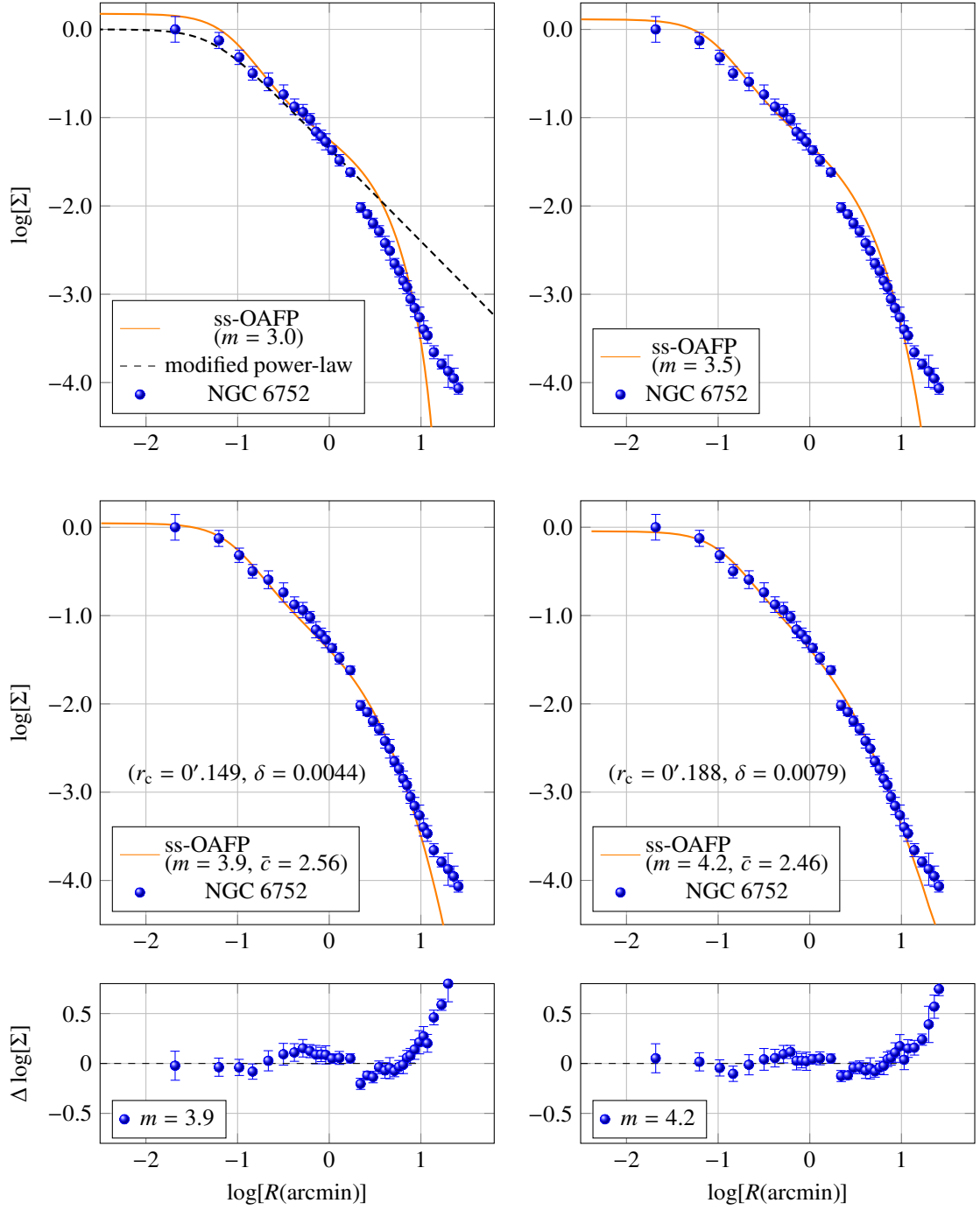

Another reason why we chose $m=3.9$ is that the energy-truncated ss-OAFP model with $m=3.9$ agreeably fits the relatively new data for NGC 6752 reported in \citep{Ferraro_2003}. \cite{Ferraro_2003} provided data points and error bars of the projected surface density for NGC 6752, which is convenient to test our model (since we do not have to artificially extract data from their graph.). In their work, the King model does not well fit the central part of the projected density profile since the cluster is one of (possible) PCC clusters with a power-law profile like core. Hence, following  \citep{Lugger_1995}, they employed modified power law profile like $\sim(1+(r/3.1)^{2})^{-0.525}$ where $r$ is measured in $\log\left[\text{arcsec}\right]$. This well fits the central part as shown in Figure \ref{fig:NGC6752} (left top). On one hand, our model with $m=3.0$ is not close to the morphology of the cluster at all on the figure.  We, however, can more reasonably fit our model to the same data with greater $m$; especially, for $m=4.2$ the model well fits the data except in the tail of the cluster. Even with $m=3.9$, one can find a reasonable fit to the data, hence the present work chose $m=3.9$ to consistently accumulate the data for both KM clusters and PPC clusters.

\section{Fitting of the ss-OAFP model to PCC clusters}\label{sec:Fit_OAFP_PCC_example}

\begin{figure}[H]
	\begin{tikzpicture}
	\begin{axis}[width=8cm,height=7cm, grid=major,xmin=-2.5,xmax=1,ymin=-4.0,ymax=0.1,ylabel=\normalsize{$\text{SB}_\text{o}-\text{SB}$},legend style={cells={align=left}},legend entries={\normalsize{ss-OAFP}\\\normalsize{$(\bar{c}=1.62)$}, \normalsize{ NGC 6388 (n)}},yticklabel style={/pgf/number format/.cd,fixed zerofill,precision=1},legend pos=south west]
	\addlegendimage{no markers, orange};
	\addlegendimage{only marks, mark=triangle*, red};
	\addplot [color = orange ,mark=no,thick,solid] table[x index=0, y index=1]{R_Sig_NGC6388.txt};
	\addlegendimage{only marks, mark=triangle*, red};
	\addplot[color = red, only marks, mark=triangle*, mark options={red}, error bars,y dir=both, y explicit] table [x index=0, y index=1, y error index=2]{Rd_Sigd_Errbar_Err_NGC6388.txt};
	\node[black] at (-1.2,-2.2) {(\normalsize{$r_\text{c}=0\textquotesingle.25, \, \delta=0.047$})} ; 
	\end{axis}
	\end{tikzpicture}\hspace{0.3cm}
	\begin{tikzpicture}
	\begin{axis}[width=8cm,height=7cm, grid=major,xmin=-2.5,xmax=1.0,ymin=-4.0,ymax=0.1,legend style={cells={align=left}},legend entries={\normalsize{ss-OAFP}\\\normalsize{$(\bar{c}=2.60)$}, \normalsize{ Trz 2 (c)}},yticklabel style={/pgf/number format/.cd,fixed zerofill,precision=1},legend pos=south west]
	\addlegendimage{no markers, orange};
	\addlegendimage{only marks, mark=ball, blue};
	\addplot [color = orange ,mark=no,thick,solid] table[x index=0, y index=1]{R_Sig_Trz2.txt};
	\addplot[color = blue, only marks, mark=ball, mark options={blue}, error bars,y dir=both, y explicit] table [x index=0, y index=1, y error index=2]{Rd_Sigd_Errbar_Err_Trz2.txt};
	\node[black] at (-1.2,-2.2) {(\normalsize{$r_\text{c}=0\textquotesingle.078, \, \delta=0.0064$})} ; 
	\end{axis}
	\end{tikzpicture}
	
	\vspace{0.3cm}
	
	\begin{tikzpicture}
	\begin{axis}[width=8cm,height=4cm, grid=major,xmin=-2.5,xmax=1,ymin=-0.8,ymax=0.8,xlabel=\normalsize{$\log[R \text{(arcmin)}]$ },ylabel=\normalsize{$\Delta(\text{SB}_\text{o}-\text{SB})$},legend style={cells={align=left}},legend entries={\normalsize{ NGC 6388(n)}},legend pos=south west]
	\addlegendimage{only marks, mark=triangle*, red};
	\addplot[color = red, only marks, mark=triangle*, mark options={red}, error bars,y dir=both, y explicit] table [x index=0, y index=3, y error index=2]{Rd_Sigd_Errbar_Err_NGC6388.txt};
	\draw[dashed] (-2.5,0)--(1.0,0);
	\end{axis}
	\end{tikzpicture}\hspace{0.3cm}
	\begin{tikzpicture}
	\begin{axis}[width=8cm,height=4cm, grid=major,xmin=-2.5,xmax=1.0,ymin=-0.8,ymax=0.8,xlabel=\normalsize{$\log[R \text{(arcmin)}]$ },legend entries={\normalsize{ Trz 2 (c)}},legend pos=south west]
	\addlegendimage{only marks, mark=ball, blue};
	\addplot[color = blue, only marks, mark=ball, mark options={blue}, error bars,y dir=both, y explicit] table [x index=0, y index=3, y error index=2]{Rd_Sigd_Errbar_Err_Trz2.txt};
	\draw[dashed] (-2.5,0)--(1.0,0);
	\end{axis}
	\end{tikzpicture}
	\caption{Fitting of the energy-truncated ss-OAFP model ($m=3.9$) to the surface brightness profiles of Terzan 2 and NGC 6388 reported in \citep{Djorgovski_1986}.  The unit of the surface brightness (SB) is B magnitude per square of arcseconds. The brightness is normalized by the magnitude $\text{SB}_\text{o}$ at the smallest radius point. In the legends, (n) means normal or KM cluster and (c) means PCC cluster as judged so in \citep{Djorgovski_1986}. $\Delta(\text{SB}_\text{o}-\text{SB})$ is the corresponding deviation of $\text{SB}_\text{o}-\text{SB}$ from the model.}
	\label{fig:fitting_PCC_Djorgovski}
\end{figure}
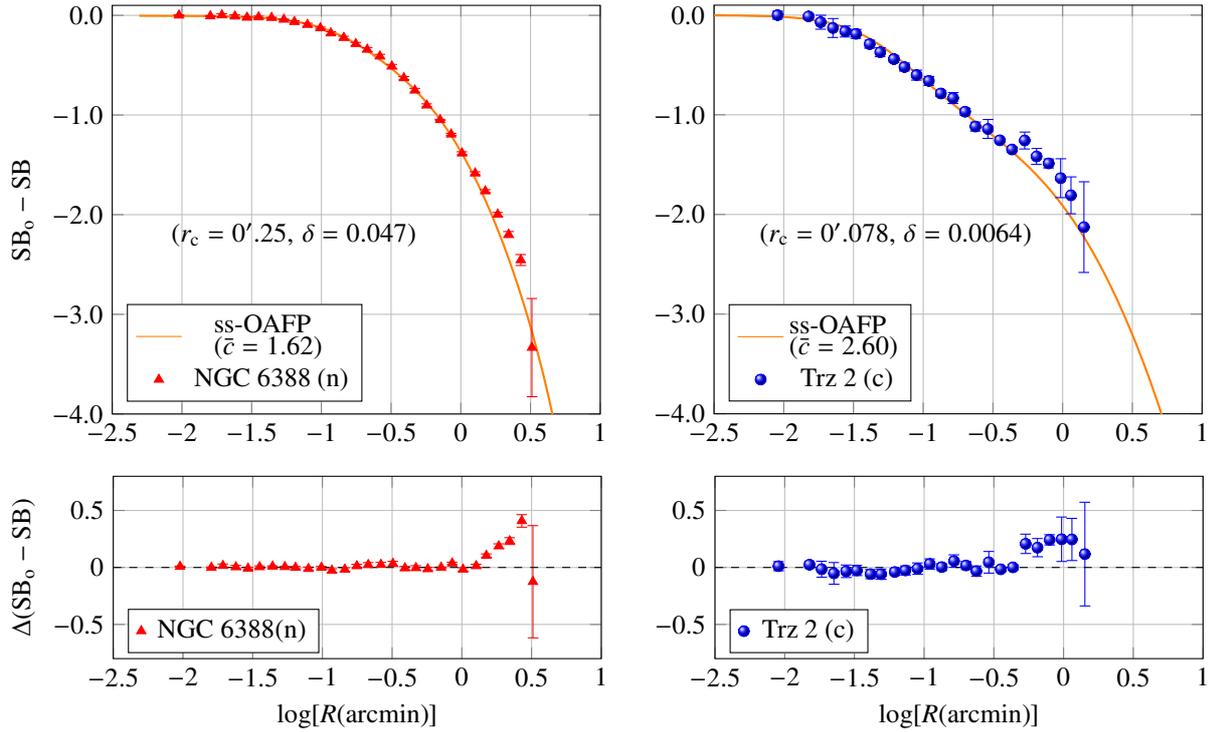

The energy-truncated ss-OAFP model with $m=3.9$ reasonably fits to the projected structural profiles of PCC clusters with resolved cores at $\bigr\lessapprox1$ arcminute. Other than NGC 6752, we also have access to numerical data of the projected density profile for NGC 6397 from \citep{Drukier_1995}. Our model reasonably fits to the density for NGC 6397 with $\chi_{\nu}^{2}=1.52$ (Figure \ref{fig:NGC6397}) where the reduced chi-square is defined as follows
\begin{align}
\chi^{2}_\nu=\sum\frac{\chi^{2}}{n_\text{d.f.}},
\end{align}
where $\chi^{2}$ is the chi-square value between the observed data and our model, and $n_\text{d.f.}$ is the degree of freedom. We chose $n_\text{d.f.}=3$ in the same way as the King model since index $m$ of our model is fixed to 3.90. Since we did not have access to numerical values for the rest of the projected density profiles for PCC clusters reported in \citep{Djorgovski_1986,Lugger_1995}, our error analysis becomes less trustful hereafter.  Yet it appears enough to capture the applicability of our model to the PCC clusters. For example, \cite{Meylan_1997} introduced NGC 6388 and Terzan 2 as an example for a KM cluster and PCC cluster by citing the surface brightness profiles of the clusters from \cite{Djorgovski_1986}'s work though, the energy-truncated ss-OAFP model reasonably fits both density profiles at radii $R\lessapprox1$ arcminute (Figure \ref{fig:fitting_PCC_Djorgovski}). (As discussed in Section \ref{sec:discuss_finite_OAFP}, to fit our model to densities at $10\sim100$ arcminutes the value of $m$ is to be $m\gtrapprox4.2$.) In a similar way to NGC 6752 and NGC 6397, we applied our model to PCC clusters with resolved cores and some clusters with unresolved cores reported in \citep{Djorgovski_1986,Lugger_1995} (See Appendix \ref{Appendix_OAFP_Fit_PCC} in which the 'possible' PCC clusters reported in \citep{Kron_1984} are also fitted).  Table \ref{table:chi} shows the values of $\chi_{\nu}^{2}$ for both KM- and PCC- clusters that we obtained the uncertainties in observed densities from the numerical values or graphs. The result is obvious that the profiles of the KM clusters are well fitted by the ss-OAFP model for all the data points given. On one hand, for PCC clusters, the model fits to only clusters with resolved cores reported in \citep{Lugger_1995}. For example, the PCC clusters with partially-resolved cores (NGC 6453, NGC 6522 and NGC 7099) and resolved cores (NGC 6397 and NGC 6752) are reasonably fitted at $\bigr\lessapprox$1  arcminumte with $\chi_{\nu}^{2}\lessapprox2$. Even a PCC cluster with unresolved core (NGC 6342) can be reasonably fitted in a similar way, while the present work does not account for the `seeing-effect' that comes from the finiteness of the seeing-disk. The structural profiles of the rest of PCC clusters with unresolved cores (e.g. NGC 5946 and NGC 6624) were hopeless to be fitted even only for the cores, whose central parts have steeper power-law profiles compared to our model. This was expected since the present model does not correctly include the effect of binaries whose heating effect is possibly inefficient to provide resolved cores.

\begin{figure}[H]
	\centering
	\begin{tikzpicture}
	\begin{axis}[width=7.5cm,height=7cm, grid=major,xmin=-1,xmax=1.5,ymin=-4.5,ymax=0.5,xlabel=\normalsize{$\log[R\text{(arcmin)}]$},ylabel=\normalsize{$\log[\Sigma]$ }, legend style={cells={align=left}},legend entries={\normalsize{ss-OAFP}\\\normalsize{$(\bar{c}=2.19)$}, \normalsize{NGC 6397 (c)}},yticklabel style={/pgf/number format/.cd,fixed zerofill,precision=1},legend pos=south west]
	\addlegendimage{no markers, orange};
	\addlegendimage{only marks, mark=ball, blue};
	\addplot [color = orange ,mark=no,thick,solid] table[x index=0, y index=1]{R_Sig_NGC6397.txt};
	\addplot[color = blue, only marks, mark=ball, mark options={blue}, error bars,y dir=both, y explicit] table [x index=0, y index=1, y error index=2]{Rd_Sigd_Errbar_Err_NGC6397.txt};
	\node[black] at (-0.1,-2.8) {(\normalsize{$r_\text{c}=0\textquotesingle.085, \, \delta=0.0012$})} ; 
	\end{axis}
	\end{tikzpicture}\hspace{0.2cm}
	\begin{tikzpicture}
	\begin{axis}[width=7cm,height=4cm, grid=major,xmin=-1,xmax=1.5,ymin=-0.8,ymax=0.8,xlabel=\normalsize{$\log[R \text{(arcmin)}]$ },ylabel=\normalsize{$\Delta\log[\Sigma]$ },legend style={cells={align=left}},legend entries={\normalsize{ NGC 6397 (c)}},legend pos=south west]
	\addlegendimage{only marks, mark=ball, blue};
	\addplot[color = blue, only marks, mark=ball, mark options={blue}, error bars,y dir=both, y explicit] table [x index=0, y index=3, y error index=2]{Rd_Sigd_Errbar_Err_NGC6397.txt};
	\draw[dashed] (-2.5,0)--(1.5,0);
	\end{axis}
	\end{tikzpicture}
	\caption{Fitting of energy-truncated ss-OAFP model ($m=3.9$) to the projected surface density of NGC 6397 reported in \citep{Drukier_1993}. The unit of $\Sigma$ is number per square of arcminutes and $\Sigma$ is normalized so that the density is unity at smallest radius for data. In the legends, (c) means PCC cluster as judged so in \citep{Djorgovski_1986}. $\Delta\log[\Sigma]$ is the corresponding deviation of $\Sigma$ from the model on log scale.}
	\label{fig:NGC6397}
\end{figure}
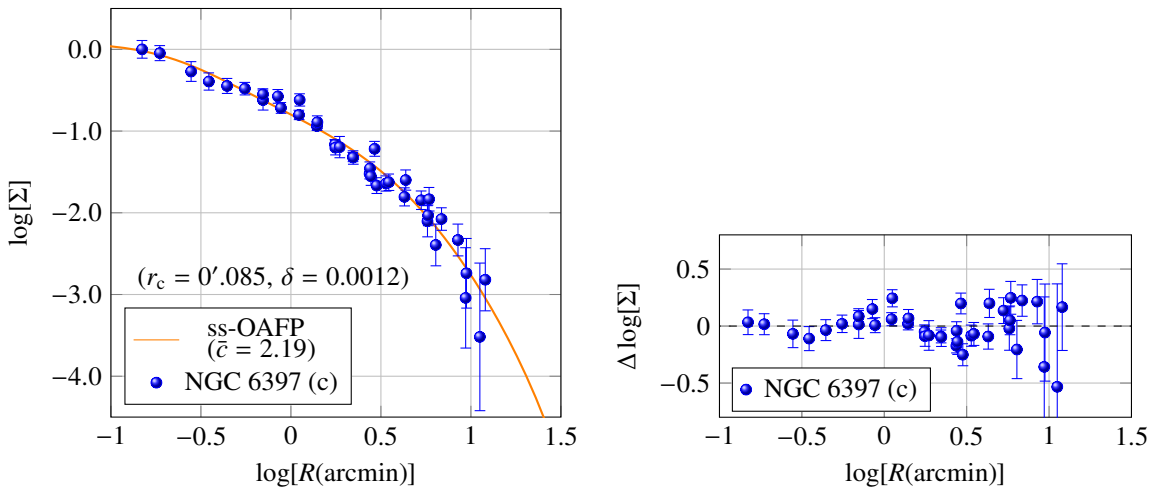

\begin{table*}\centering
	\ra{1.3}
	\begin{tabular}{@{}lll@{}}\toprule
		KM cluster  & $\chi^{2}_{\nu}$ & $N_\text{b}$\\ 
		\midrule
		NGC 288    &0.45 & 0 \\
		NGC 1851  &0.56 & 0 \\
		NGC 5466  &2.07 & 0\\
		NGC 6121  &0.72 & 0\\
		NGC 6205  &1.05 & 0\\
		NGC 6254  & 0.57 & 0\\
		NGC 6626  & 0.47 & 0\\
		NGC 6809  & 0.44 & 0\\
		Pal 3  & 0.06 & 0\\
		Pal 4  & 0.34 & 0\\
		Pal 14  & 0.31 & 0\\
		Trz 5  & 2.23 & 0\\
		\bottomrule
	\end{tabular}\hspace{0.6cm}
	\begin{tabular}{@{}lll@{}}\toprule
		PCC cluster   & $\chi^{2}_{\nu}$ & $N_\text{b}$\\ 
		\midrule
		NGC 6342  & 1.73 & 3\\
		NGC 6397   &1.52 & 0 \\
		NGC 6453  &1.89 & 5\\
		NGC 6522  & 2.52 & 5\\
		NGC 6558  & 2.17 & 5\\
		NGC 6752   &2.00 & 6 \\
		NGC 7099  & 2.12 & 2\\
		Trz 1         & 2.41 & 5\\
		Trz 2         &1.94 & 0\\
		\bottomrule
	\end{tabular}
	\begin{tabular}{@{}lll@{}}\toprule
		PCC cluster   & $\chi^{2}_{\nu}$ & $N_\text{b}$\\ 
		\midrule
		NGC 5946 &6.75 & 5\\
		NGC 6624 &7.18 & 5\\
		\bottomrule
	\end{tabular}
	\caption{Values of $\chi^{2}_{\nu}$ and number of points discarded from the calculation. The data used for fitting to KM clusters are from \citep{Miocchi_2013}, to NGC 6397 from \citep{Drukier_1993}, to Terzan 2 from \citep{Djorgovski_1986}, to NGC 6752 from \citep{Ferraro_2003} and to the rest of PCC clusters from \citep{Lugger_1995}. $N_\text{b}$ is the number of data points at large radii excluded from calculation.}
	\label{table:chi}
\end{table*}

\section{(Main result) Relaxation time and completion rate of core collapse against concentration including PCC clusters with resolved cores}\label{sec:Tr_compl}

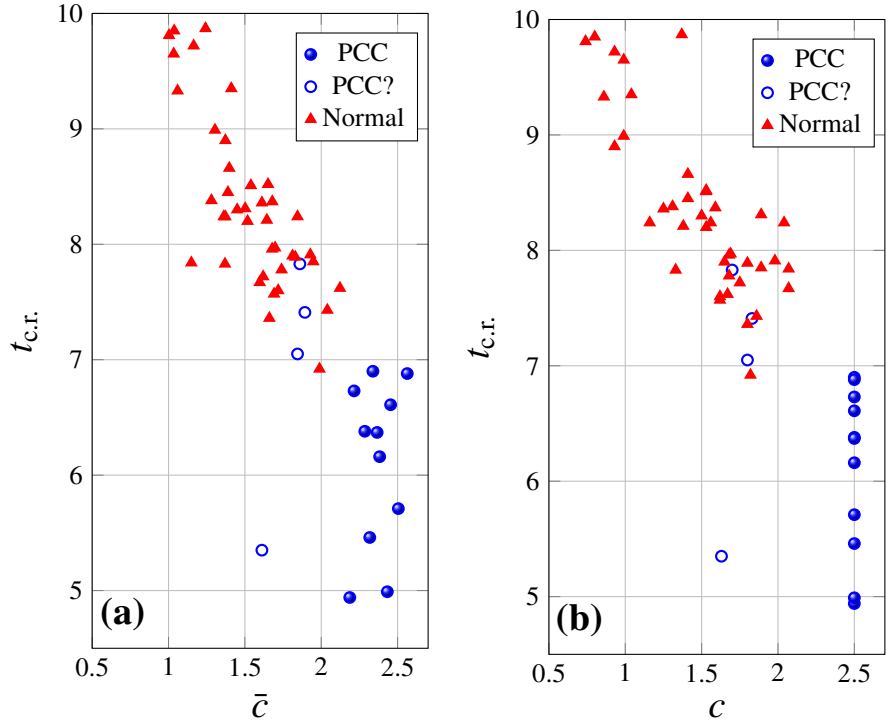
\begin{figure}[H]
	\centering
	\begin{tikzpicture}
	\begin{axis}[width=6cm,height=10cm, grid=major,xlabel=\Large{$\bar{c}$ },ylabel=\Large{$t_\text{c.r.}$},xmin=0.5,xmax=2.7,ymin=4.5,ymax=10,legend pos=north east]
	\addplot [only marks, color = blue, mark=ball, mark size =2pt,thick,solid]  table[x index=0, y index=2]{cPCC_cPCCKing_logTcPCC.txt};
	\addlegendentry{\normalsize{PCC}} 
	\addplot [only marks, color = blue, mark=o, mark size =2pt,thick,solid]  table[x index=0, y index=2]{cPC_cPCKing_logTcPC.txt};
	\addlegendentry{\normalsize{PCC?}} 
	\addplot [only marks ,color = red ,mark=triangle*,thick, solid, mark size =2pt] table[x index=0, y index=2]{c_cKing_logTc.txt};
	\addlegendentry{\normalsize{Normal}} 
	\addplot [only marks ,color = red ,mark=triangle*,thick, solid, mark size =2pt] table[x index=0, y index=2]{cNO_cNOKing_logTcNO.txt};
	\node[black] at (0.7,4.8) {\Large{\textbf{(a)} }};
	\end{axis}
	\end{tikzpicture}\hspace{0.3cm}
	\begin{tikzpicture}
	\begin{axis}[width=6cm,height=10cm, grid=major,xlabel=\Large{$c$ },ylabel=\Large{$t_\text{c.r.}$},xmin=0.5,xmax=2.7,ymin=4.5,ymax=10,legend pos=north east]
	\addplot [only marks, color = blue, mark=ball, mark size =2pt,thick,solid] table[x index=1, y index=2]{cPCC_cPCCKing_logTcPCC.txt};
	\addlegendentry{\normalsize{PCC}} 
	\addplot [only marks, color = blue, mark=o, mark size =2pt,thick,solid] table[x index=1, y index=2]{cPC_cPCKing_logTcPC.txt};
	\addlegendentry{\normalsize{PCC?}} 
	\addplot [only marks ,color = red ,mark=triangle*,thick, solid, mark size =2pt] table[x index=1, y index=2]{c_cKing_logTc.txt};
	\addlegendentry{\normalsize{Normal}} 
	\addplot [only marks ,color = red ,mark=triangle*,thick, solid, mark size =2pt] table[x index=1, y index=2]{cNO_cNOKing_logTcNO.txt};
	\node[black] at (0.7,4.8) {\Large{\textbf{(b)} }};
	\end{axis}
	\end{tikzpicture}
	\caption{ Core relaxation time against  \textbf{(a)} concentration  $\bar{c}$ obtained from the energy-truncated ss-OAFP model ($m=3.9$) applied to PPC- and KM- clusters and \textbf{(b)} concentration $c$ based on the King model reported in \citep[][(2010 edition)]{Harris_1996}.}
	\label{fig:c_Tr}
\end{figure}

Concentration $\bar{c}$ is a possible measure to characterize the states of globular clusters in the relaxation evolution especially for the cores, hence the present section compares $\bar{c}$ to the core relaxation time and completion rate of core collapse. Since the energy-truncated ss-OAFP model can reasonably apply not only to KM clusters (Appendix \ref{Appendix_OAFP_Fit}) but also to PCC clusters (Appendix \ref{Appendix_OAFP_Fit_PCC}), one may systematically discuss their relationship. We first discuss here how the core relaxation time depends on the concentration. Figure \ref{fig:c_Tr}\textbf{(a)} depicts the characteristics of the core relaxation time $t_\text{c.r.}$ against the concentration $\bar{c}$ and  Figure \ref{fig:c_Tr}\textbf{(b)} the corresponding characteristics based on King model reported in \citep[][(2010 edition)]{Harris_1996}. All the relaxation times on both figures for PCC- and KM- clusters are the values reported in \cite[][(2010 edition)]{Harris_1996}'s catalog.  In the catalog, some of concentrations are depicted as `2.50c' for clusters whose projected structural profiles are not well fitted by the King model. Hence, we simply assumed in Figure \ref{fig:c_Tr} \textbf{(b)} the concentration of such clusters is 2.50. In Figure \ref{fig:c_Tr}\textbf{(b)} the relaxation time decreases with increasing concentration $c$ for KM clusters, yet it is not clear if PCC clusters have the same tendency. On one hand, Figure \ref{fig:c_Tr}\textbf{(a)} shows not only the relaxation time decreases with concentration $\bar{c}$ for KM clusters but also the time drops down almost vertically for PCC clusters for long relaxation time. This tendency well captures the nature of PCC clusters whose projected profiles can be close to the ss-OAFP model in complete-collapsed state but still similar to the King model (KM clusters) in expansion phase after their cores collapse.

\begin{figure}[H]
	\centering
	\begin{tikzpicture}
	\begin{axis}[width=7cm,height=11cm, grid=major,xlabel=\Large{$\bar{c}$ },ylabel=\large{completion rate},xmin=0.5,xmax=2.7,ymin=0,ymax=1.1,legend pos=south east]
	\addplot [only marks,color = blue, mark=ball, mark size =2pt,thick,solid]  table[x index=0, y index=2]{cPCC_cPCCKing_complPCC.txt};
	\addlegendentry{\normalsize{PCC}} 
	\addplot [only marks,color = blue, mark=o, mark size =2pt,thick,solid]  table[x index=0, y index=2]{cPC_cPCKing_complPC.txt};
	\addlegendentry{\normalsize{PCC?}} 
	\addplot [only marks,  color = red ,mark=triangle*,thick, solid] table[x index=0, y index=2]{c_cKing_compl.txt};
	\addlegendentry{\normalsize{Normal}} 
	\addplot [color = black ,mark=no,thick, densely dashed]  table[x index=0, y index=1]{x_guidelines_c_compl.txt};
	\addplot [color = black ,mark=no,thick, densely dashed]  table[x index=0, y index=2]{x_guidelines_c_compl.txt};
	\addplot [only marks , color = red ,mark=triangle*,thick, solid] table[x index=0, y index=2]{cNO_cNOKing_complNO.txt};
	\node[rotate=68] at (1.43,0.76) {\normalsize{$0.75(\bar{c}-2.0)+1.05$}};
	\node[rotate=68] at (2.22,0.45) {\normalsize{$0.75(\bar{c}-2.0)+0.4$}};
	\node at (1.55,1.02) {\tiny{NGC6544}};
	\node at (2.0,0.89) {\tiny{NGC6517}};
	\node at (2.17,0.73) {\tiny{NGC1851}};
	\node at (2.1,0.65) {\tiny{NGC6626}};
	\node[black] at (0.7,1.05) {\Large{\textbf{(a)}}};
	\end{axis}
	\end{tikzpicture}\hspace{0.4cm}
	\begin{tikzpicture}
	\begin{axis}[width=7cm,height=11cm, grid=major,xlabel=\Large{$c$ },ylabel=\large{completion rate},xmin=0.5,xmax=2.7,ymin=0,ymax=1.1,legend pos=south east]
	\addplot [only marks,color = blue, mark=ball, mark size =2pt,thick,solid,mark options={fill=blue}] table[x index=1, y index=2]{cPCC_cPCCKing_complPCC.txt};
	\addlegendentry{\normalsize{PCC}} 
	\addplot [only marks,color = blue, mark=o, mark size =2pt,thick,solid,mark options={fill=blue}] table[x index=1, y index=2]{cPC_cPCKing_complPC.txt};
	\addlegendentry{\normalsize{PCC?}} 
	\addplot [only marks , color = red ,mark=triangle*,thick, solid] table[x index=1, y index=2]{c_cKing_compl.txt};
	\addlegendentry{\normalsize{Normal}} 
	\addplot [color = black ,mark=no,thick, densely dashed]  table[x index=0, y index=1]{x_guidelines_c_compl.txt};
	\addplot [color = black ,mark=no,thick, densely dashed]  table[x index=0, y index=2]{x_guidelines_c_compl.txt};
	\addplot [only marks , color = red ,mark=triangle*,thick, solid] table[x index=1, y index=2]{cNO_cNOKing_complNO.txt};
	\node at (1.55,1.03) {\tiny{NGC6544}};
	\node at (2.02,0.88) {\tiny{NGC6517}};
	\node[black] at (0.7,1.05) {\Large{\textbf{(b)}}};
	\end{axis}
	\end{tikzpicture}
	\caption{Completion rate of core collapse against \textbf{(a)} concentration $\bar{c}$ based on the energy-truncated ss-OAFP model with $m=3.9$ and \textbf{(b)} concentration $c$ based on the King model reported in \citep[][(2010 edition)]{Harris_1996}.}
	\label{fig:c_complete}
\end{figure}
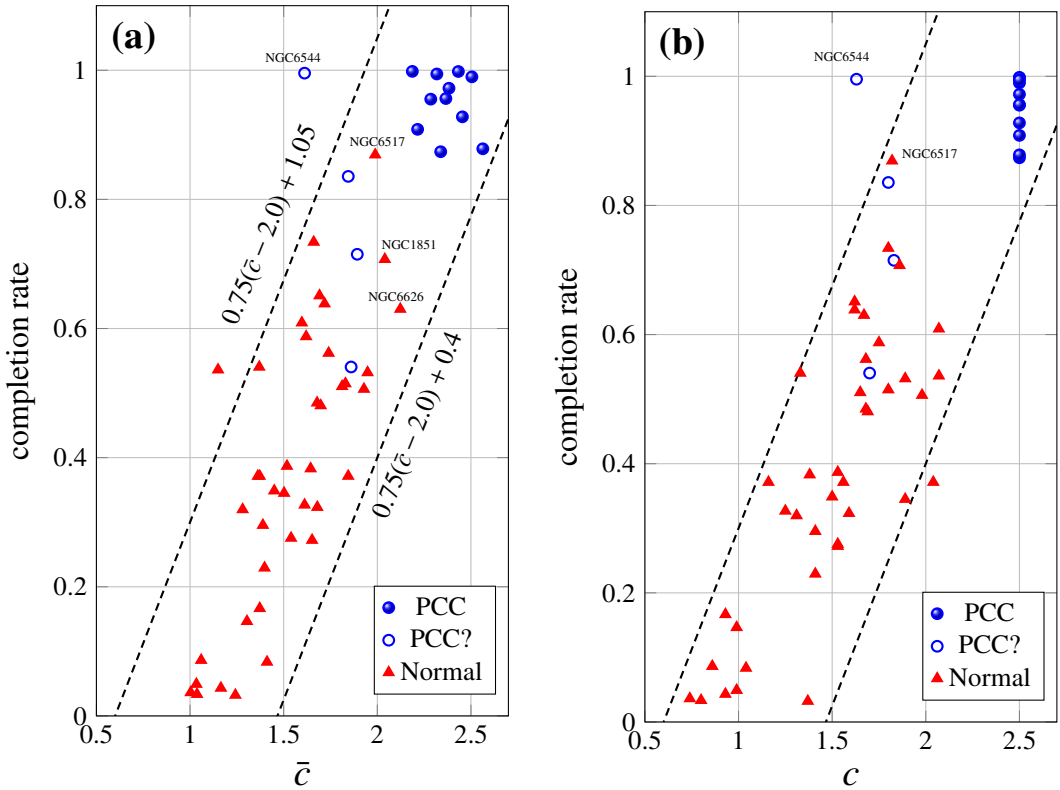

To see the completion rate of core collapse, we use the formula employed in \citep{Lightman_1982}   
\begin{align}
\eta_\text{c}\equiv\frac{t_\text{o,age}}{t_\text{c.r.o}}\equiv\frac{t_\text{o,age}}{t_\text{c.r.}}\frac{-(1+Aq_\text{o})+\sqrt{(1+Aq_\text{o})^{2}+4ABq_\text{o}}}{2Bq_\text{o}}
\end{align}\label{Eq.compl}
where $A=35$, $B=4.8$ and $q_\text{o}=t_\text{o,age}/t_\text{c.r.}$ with $t_\text{o,age}$ being the order of age of clusters $\sim10^{10}$ years. The time $t_\text{c.r.o}$ is the estimated relaxation time at the beginning of evolution of each cluster based on $N$-body simulation. Figure \ref{fig:c_complete} (\textbf{a}) shows the completion rate against concentration $\bar{c}$ obtained from the energy-truncated ss-OAFP model. The majority of data plots are within the region between lines $\eta_\text{c}=0.75(\bar{c}-2.0)+1.05$ and $\eta_\text{c}=0.75(\bar{c}-2.0)+0.40$ that are empirical lines of equation, not based on physical arguments. Figure \ref{fig:c_complete} (\textbf{b}) shows that the corresponding characteristics of $\eta_\text{c}$ against concentration $c$ based on the King model, and the same two lines reasonably include the majority of data plots between them. From Figure \ref{fig:c_complete} (\textbf{a}),  one can find several conclusion. (i) The criterion explained in \citep{Meylan_1997} still works that clusters with $c>2.0$ are PCC clusters but if the completion rate is above 0.8 (ii) Clusters with over a completion rate of 0.8 are PPC clusters except for a cluster NGC 6517. (iii) KM clusters with high concentration ($c\geq2.0$) in \cite[][(2010 edition)]{Harris_1996}'s catalog are reasonably close to the other KM stars in the figure and their concentration are lowered to values smaller than 2.0, while our model suggests two KM clusters (NGC 1851 and NGC 6626) with high concentration ($\bar{c}\geq2.0$) have morphology close to complete core-collapse state. (iv) A PCC cluster (NGC 6544) differentiates itself from the KM- and PCC- clusters in the sense that NGC 6544 has a high completion rate (0.989) compared to the KM clusters and a \emph{very} low concentration  ($\bar{c}=1.61$) compared to the rest of PCC clusters. Hence, the cluster may be a good candidate for search of a PPC cluster that may have one of most expanded cores, while the cluster was judged only as `possible PCC' in \citep{Djorgovski_1986}. (v) Our model with $\bar{c}\approx1$ fits the projected structural profiles of low-concentration clusters, which infers the clusters may have structures similar to polytrope of $m\approx3.9$. The conclusion (i) simply confirmed an expected property of core collapse process and conclusions (ii) through (iv) require a detail case study for each cluster, which is out of scope in the present work. Hence, we further discuss only the conclusion (v) in detail in Section \ref{sec:discuss_finite_OAFP}. 

\section{Discussion}\label{sec:discuss_finite_OAFP}

The present section discusses the three topics that we could not detail in the main results; (i) a possible use of the energy-truncated ss-OAFP model with high index $m$ (higher than $m=3.9$) to the structural profiles of globular clusters in a broad range of radii $10^{-2}\sim10^{1}$  arcminutes (Section \ref{sec:whole_fit}), (ii) an application limit and an approximated form of our model (Section \ref{sec:apprx_FssOAFP}) and (iii) why our model can fit the projected structural profiles even for low-concentration ($\bar{c}\approx1$) globular clusters (Section \ref{sec:why_FssOAFP}).

\subsection{Fitting of the energy-truncated OAFP model to`whole' projected density with higher index $m$}\label{sec:whole_fit}

\begin{figure}
	\begin{tikzpicture}
	\begin{axis}[width=8cm,height=8cm, grid=major,xmin=-3.5,xmax=1.5,ymin=-14,ymax=0.5,ylabel=\normalsize{$\text{SB}_\text{o}-\text{SB}$},xlabel=\normalsize{$\log[R\text{(arcmin)}]$},legend style={cells={align=left}},legend entries={\normalsize{ss-OAFP}\\\normalsize{$(\bar{c}=2.16,m=4.2)$}, \normalsize{NGC 6341(n)}, \normalsize{Cheb.}},yticklabel style={/pgf/number format/.cd,fixed zerofill,precision=1},legend pos=south west]
	\addlegendimage{no markers, orange};
	\addlegendimage{only marks, mark=triangle*, red};
	\addlegendimage{no markers, red, densely dashed};
	\addplot [color = orange ,mark=no,thick,solid] table[x index=0, y index=1]{R_Sig_NGC6341_Noyora_whole.txt};
	\addplot[color = red, only marks, mark=triangle*, mark options={red}, error bars,y dir=both, y explicit] table [x index=0, y index=1, y error index=2]{Rd_Sigd_Errbar_Err_NGC6341_Noyora_whole.txt};
	\addplot [color = red ,mark=no,thick,densely dashed] table[x index=0, y index=1]{Rc_Sigc_NGC6341_Trager_whole.txt};
	\node[black] at (-1.9,-5.0) {(\normalsize{$r_\text{c}=0\textquotesingle.25, \, \delta=0.038$})} ; 
	\end{axis}
	\end{tikzpicture}\hspace{0.3cm}
	\begin{tikzpicture}
	\begin{axis}[width=8cm,height=8cm, grid=major,xmin=-3.5,xmax=0.6,ymin=-8.5,ymax=0.5,xlabel=\normalsize{$\log[R(\text{arcmin})]$},legend style={cells={align=left}},legend entries={\normalsize{ss-OAFP}\\\normalsize{$(\bar{c}=3.28, m=4.2)$}, \normalsize{NGC 6284(c)},\normalsize{Cheb.}},yticklabel style={/pgf/number format/.cd,fixed zerofill,precision=1},legend pos=south west]
	\addlegendimage{no markers, orange};
	\addlegendimage{only marks, mark=ball, blue};
	\addlegendimage{no markers, blue, densely dashed};
	\addplot [color = orange ,mark=no,thick,solid] table[x index=0, y index=1]{R_Sig_NGC6284_Noyora_whole.txt};
	\addplot[color = blue, only marks, mark=ball, mark options={blue}, error bars,y dir=both, y explicit] table [x index=0, y index=1, y error index=2]{Rd_Sigd_Errbar_Err_NGC6284_Noyora_whole.txt};
	\addplot [color = blue ,mark=no,thick,densely dashed] table[x index=0, y index=1]{Rc_Sigc_NGC6284_Trager_whole.txt};
	\node[black] at (-2.0,-4.7) {(\normalsize{$r_\text{c}=0\textquotesingle.0040, \, \delta=0.0055$})} ; 
	\end{axis}
	\end{tikzpicture}
	\caption{Fitting of the energy-truncated ss-OAFP model to the surface brightness profiles of NGC 6284 and  NGC 6341 reported in \citep{Noyola_2006}. `Cheb.' means the Chebyshev approximation of the surface brightness reported in \citep{Trager_1995}.  The unit of the surface brightness (SB) is V magnitude per square of arcseconds. The brightness is normalized by the magnitude $\text{SB}_\text{o}$ at the smallest radius point. In the legends, (n) means normal or KM cluster and (c) means PCC cluster as judged so in \citep{Djorgovski_1986}.}
	\label{fig:fitting_whole_ex}
\end{figure}
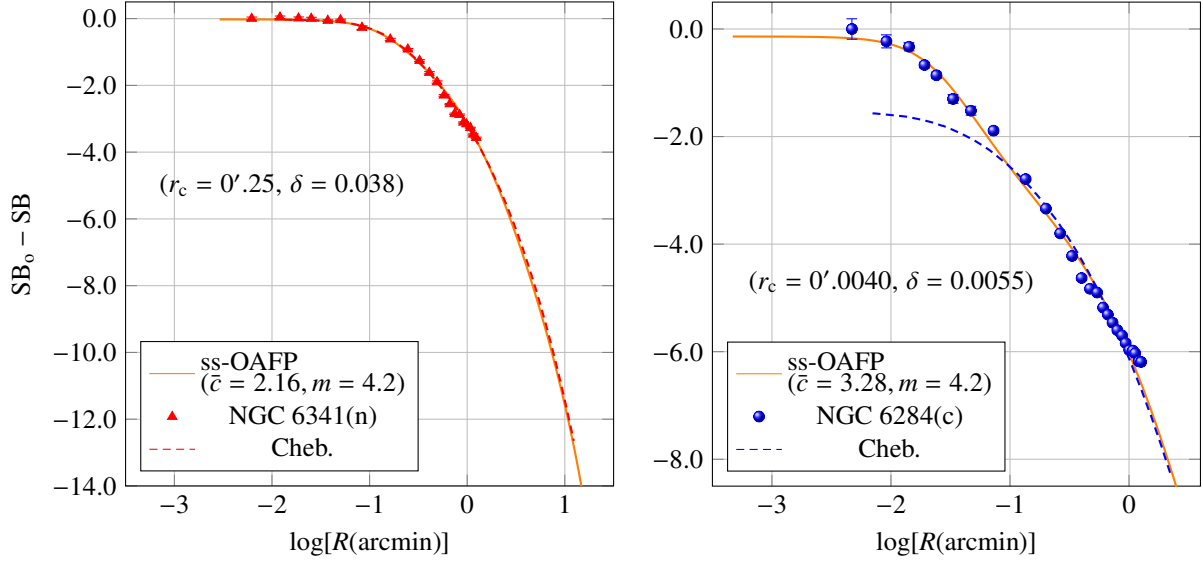

In the present section, we show that the energy-truncated ss-OAFP model with $m=4.2$ well fits to the surface brightness profiles of some PCC clusters and KM clusters in a broad range of radii (approximately between 0.01 and 10 arcminutes). In the main result (Section \ref{sec:Fit_OAFP_PCC_example}), we employed $m=3.9$ to consistently apply the model to various clusters though, as explained in Section \ref{sec:choice_m}, index $m$ higher than 3.9 (e.g. $m=4.2$) may provide a better fitting. Also, more recent surveys such as \emph{Gaia} 2 provided structural profiles with elongated outer halos for some clusters that can not be well fitted by the King model, rather the Wilson model showed a better fitting \citep{de_Boer_2019}. A similar situation was reported also in \citep{Miocchi_2013}. Under the circumstances, we applied our model to the $V$-band magnitudes of globular clusters reported in \citep{Noyola_2006} combined with those of \citep{Trager_1995}\footnote{Unfortunately, we did not have an access to data of \citep{de_Boer_2019} and were not able to employ \emph{WebPlotDigitlizer} to extract their data since many of their data plots and error bars are overlapped each other.}. Following \citep{Noyola_2006}, we simply overlapped their magnitudes to those of \citep{Trager_1995}. For example, Figure \ref{fig:fitting_whole_ex} shows the fitting of our model with $m=4.2$ to the surface brightness profile of NGC 6341 and NGC 6284; the former is a KM cluster and the latter PCC. Following the fitting done in  \citep{Noyola_2006},  we fitted our model to the surface brightness so that in the outer halo our model fits the data plots (Chebyshev approximation) of \citep{Trager_1995} rather than the corresponding plots of \citep{Noyola_2006}.   We need more realistic effects (e.g. mass function) to more reasonably fit our model to NGC 6284 though, our model appears capable to capture the structure (core and tidal radius) of the cluster. For the rest of fitting, see Appendix \ref{Fitting_KM_Noyola} for KM clusters and Appendix \ref{sec:PCC_OAFP_noyola} for PCC clusters in which we did not include some clusters whose central cores have a negative slope reported in \citep{Noyola_2006}. For PCC clusters, we needed our model only with $m=4.2$ while for KM clusters we needed indexes higher than $m=4.2$. 

\subsection{Application limit and an approximated form of the energy-truncated ss-OAFP model}\label{sec:apprx_FssOAFP}

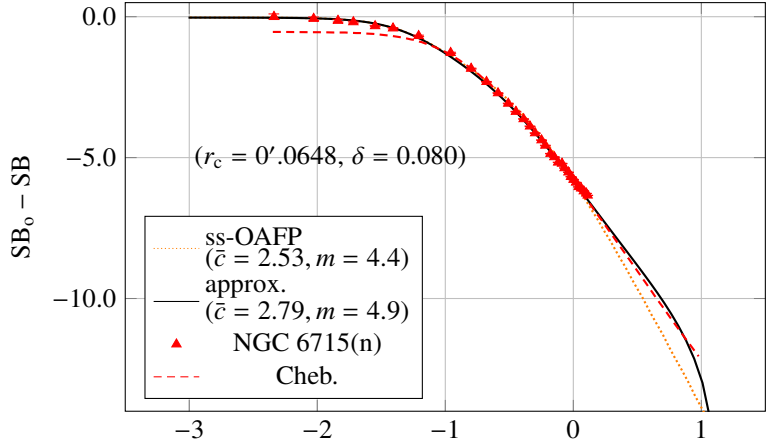
\begin{figure}\centering
	\begin{tikzpicture}
	\begin{axis}[width=10cm,height=7cm, grid=major,xmin=-3.5,xmax=1.5,ymin=-14, ymax=0.5,label=\normalsize{$\log[R(\text{arcmin})]$},ylabel={$\text{SB}_\text{o}-\text{SB}$},legend style={cells={align=left}},legend entries={\normalsize{ss-OAFP}\\ \normalsize{$(\bar{c}=2.53,m=4.4)$},\normalsize{approx. } \\ \normalsize{$(\bar{c}=2.79, m=4.9)$}, \normalsize{NGC 6715(n)}, \normalsize{Cheb.}},yticklabel style={/pgf/number format/.cd,fixed zerofill,precision=1},legend pos=south west]
	\addlegendimage{no markers, orange, densely dotted};
	\addlegendimage{no markers, black};
	\addlegendimage{only marks, mark=triangle*, red};
	\addlegendimage{no markers, red, densely dashed};
	\addplot [color = orange ,mark=no,thick,densely dotted] table[x index=0, y index=1]{R_Sig_NGC6715_Noyora_whole.txt};
	\addplot [color = black ,mark=no,thick,solid] table[x index=0, y index=1]{R_Sig_NGC6715_Noyora_whole_apprx.txt};
	\addplot[color = red, only marks, mark=triangle*, mark options={red}, error bars,y dir=both, y explicit] table [x index=0, y index=1, y error index=2]{Rd_Sigd_Errbar_Err_NGC6715_Noyora_whole.txt};
	\addplot [color = red ,mark=no,thick,densely dashed] table[x index=0, y index=1]{Rc_Sigc_NGC6715_Trager_whole.txt};
	\node[black] at(-1.9,-5){(\normalsize{$r_\text{c}=0\textquotesingle.0648, \, \delta=0.080$})} ; 
	\end{axis}
	\end{tikzpicture}
	\caption{Fitting of the energy-truncated ss-OAFP model and its approximate form to the surface brightness of NGC 6715 reported in \citep{Noyola_2006}. `Cheb.' means the Chebyshev approximation of the surface brightness reported in \citep{Trager_1995}. The core radius $r_\text{c}$ and $\delta$ in the figure were acquired from the approximated form.  The unit of the surface brightness (SB) is V magnitude per square of arcseconds.  The brightness is normalized by the magnitude $\text{SB}_\text{o}$ at the smallest radius point. In the legends, (n) means normal or KM cluster as judged so in \citep{Djorgovski_1986}.}
	\label{fig:fitting_fail_approx}
\end{figure}

The energy-truncated ss-OAFP model suffers from unrealistically large tidal radii ($\sim 10^{5}  $ arcminutes) beyond $m=4.4$ and can not approach $m=5$, which prevents one from fitting the model to some globular clusters, hence the present section introduces an approximated form of the model as a remedy. Sections \ref{sec:choice_m}, \ref{sec:Fit_OAFP_PCC_example} and \ref{sec:whole_fit} showed the applicability of the energy-truncated ss-OAFP model to many globular clusters for $m=3.9$ and $m=4.2\sim4.4$ though, it turns out that some globular clusters have more elongated structures in the outer halos that our model can not reach. For example, Figure \ref{fig:fitting_fail_approx} shows the projected density of NGC 6715 (reported in \citep{Noyola_2006}) fitted by the model with $m=4.4$. The observed density has a calmer slope in the halo compared to our model. Our model can reach further large radii by increasing $m$, yet it turns out the tidal radius unrealistically diverges soon, for example, the tidal radius is approximately $10^{5}$ arcminutes for $m=4.4$ and $\delta=0.01$. (Even the tidal radius of polytrope of $m=4.5$ is approximately only 31.54.)

\begin{figure}[H]
	\centering
	\begin{tikzpicture}
	\begin{semilogyaxis}[width=7cm,height=9cm, grid=major,xlabel=\normalsize{$-E$ },xmin=0.6,xmax=1.0,ymin=1e-3,ymax=1.1,legend pos=south east]
	\addplot [color = orange, mark=no, thick,densely dashed] table[x index=0, y index=1]{Epo_regDo_regDpoly_apprx.txt};
	\addlegendentry{\normalsize{ $D_\text{o}/D_\text{o}(-1)$}} 
	\addplot [color = red ,mark=no,thick, densely dotted] table[x index=0, y index=2]{Epo_regDo_regDpoly_apprx.txt};
	\addlegendentry{\normalsize{$\delta B(3.4,1.5)(-E)^{3.9}/D_\text{o}(-1)$}} 
	\addplot [color = black ,mark=no,thick,solid]  table[x index=0, y index=3]{Epo_regDo_regDpoly_apprx.txt};
	\addlegendentry{\normalsize{$\exp[-13.88(1+E)]$}} 
	\node at (0.65,0.8) {\normalsize{\textbf{(a)}}};
	\end{semilogyaxis}
	\end{tikzpicture}\hspace{0.3cm}
	\begin{tikzpicture}
	\begin{axis}[width=7cm,height=9cm, grid=major,xlabel=\normalsize{$\log [\bigr]$},ylabel=\normalsize{$\log[\Sigma]$ },xmin=-3,xmax=2.8,ymin=-5.3,ymax=0.05,legend pos=south west]
	\addplot [color = blue, mark=no, thick,solid] table[x index=0, y index=1]{R_Sig_OAFP_m39_del_0001.txt};
	\addlegendentry{\normalsize{OAFP ($\delta=10^{-3}$)}}; 
	\addplot [ color = red ,mark=no,thick, solid, densely dashdotted] table[x index=0, y index=1]{R_Sig_OAFP_apprx_m39_del_0001.txt};
	\addlegendentry{\normalsize{approx. ($\delta=10^{-3}$)}}; 
	\addplot [color = black ,mark=no,thick,densely dashed]  table[x index=0, y index=1]{R_Sig_OAFP_m39_del_01.txt};
	\addlegendentry{\normalsize{OAFP ($\delta=0.1$)}}; 
	\addplot [color = orange ,mark=no,thick,solid, dotted]  table[x index=0, y index=1]{R_Sig_OAFP_apprx_m39_del_01.txt};
	\addlegendentry{\normalsize{approx. ($\delta=0.1$)}}; 
	\node at (-2.5,-0.2) {\normalsize{\textbf{(b)}}};
	\end{axis}
	\end{tikzpicture}
	\caption{\textbf{(a)}Dimensionless density profiles $D_\text{o}$, $\exp[-13.88(1+E)]$ and $\delta B(3.4,1.5)(-E)^{3.9}$ for $\delta=0.01$.\textbf{(b)} Projected density profiles for the energy-truncated ss-OAFP model and its approximation for $\delta=0.1$ and $\delta=0.001$. For the both, $m$ is fixed to $3.9$.}
	\label{fig:Do_apprx}
\end{figure}
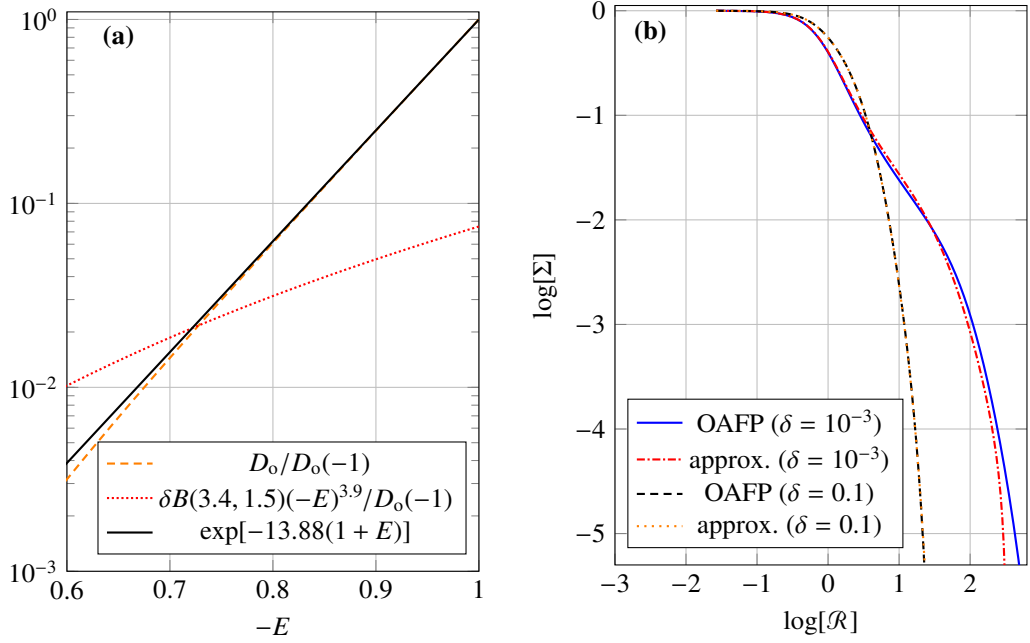

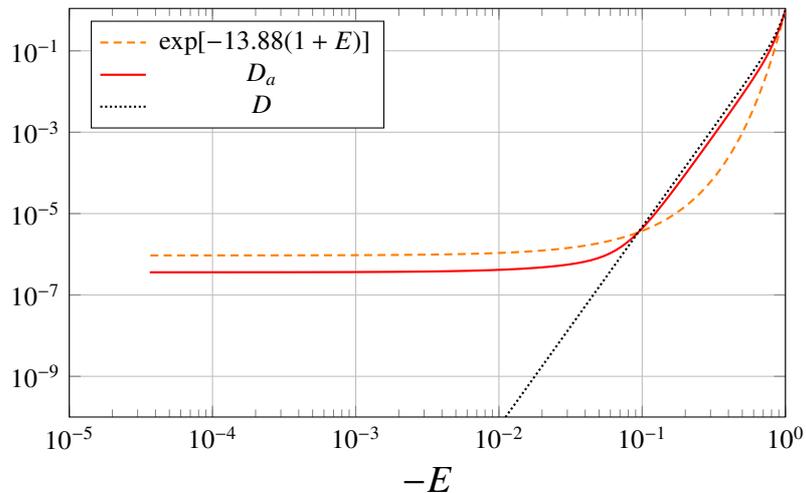
\begin{figure}[H]\centering
	\begin{tikzpicture}
	\begin{loglogaxis}[width=11cm,height=7cm, grid=major,xlabel=\Large{$-E$ },xmin=1e-5,xmax=1.0,ymin=1e-10,ymax=1.1,legend pos=north west]
	\addplot [color = orange, mark=no, thick,densely dashed] table[x index=0, y index=1]{Ep_exp_Dapprx_D.txt};
	\addlegendentry{\normalsize{ $\exp[-13.88(1+E)]$}} 
	\addplot [color = red ,mark=no,thick, solid] table[x index=0, y index=2]{Ep_exp_Dapprx_D.txt};
	\addlegendentry{\normalsize{$D_a$}} 
	\addplot [color = black ,mark=no,thick,densely dotted]  table[x index=0, y index=3]{Ep_exp_Dapprx_D.txt};
	\addlegendentry{\normalsize{$D$}} 
	\end{loglogaxis}
	\end{tikzpicture}
	\caption{Density profiles $D_\text{o}$ of the energy-truncated ss-OAFP model, $D_\text{a}$ of the approximated model and $\exp[-13.88(1+E)]$.}
	\label{fig:Do_apprx_2}
\end{figure}

To resolve the unrealistically large tidal radius of the ss-OAFP model with high indexes $m$, we introduce an approximated form of the energy-truncated ss-OAFP model that well fits the structural profiles of Galactic globular clusters. First, one may see that the values of $\delta$ that fit to globular clusters are relatively high $(\delta>0.004)$ and they would not show a distinctive sign of the power-law profile of $\Sigma\propto r^{-1.23}$ in the inner halo (Recall Figure \ref{fig:potential_fssOAFP}; the power law $r^{-2.23}$ in density profile is recognizable when the value of $\delta$ is small $\sim10^{-3}$.). This infers we may approximate the model by excluding the contribution from the asymptotic power-law profile ($\Sigma\propto r^{-1.23}$) in the halo. In fact, the density profile for the energy-truncated ss-OAFP model can be well fitted by the exponential profile $\exp[-13.88(1+E)]$ for low energies $E<-0.7$ as shown in Figure \ref{fig:Do_apprx} \textbf{(a)}. In the figure the exact model switches from the exponential decay (isothermal sphere) to the power law decay (polytropic sphere) around $E=-0.72$ at which their values are still the same order.  Accordingly we may define an approximate form of DF for our model as follows
\begin{align}
\tilde{F}(E)=\frac{\exp\left[-13.88(1+E)\right]+\delta\, (-E)^{m}}{\exp\left[-13.88\right]+\delta}.
\end{align}
where 13.88 is the value of the scaled escape energy $\chi_\text{esc}$ for complete core collapse. One needs to know the advantage and disadvantage of using the approximated model. This approximation would be handy in the sense that one does not have to resort to the inverse form of the Poisson equation unlike our exact model. Also, one can employ the scaled escape energy other than the 13.88 if considering $\chi_\text{esc}$ as a new parameter allowing four degrees of freedom \footnote{We have tried to employ the four parameters model though, the useful values of $\chi_\text{esc}$ with which the approximated model well fits globular clusters were $\chi_\text{esc}\gtrapprox 9$ holding the index $m$ to 3.9. Less than those values, the morphology of the approximated model perfectly changed from the exact model and especially the approximated model with low $\chi_\text{esc}(\lessapprox9)$ could not fit PCC cluster at all; perhaps, this reflects the result of \citep{Cohn_1980} in which the signature of self-similarity and core collapse appears for $\chi_\text{esc}(\gtrapprox9)$.}. One one hand, the approximated model should not be applied to PCC clusters at first place since the approximation can be reasonable when $\delta>>10^{-3}$ as shown in Figure \ref{fig:Do_apprx} \textbf{(b)}. In the figure, the projected density profiles of the exact and approximated models are almost identical for large $\delta=0.1$ while they obviously deviate from each other for small $\delta$. More importantly, the approximated model loses the physical significance to examine how close the states of globular clusters are to complete core collapse state.  Yet, the approximated form well fits even for the elongated structure in the outer halo of NGC 6715 (Figure \ref{fig:fitting_fail_approx}). The reason why the approximated model may take higher $m (>4.4)$ is that the density profile follows the exponential decay $\exp[-\chi_\text{esc}(1+E)]$ at large $E$, which shortens the tidal radius in a similar way that Woolley's model \citep{Woolley_1961} does. This can be seen in Figure \ref{fig:Do_apprx_2} in which density profile $D_\text{a}$ of the approximated model behaves like exponential function at large $E$. See Appendices \ref{Fitting_KM_Noyola} and \ref{sec:PCC_OAFP_noyola} to find the rest of application of the approximated model to KM- and PCC- clusters in a broad range of radii.

\subsection{Are low-concentration globular clusters like spherical polytropes?}\label{sec:why_FssOAFP}

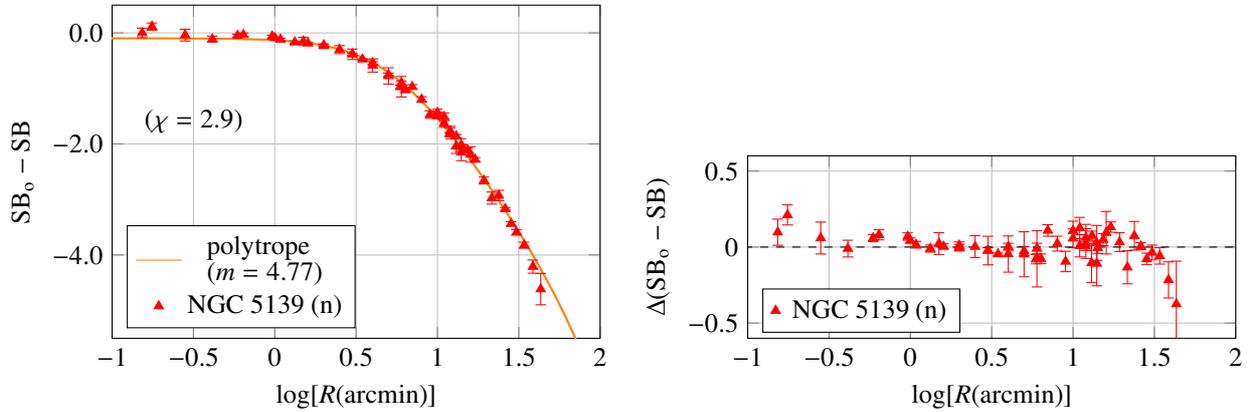
\begin{figure}[H]
	\begin{tikzpicture}
	\begin{axis}[width=8cm,height=6cm, grid=major,xmin=-1.0,xmax=2.0,ymin=-5.5,ymax=0.5,xlabel=\normalsize{$\log[R\text{(arcmin)}]$ },ylabel=\normalsize{$\text{SB}_\text{o}-\text{SB}$},legend style={cells={align=left}},legend entries={\normalsize{polytrope}\\\normalsize{$(m=4.77)$}, \normalsize{NGC 5139 (n)}},yticklabel style={/pgf/number format/.cd,fixed zerofill,precision=1},legend pos=south west]
	\addlegendimage{no markers, orange};
	\addlegendimage{only marks, mark=triangle*, red};
	\addplot [color = orange ,mark=no,thick,solid] table[x index=0, y index=1]{R_Sig_NGC5139_poly.txt};
	\addplot[color = red, only marks, mark=triangle*, mark options={red}, error bars,y dir=both, y explicit] table [x index=0, y index=1, y error index=2]{Rd_Sigd_Errbar_Err_NGC5139_poly.txt};
	\node[black] at (-0.5,-1.6) {(\normalsize{$\chi=2.9$})} ; 
	\end{axis}
	\end{tikzpicture}\hspace{0.2cm}
	\begin{tikzpicture}
	\begin{axis}[width=8cm,height=4cm, grid=major,xmin=-1.0,xmax=2.0,ymin=-0.6,ymax=0.6,xlabel=\normalsize{$\log[R\text{(arcmin)}]$ },ylabel=\normalsize{$\Delta(\text{SB}_\text{o}-\text{SB})$},legend style={cells={align=left}},legend entries={\normalsize{ NGC 5139 (n)}},legend pos=south west]
	\addlegendimage{only marks, mark=triangle*, red};
	\addplot[color = red, only marks, mark=triangle*, mark options={red}, error bars,y dir=both, y explicit] table [x index=0, y index=3, y error index=2]{Rd_Sigd_Errbar_Err_NGC5139_poly.txt};
	\draw[dashed] (-2.5,0)--(2.0,0);
	\end{axis}
	\end{tikzpicture}
	\caption{Fitting of the polytropic sphere of index $m$ to the surface brightness  of NGC 5139 from \citep{Meylan_1987}. The unit of the surface brightness (SB) is V magnitude per square of arcminutes. The brightness is normalized by the magnitude $\text{SB}_\text{o}$ at the smallest radius point. In the legends, (n) means 'normal' or KM cluster as judged so in \citep{Djorgovski_1986}. $\Delta(\text{SB}_\text{o}-\text{SB})$ is the corresponding deviation of $\text{SB}_\text{o}-\text{SB}$ from the model.}
	\label{fig:fitting_poly_3}
\end{figure}

The result of Section \ref{sec:Tr_compl} indicates that low-concentration cluster may have projected structural profiles of polytropic spheres.  This originates from the fact that some clusters (such as Palomar 3 and Palomar 4) have concentrations $\bar{c}$ close to one, which means the projected structural profiles are like those of polytrope of $m=3.9$ rather than isothermal sphere. Yet, fitting of the energy-truncated ss-OAFP model to those clusters are overfitting due to low number of data points and large size of error bars. Hence, to see whether the projected structural profiles of  low-concentration globular clusters are polytropic, Appendix \ref{Appendix_poly_Fit} shows the structural profile data \citep{Kron_1984,Trager_1995,Miocchi_2013} fitted by the polytropic sphere model. We found 18 polytropic globular clusters whose projected density and surface brightness profiles are well fitted by polytropic-sphere model. Here, we show the example for NGC 288 and NGC 6254; their concentrations based on the energy-truncated ss-OAFP model are 1.30 and 1.64 and those based on the King model is 1.0\footnote{Here, we consider the concentration of NGC 288 is 1.0 based on our fitting of the King model. We confirmed the concentrations $c$ and values of $\chi_{\nu}$ reported in Table 2 of \citep{Miocchi_2013} based on our calculation for the clusters that we cited here, but not for NGC 288. We actually found the same result ($\chi_{\nu}=1.7$ with $W_\text{o}=5.8$) for NGC 288 as their result though, we use the result of our calculation and consider NGC 288 is a low concentration cluster. This is since we found the concentration for NGC 288 is $c=1.0$ for $W_\text{o}=5.0$ that provides $\chi_{\nu}=0.48$ which is smaller than their value  $\chi_{\nu}=1.7$ with $W_\text{o}=5.8$ and close to unity. }  and 1.41. NGC 288 is a good example for a polytropic globular cluster while NGC 6254 for a non-polytropic cluster. Figure \ref{fig:fitting_poly_3} also depicts another example of polytropic-like globular cluster NGC 5139 whose surface brightness profile was reported in \citep{Meylan_1987}. The central part of the cluster deviate from polytrope model due to the weak cusp though the inner- and outer- halos are well fitted by the polytropic sphere. In the rest, we consider possible physics arguments that the low-concentration globular clusters may have structures like polytropic spheres (Section \ref{sec:posible_poly}) and its criticism (Section \ref{sec:critic_poly}).

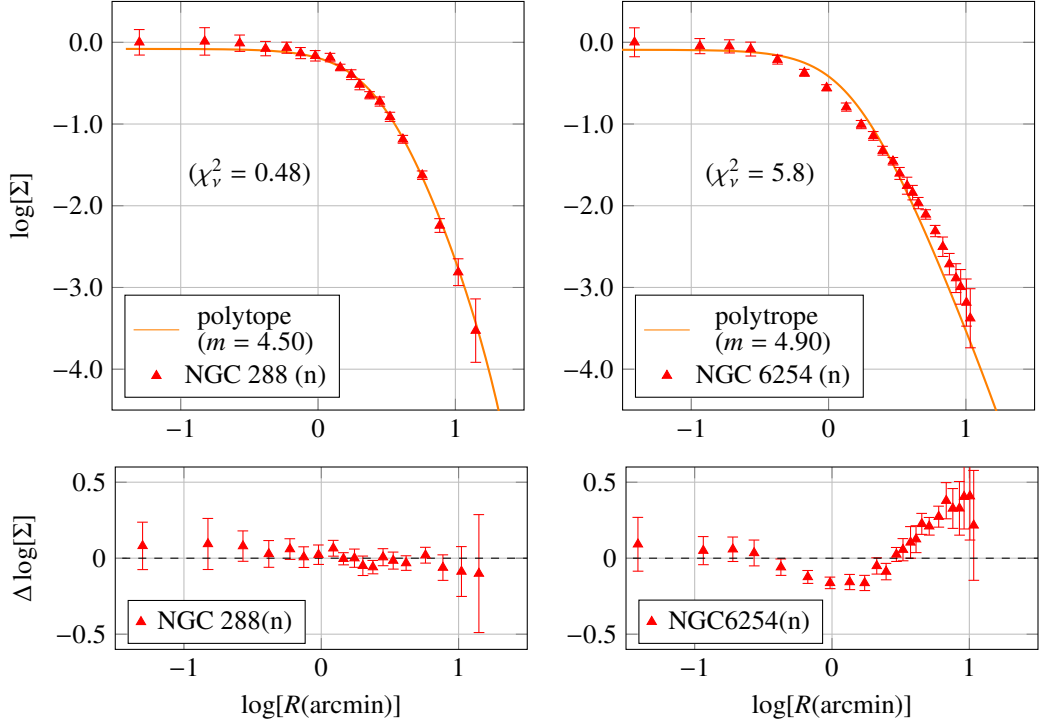
\begin{figure}[H]\centering
	\begin{tikzpicture}
	\begin{axis}[width=7cm,height=7cm, grid=major,xmin=-1.5,xmax=1.5,ymin=-4.5,ymax=0.5,ylabel=\normalsize{$\log[\Sigma]$},legend style={cells={align=left}},legend entries={\normalsize{polytope}\\\normalsize{$(m=4.50)$}, \normalsize{NGC 288 (n)}},yticklabel style={/pgf/number format/.cd,fixed zerofill,precision=1},legend pos=south west]
	\addlegendimage{no markers, orange};
	\addlegendimage{only marks, mark=triangle*, red};
	\addplot [color = orange ,mark=no,thick,solid] table[x index=0, y index=1]{R_Sig_NGC288_poly.txt};
	\addplot[color = red, only marks, mark=triangle*, mark options={red}, error bars,y dir=both, y explicit] table [x index=0, y index=1, y error index=2]{Rd_Sigd_Errbar_Err_NGC288_poly.txt};
	\node[black] at (-0.5,-1.6) {(\normalsize{$\chi^{2}_{\nu}=0.48$})} ; 
	\end{axis}
	\end{tikzpicture}\hspace{0.3cm}
	\begin{tikzpicture}
	\begin{axis}[width=7cm,height=7cm, grid=major,xmin=-1.5,xmax=1.5,ymin=-4.5,ymax=0.5,legend style={cells={align=left}},legend entries={\normalsize{polytrope}\\\normalsize{$(m=4.90)$}, \normalsize{NGC 6254 (n)}},yticklabel style={/pgf/number format/.cd,fixed zerofill,precision=1},legend pos=south west]
	\addlegendimage{no markers, orange};
	\addlegendimage{only marks, mark=triangle*, red};
	\addplot [color = orange ,mark=no,thick,solid] table[x index=0, y index=1]{R_Sig_NGC6254_poly.txt};
	\addplot[color = red, only marks, mark=triangle*, mark options={red}, error bars,y dir=both, y explicit] table [x index=0, y index=1, y error index=2]{Rd_Sigd_Errbar_Err_NGC6254_poly.txt};
	\node[black] at (-0.5,-1.6) {(\normalsize{$\chi^{2}_{\nu}=5.8$})} ; 
	\end{axis}
	\end{tikzpicture}
	\vspace{0.2cm}
	
	\begin{tikzpicture}
	\begin{axis}[width=7cm,height=4cm, grid=major,xmin=-1.5,xmax=1.5,ymin=-0.6,ymax=0.6,xlabel=\normalsize{$\log[R(\text{arcmin})]$ },ylabel=\normalsize{$\Delta\log[\Sigma]$},legend style={cells={align=left}},legend entries={\normalsize{ NGC 288(n)}},legend pos=south west]
	\addlegendimage{only marks, mark=triangle*, red};
	\addplot[color = red, only marks, mark=triangle*, mark options={red}, error bars,y dir=both, y explicit] table [x index=0, y index=3, y error index=2]{Rd_Sigd_Errbar_Err_NGC288_poly.txt};
	\draw[dashed] (-2.5,0)--(1.5,0);
	\end{axis}
	\end{tikzpicture}\hspace{0.3cm}
	\begin{tikzpicture}
	\begin{axis}[width=7cm,height=4cm, grid=major,xmin=-1.5,xmax=1.5,ymin=-0.6,ymax=0.6,xlabel=\normalsize{$\log[R(\text{arcmin})]$ },legend style={cells={align=left}},legend entries={\normalsize{ NGC6254(n)}},legend pos=south west]
	\addlegendimage{only marks, mark=triangle*, red};
	\addplot[color = red, only marks, mark=triangle*, mark options={red}, error bars,y dir=both, y explicit] table [x index=0, y index=3, y error index=2]{Rd_Sigd_Errbar_Err_NGC6254_poly.txt};
	\draw[dashed] (-2.5,0)--(1.5,0);
	\end{axis}
	\end{tikzpicture}
	\caption{Fitting of the polytropic sphere of index $m$ to the projected density $\Sigma$ of NGC 288 and NGC 6254 reported in \citep{Miocchi_2013}. The unit of $\Sigma$ is number per square of arcminutes and $\Sigma$ is normalized so that the density is unity at smallest radius for data. In the legends, (n) means `normal' or KM cluster as judged so in \citep{Djorgovski_1986}. $\Delta\log[\Sigma]$ is the corresponding deviation of $\Sigma$ from the model.}
	\label{fig:fitting_poly}
\end{figure}

\subsubsection{Discussion of low-concentration clusters being polytropic stellar polytrope}\label{sec:posible_poly}

The low-concentration clusters may have cores that are in states of non-equilibrium, possibly modeled by polytropic spheres, rather than a state of (local) thermodynamical equilibrium, \emph{if} the mass loss from the clusters is less significant. Considering that the King- and energy-truncated-ss-OAFP- models also fit the projected structural profiles of low-concentration globular clusters, the cores of such clusters are apparently well relaxed.  Yet, it is not clear whether their DFs actually have reached a local Maxwellian DF considering that their core relaxation times $t_\text{c.r.}$ are long ($1\gtrapprox$Gyr) and even the self-similar solution to the OAFP equation can not strictly reach a Maxwellian DF as shown in \citep{Ito_2020_2}.  

In the initial stage of relaxation evolution in globular clusters, the relaxation process may be dominated by its non-dominant effect on order of secular evolution time scale $t_\text{sec}\sim Nt_\text{cross}$ where $t_\text{cross}$ is the crossing time. The scale of half-mass relaxation time $t_\text{r}/t_\text{cross}\sim N/\ln[N]$ is the case when stars can orbit at possible maximum apocenter that is the tidal radius $r_\text{tid}$ and minimum pericenter that is order of $r_\text{tid}/N$. Yet, in the early stage of evolution the pericenter may be much larger on average. The extreme cases were discussed and mathematical formulated in \citep{Kandrup_1981,Kandrup_1988}. When stars are enough far from each other, ideally many of them separate from the others at order of $1/n^{1/3} \sim r_\text{tid}$. The dominant effect discussed in \citep{Chandra_1943c}, that corresponds with order of stellar separation from $ r_\text{tid}/N$ up to $1/n^{1/3}$, may not be important and (non-dominant) many-body effect can be more effective. For homogeneous self-gravitating systems, we can directly use the (local) relaxation time as the measure of relaxation, while in case of inhomogenous system we also need to consider the non-dominant (many-body relaxation) effect which is effective on time scales longer than $t_\text{c.r}$ by a factor of $\sim\ln[0.1N]$. The mathematical formulation including many-body effect for relaxation time is no longer logarithmic but it mitigates in collision integral as summation of Fourier series expansion under orbit-averaging of kinetic equation \citep{Polyachenko_1982,Ito_2018_4} and order of the many-body relaxation time is $\sim N t_\text{cross}$ rather than $N/\ln[N] t_\text{cross}$ in the early stage. \cite{Kandrup_1985} discussed some simple examples for this matter by neglecting the effect of evaporation and gravothermal instability, which is the case when the two-body relaxation processes are not dominant yet. To avoid evaporation, \cite{Kandrup_1985} considered the self-gravitating system is confined in a box and discussed a secular evolution by terming relaxation process 'anomalous' collision that may cause a deviation of stellar DF from Maxwellian DF on the secular evolution time scale. 

\begin{figure}[H]\centering
	\begin{tikzpicture}
	\begin{axis}[width=6.5cm,height=9cm, grid=major,xlabel=\normalsize{$c$},ylabel=\normalsize{$\eta_\text{M}$},xmin=0,xmax=2.8,ymin=-0.3,ymax=4,legend pos=north west]
	\addplot [only marks, color = red, mark=*, thick] table[x index=0, y index=1]{c_eta.txt};
	\addlegendentry{\normalsize{polytropic}} 
	\addplot [only marks, color = orange ,mark=o,thick] table[x index=0, y index=1]{cq_etaq.txt};
	\addlegendentry{\normalsize{polytropic?}} 
	\addplot [only marks, color = blue ,mark=triangle,thick]  table[x index=0, y index=1]{cno_etano.txt};
	\addlegendentry{\normalsize{non-polytropic}} 
	\draw [black, dashed] (0,1) -- (3,1);
	\end{axis}
	\end{tikzpicture}
	\caption{Secular relaxation parameter $\eta_\text{M}$ against concentration $c$. The values of concentration are adapted from \citep[][(2010 edition)]{Harris_1996}.}
	\label{fig:c_eta}
\end{figure}
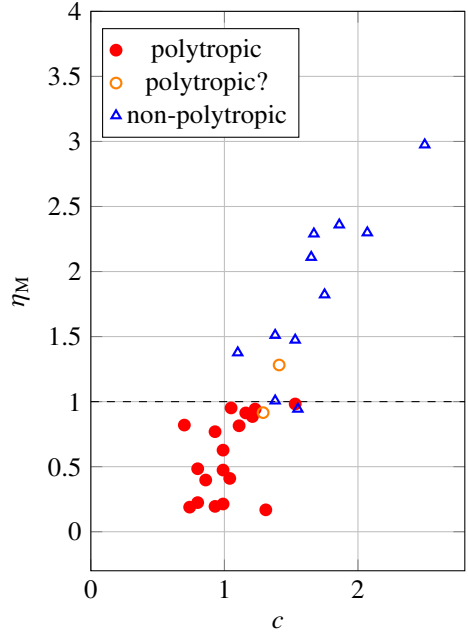

Table \ref{table:tcr_age_poly} shows the time-scales of dynamics; current and estimated initial relaxation times $t_\text{c.r.}$ and $t_\text{c.r.o}$ and age $t_\text{age}$ of clusters with the total mass $M$. We estimated the values of $t_\text{c.r.o}$ using the analysis of \citep{Lightman_1982}. Here what we would like to know is how many initial secular evolution times have already passed during their cluster ages. This may be measured by the following 'secular-evolution' parameter
\begin{align}
\eta_\text{M}\equiv\frac{t_\text{age}}{t_\text{sec}},
\end{align}      
where the secular evolution time is defined as
\begin{align}
t_\text{sec}\equiv \ln\left[0.11\frac{M}{M_\odot}\right]t_\text{c.r.o},
\end{align}
where $M_\odot$ is the solar mass and $M$ the dynamical mass for each cluster reported in \citep{Mandushev_1991}. The natural log and factor $0.11$ are introduced so that the mathematical expression for $t_\text{c.r.o}$ follows the core relaxation time of \citep{Spitzer_1988} and quantitatively the results of $N$-body simulations \citep{Aarseth_1998}.  In \citep{Lightman_1982} to calculate the completion rate, the initial relaxation time $t_\text{c.r.o}$ was compared to the order of cluster age $t_\text{o,age}$ while we would like to compare $t_\text{sec}$ and $t_\text{age}$ to see whether the clusters could have reached a state described by (local) Maxwellian DF in their cores. If $\eta_\text{M}\gtrapprox1$, the core of a cluster may be a state described by Maxwellian DF while if $\eta_\text{M}\lessapprox1$ the cluster may be in a non-equilibrium state at present; the latter would provide some insight of polytrope model being a possible model for low-concentration clusters.  Table \ref{table:tcr_age_poly} shows that the globular clusters well fitted by polytropic spheres have small $\eta_\text{M}$ $(0.20<\eta_\text{M}\lessapprox1)$. On one hand, the parameter $\eta_\text{M}$ of the clusters that could not be fitted by polytropes are $1\lessapprox\eta_\text{M}\lessapprox3.77$ on which the maximum value was achieved by NGC 7099 (one of PCC clusters). NGC 3201 and NGC 4590 are classified into the intermediate class in which a polytrope model is apparently fitted to the projected structural profiles at part of cluster radii. Figure \ref{fig:c_eta} shows the secular evolution rate against concentration $c$; for the concentration we employed the \cite[][(2010 edition)]{Harris_1996}'s values. It appears that $\eta_\text{M}=1$ is a good threshold to separate polytropic- and non-polytropic clusters. Especially when $c\approx1.5$ and $\eta_\text{M}\approx1$, both the polytropic- and non-polytropic clusters coexist.  

The realization of polytropic clusters was discussed by \citep{Taruya_2003_3} based on $N$-body simulation. Assuming a self-gravitating system of equal-mass is enclosed in an adiabatic container, they found the simulated distribution function can be well approximated by stellar polytropes even on time scales much longer than half-mass relaxation time. This was also confirmed using isotropic time-dependent Fokker-Planck model \citep{Taruya_2004}. Especially, \cite{Taruya_2003_3} also tested the system without an adiabatic wall; of course, due to the evaporation stellar DF largely deviates from the stellar polytrope while in the early stage of evolution the simulated DF seems well fitted by the DF for polytropes; in their work $m=5.7$ at $T=50$ seems provide a DF reasonably close to the DF for a polytrope. Also, the inner part of the systems and stellar DF at low energy are well fitted by DF for polytropes regardless of the effect of escaping stars. Their results implicates that the stellar DF and structural profile of star cluster can be alike a polytrope unless the effect of evaporation is dominant.

\begin{table*}\centering
	\ra{1.3}
	\scalebox{0.7}{
		\begin{tabular}{@{}lrrrrrl|l@{}}\toprule
			\textbf{polytropic} &$c$ & $t_\text{c.r.}$& $t_\text{c.r.o}$& $\log\left[\frac{M}{M_\text{o}}\right]$&$\eta_\text{M}$& $t_\text{age}$ & Reference for $t_\text{age}$\\
			cluster &      &(Gyr)  &(Gyr)  & & &(Gyr)&\\
			\midrule
			NGC 288 &0.99  &0.98 &2.0 &4.64 &0.63&10.62 &\citep{Forbes_2010}\\
			NGC 1261&1.16 &0.39&1.15& 5.17&0.913&10.24&\citep{Forbes_2010}\\
			NGC 5053&0.74 & 6.5& 8.2  &4.41 &0.19&12.29&\citep{Forbes_2010}\\
			NGC 5139&1.31   & 4.0& 5.5  &6.38 &0.17&11.52& \citep{Forbes_2010}\\
			NGC 5466&1.04  &2.2 &3.6  &4.85 &0.41&13.57 &\citep{Forbes_2010}\\
			NGC 5897&0.86 & 2.1& 3.5  &4.83 &0.40&12.3&\citep{Forbes_2010}\\
			NGC 5986&1.23 &0.38 &1.24  &5.48 &0.94&12.16&\citep{Forbes_2010}\\
			NGC 6101&0.80&1.6 &2.9  &4.83 &0.48&12.54&\citep{Forbes_2010}\\
			NGC 6205&1.53& 0.32&1.12  &5.59 &0.98&11.65&\citep{Forbes_2010}\\
			NGC 6402&0.99 &1.14  &2.35  &5.89  &0.47 &12.6 &\citep{Santos_2004}\\
			NGC 6496&0.70 &0.87  &2.0  & 4.29  &0.82 &12.42 &\citep{Forbes_2010}\\
			NGC 6712&1.05&0.40  &1.2  & 4.98  &0.95 &10.4 &\citep{Forbes_2010}\\
			NGC 6723&1.11&0.62  &1.7  & 5.15  &0.81 &13.06 &\citep{Forbes_2010}\\
			NGC 6809&0.93  &0.72 &1.8 &5.03 &0.77&13.0$\pm$0.3&\citep{Wang_2016}\\
			NGC 6981&1.21&0.52  &1.4   & 4.80   &0.89  &10.88  &\citep{Forbes_2010}\\
			Pal 3 &0.99 & 4.5 & 5.8   &4.36 &0.21&9.7&\citep{Forbes_2010}\\
			Pal 4 &0.93 & 5.2 & 6.5   &4.21 &0.19&9.5&\citep{Forbes_2010}\\
			Pal 14&0.80  & 7.1&  8.6  &3.83 &0.22&$13.2\pm0.3$&\citep{Sollima_2010}\\
			\bottomrule
	\end{tabular}}
	
	\vspace{0.6cm}
	
	\scalebox{0.7}{
		\begin{tabular}{@{}lrrrrrl|l@{}}\toprule
			\textbf{polytropic} &$c$ & $t_\text{c.r.}$& $t_\text{c.r.o}$& $\log\left[\frac{M}{M_\text{o}}\right]$&$\eta_\text{M}$& $t_\text{age}$ & Reference for $t_\text{age}$\\
			cluster? &      &(Gyr)  &(Gyr) &  & &(Gyr)&\\
			\midrule
			NGC 3201  &1.29 &0.41 &1.2  &5.05 &0.92 &10.24 &\citep{Forbes_2010}\\
			NGC 4590  &1.41 &0.28 & 1.1 &4.95 &1.29 &13.0 $\pm$1.0 &\citep{Dotter_2009}\\
			\bottomrule
	\end{tabular}}
	
	\vspace{0.6cm}
	
	\scalebox{0.7}{
		\begin{tabular}{@{}lrrrrrl|l@{}}\toprule
			\textbf{non-polytropic} &$c$ & $t_\text{c.r.}$& $t_\text{c.r.o}$& $\log\left[\frac{M}{M_\text{o}}\right]$& $\eta_\text{M}$& $t_\text{age}$ & Reference for $t_\text{age}$\\
			cluster &      &(Gyr)  &(Gyr) & &  &(Gyr)&\\
			\midrule
			NGC 1851  &1.86 &0.027&0.38 &5.42&2.36&9.2$\pm$1.1 &\citep{Salaris_2002}\\
			NGC 5634  &2.07 &0.047&0.53 & 5.18&2.30&11.8 &\citep{Forbes_2010}\\
			NGC 6121  &1.65 &0.079&0.61 &4.83&2.11&11.5$\pm$0.4 &\citep{Wang_2016}\\
			NGC 6144  &1.55 &0.60&1.7 & 4.76&0.94&13.82 &\citep{Forbes_2010}\\
			NGC 6254  &1.38 &0.16& 0.81 &5.06&1.51&11.39$\pm$1.1 &\citep{Forbes_2010}\\
			NGC 6273  &1.53 &0.33&1.1 & 6.03&1.47&11.90 &\citep{Forbes_2010}\\
			NGC 6352  &1.10 &0.29&1.1 & 4.57&1.37&12.67 &\citep{Forbes_2010}\\
			NGC 6388  &1.75 &0.052&0.553 & 6.16&1.82&12.03 &\citep{Forbes_2010}\\
			NGC 6626  &1.67 &0.042&0.52 & 5.36&2.29&12.1$\pm$1.0&\citep{Kerber_2018}DSED method\\
			NGC 6656  &1.38 &0.34&1.2 & 5.53&1.00&12.67 &\citep{Forbes_2010}\\
			NGC 7099(c)&2.50 &0.0023&0.35 &4.91&2.97&12.93 &\citep{Forbes_2010}\\
			\bottomrule
	\end{tabular}}
	\caption{Secular parameters calculated from the core relaxation times and ages of polytropic- and non-polytropic- clusters.  The current relaxation time $t_\text{c.r.}$ and  concentration $c$ are values reported in \citep[][(2010 edition)]{Harris_1996}. The total masses $\log\left[\frac{M}{M_\text{o}}\right]$ are from \citep{Mandushev_1991} where $M_\text{o}$ is solar mass. We adapted the ages of clusters from \citep{Forbes_2010} and we resorted to other sources when we found more recent data or could not find the cluster age in \citep{Forbes_2010}.}
	\label{table:tcr_age_poly}
\end{table*}

\subsubsection{Criticism on polytropic globular clusters}\label{sec:critic_poly}

The discussion for polytropic globular clusters in Section \ref{sec:posible_poly} is oversimplified in the sense that actual globular clusters are subject to mass spectrum (segregation) with stellar evolution and tidal effects (shock), hence the clusters are supposed to have lost a significant amount of stars while the relation of mass spectrum and polytrope sphere has not been discussed, which indicates polytropic globular clusters are a phenomenological concept.  In the case of \emph{isolated} $N$-body system with \emph{equal} mass, the cluster loses a small fraction $(\sim0.1\%)$ of total stellar mass in the first five initial relaxation time scale \citep{Baumgardt_2002_2}. Yet, more realistically mass segregation and tidal effect in general make faster the process leading to core collapse \citep{Spitzer_1988,Binney_2011} which was originally discussed for both multi-mass OAFP model \citep[e.g.][]{Chernoff_1990,Takahashi_2000} and $N$-body simulation \citep[e.g.][]{Fukushige_1996,Portegies_Zwart_1998,Baumgardt_2003b} in tidal field. Especially, relatively-new observation showed an unexpected feature of low-concentration clusters; the clusters have depleted mass functions of low-mass stars compared to high-concentration ones \citep{De_Marchi_2007}.  This result means that the lower-concentration clusters have lost more stars due to evaporation or tidal stripping, which was caused by mass segregation through two-body relaxation and tidal effect from the Galaxy. The excessive loss of low-mass stars from low-concentration clusters contradicts standard stellar dynamics in which higher-concentration clusters are supposed to have lost more low-mass stars due to more frequent two-body relaxation process. Based on direct $N$-body simulation, \citep{Baumgardt_2008} explained one possible interpretation for this issue by showing that the low-concentration clusters had already undergone primordial mass segregation in the early stage of evolution due to stellar evolution. This idea was extended to a sophisticated case study for one of low-concentration clusters, Palomar 4 \citep{Zonoozi_2017}; the total mass rapidly decreases only in the first 0.1 Gyr and the mass of the cluster calmly keeps decreasing. The decrease in mass depends on the orbit of Palomar 4 though, star's total number decreases in $10$ Gyr by approximately 60 \%.  It appears that the reason why the low-concentration clusters have polytropic structures is not directly due to little loss of stars. Hence, one needs to directly discuss the relation of the DF for polytrope and globular clusters that have experienced mass segregation, which has not been detailed.

Also, the present work does not discuss the projected line-of-sight velocity dispersion profiles of the energy-truncated ss-OAFP model. Many of the polytropic clusters are low-concentration clusters, which implicates accurate observational data are hard to be obtained compared to high-concentration clusters \citep[e.g.][]{Meylan_1997}. Perhaps, more recent data from \emph{Gaia 2} \citep[e.g.][]{Baumgardt_2018}, the ESO Multi-instrument Kinematic Survey (MIKiS)  \citep{Ferraro_2018} and more accurate kinematic data may differentiate the King model from the energy-truncated ss-OAFP model.

As a conclusion of the present section, we consider the polytropic globular clusters may 'phenomenologically' apply to the structure of low-concentration clusters in a similar way that the energy truncation for ss-OAFP model based on a polytrope is not endorsed by a strict first principle. Since the discussion is not matured on the relation between mass segregation (spectrum) and stellar DF for polytrope, we need further examine this topic in future with more accurate observed structural- and kinematic- data and realistic numerical simulation. We believe especially the kinematic data can draw a line in applicability between the King and our model.

\section{Conclusion}\label{sec:conclusion_finite_OAFP}

The present work introduced a phenomenological model i.e. the energy-truncated ss-OAFP model that can fit the projected density profiles for at least half of Milky Way globular clusters including PCC clusters with resolved cores, aiming at establishing a model with wide applicability compared to the classical isotropic one-component King model. The present model is a linear algebraic combination of the DFs for the ss-OAFP model and polytropic sphere of $m$; the latter was weighted by a factor of $\delta$. The optimal value of $m$ was identified as 3.9 by comparing the concentrations and tidal (limiting) radii of the King- and our models. After this procedure, this new model has only three degrees of freedom that are the same as those of the King model while our model can even fit to the projected structural profiles of PCC clusters with resolved core in addition to those of KM clusters. The fitting results provided a completion rate of core collapse against concentration including PCC clusters with resolved-core that is consistent to standard stellar dynamics. Also, our model is more useful compared to the King model to single out KM clusters whose morphology is close to the core-collapsing cluster i.e. high concentration $\bar{c}\geq2.0$; examples are NGC 1851, NGC 6626 and NGC 6517.

Our model also can apply to globular clusters in a broad range of radii $0.01\sim 10$ arcminutes with higher $m(\geq4.2)$, however the energy-truncation based on polytrope provides unrealistically large tidal radius. Hence, we also proposed an approximated form of the energy-truncated ss-OAFP model to avoid unrealistic tidal radii. Lastly, we discussed that the low-concentration globular clusters may be polytropic in the sense that their projected structural profiles are well fitted by those of polytropes. Yet, the physical arguments and previous numerical results for the polytropic globular clusters are not well established. Hence, we consider the polytropic clusters are a heuristic idea for now, which intrigues us to work in future on examining (i) the relation of mass spectrum (segregation) of star clusters and stellar DF for polytropes and (ii) the relation of the low-concentration clusters with stellar polytrope that obeys the generalized statistical mechanics based on Tsallis entropy and (iii) applicability of the King- and our models to kinematic profiles.

\section*{Acknowledgements}
The present work is partial fulfillment of the degree of Philosophy at CUNY graduate center. 
\vspace{1cm}
\appendix
\begin{appendices}
	\section{Fitting of the finite ss-OAFP model to King-model clusters}\label{Appendix_OAFP_Fit}
	The present appendix shows fitting of the energy-truncated ss-OAFP model to projected structural profiles of Galactic KM clusters reported in \citep{Kron_1984,Noyola_2006,Miocchi_2013}. For fitting to \citep{Miocchi_2013}'s data we chose the fitting parameters in the model so that $\chi_{\nu}^{2}$ is minimized. For fitting to \citep{Kron_1984, Noyola_2006}'s data, the deviation of the model from the data plots are minimized. Section \ref{sec:KM_OAFP_Miocchi}, \ref{sec:KM_OAFP} and \ref{Fitting_KM_Noyola} show fitting of our model to the KM stars reported in \citep{Miocchi_2013}, \citep{Kron_1984} and \citep{Noyola_2006}.

\subsection{KM cluster \citep{Miocchi_2013}}\label{sec:KM_OAFP_Miocchi}

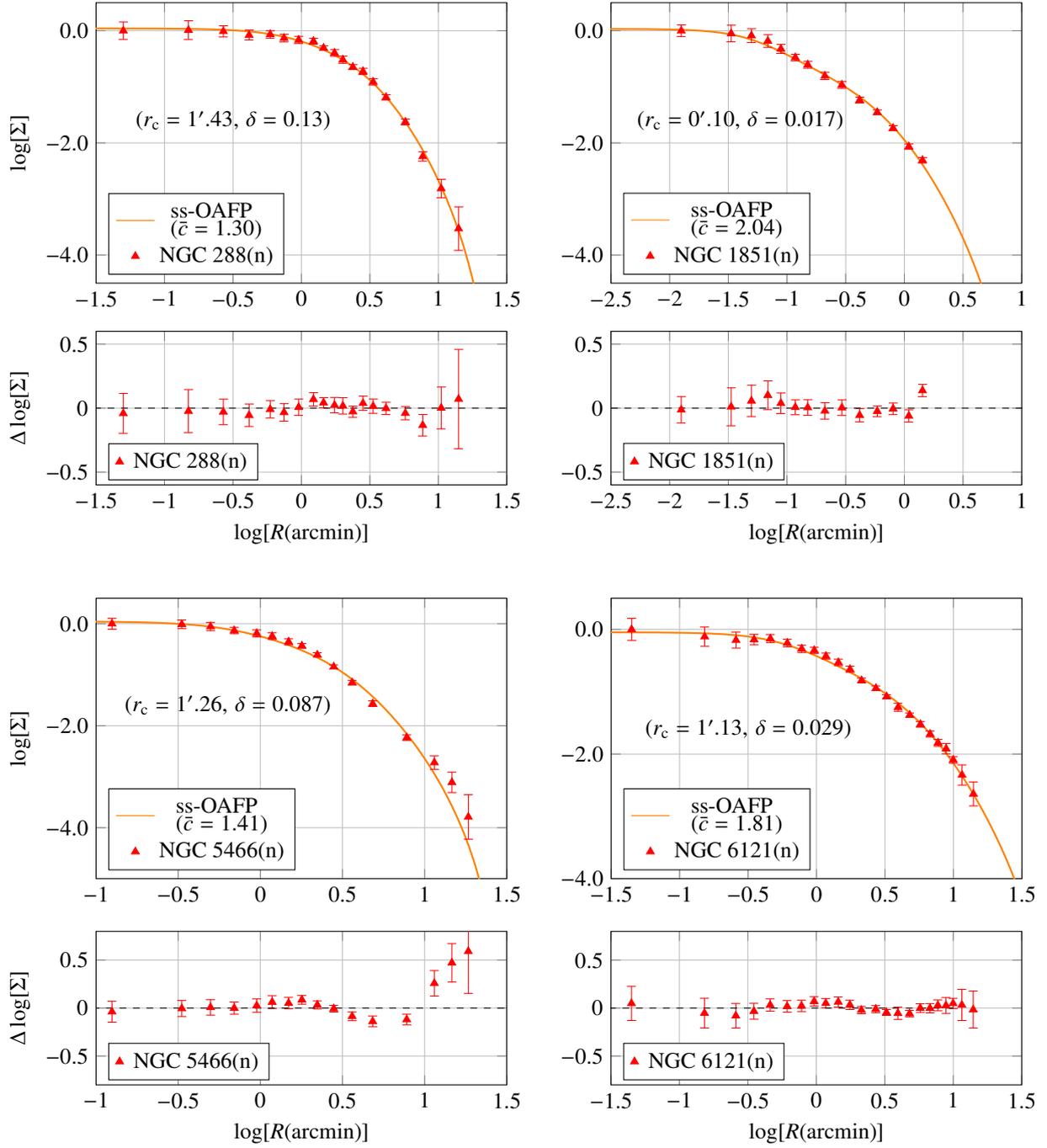
\begin{figure}[H]
	\begin{tikzpicture}
	\begin{axis}[width=8cm,height=6cm, grid=major,xmin=-1.5,xmax=1.5,ymin=-4.5,ymax=0.5,ylabel=\normalsize{$\log[\Sigma]$},legend style={cells={align=left}},legend entries={\normalsize{ss-OAFP}\\\normalsize{$(\bar{c}=1.30)$}, \normalsize{NGC 288(n)}},yticklabel style={/pgf/number format/.cd,fixed zerofill,precision=1},legend pos=south west]
	\addlegendimage{no markers, orange};
	\addlegendimage{only marks, mark=triangle*, red};
	\addplot [color = orange ,mark=no,thick,solid] table[x index=0, y index=1]{R_Sig_NGC288.txt};
	\addplot[color = red, only marks, mark=triangle*, mark options={red}, error bars,y dir=both, y explicit] table [x index=0, y index=1, y error index=2]{Rd_Sigd_Errbar_Err_NGC288.txt};
	\node[black] at (-0.5,-1.6) {(\normalsize{$r_\text{c}=1\textquotesingle.43, \, \delta=0.13$})} ; 
	\end{axis}
	\end{tikzpicture}\hspace{0.3cm}
	\begin{tikzpicture}
	\begin{axis}[width=8cm,height=6cm, grid=major,xmin=-2.5,xmax=1.0,ymin=-4.5,ymax=0.5,legend style={cells={align=left}},legend entries={\normalsize{ss-OAFP}\\\normalsize{$(\bar{c}=2.04)$}, \normalsize{NGC 1851(n)}},yticklabel style={/pgf/number format/.cd,fixed zerofill,precision=1},legend pos=south west]
	\addlegendimage{no markers, orange};
	\addlegendimage{only marks, mark=triangle*, red};
	\addplot [color = orange ,mark=no,thick,solid] table[x index=0, y index=1]{R_Sig_NGC1851.txt};
	\addplot[color = red, only marks, mark=triangle*, mark options={red}, error bars,y dir=both, y explicit] table [x index=0, y index=1, y error index=2]{Rd_Sigd_Errbar_Err_NGC1851.txt};
	\node[black] at (-1.4,-1.6) {(\normalsize{$r_\text{c}=0\textquotesingle.10, \, \delta=0.017$})} ; 
	\end{axis}
	\end{tikzpicture}
	
	\vspace{0.2cm}
	
	\begin{tikzpicture}
	\begin{axis}[width=8cm,height=4cm, grid=major,xmin=-1.5,xmax=1.5,ymin=-0.6,ymax=0.6,xlabel=\normalsize{$\log[R\text{(arcmin)}]$ },ylabel=\normalsize{$\Delta\log[\Sigma]$},legend style={cells={align=left}},legend entries={\normalsize{ NGC 288(n)}},legend pos=south west]
	\addlegendimage{only marks, mark=triangle*, red};
	\addplot[color = red, only marks, mark=triangle*, mark options={red}, error bars,y dir=both, y explicit] table [x index=0, y index=3, y error index=2]{Rd_Sigd_Errbar_Err_NGC288.txt};
	\draw[dashed] (-2.5,0)--(1.5,0);
	\end{axis}
	\end{tikzpicture}\hspace{0.3cm}
	\begin{tikzpicture}
	\begin{axis}[width=8cm,height=4cm, grid=major,xmin=-2.5,xmax=1.0,ymin=-0.6,ymax=0.6,xlabel=\normalsize{$\log[R\text{(arcmin)}]$ },legend style={cells={align=left}},legend entries={\normalsize{ NGC 1851(n)}},legend pos=south west]
	\addlegendimage{only marks, mark=triangle*, red};
	\addplot[color = red, only marks, mark=triangle*, mark options={red}, error bars,y dir=both, y explicit] table [x index=0, y index=3, y error index=2]{Rd_Sigd_Errbar_Err_NGC1851.txt};
	\draw[dashed] (-2.5,0)--(1.5,0);
	\end{axis}
	\end{tikzpicture}
	
	\vspace{0.7cm}
	
	\begin{tikzpicture}
	\begin{axis}[width=8cm,height=6cm, grid=major,xmin=-1.0,xmax=1.5,ymin=-5.0,ymax=0.5,ylabel=\normalsize{$\log[\Sigma]$},legend style={cells={align=left}},legend entries={\normalsize{ss-OAFP}\\\normalsize{$(\bar{c}=1.41)$}, \normalsize{NGC 5466(n)}},yticklabel style={/pgf/number format/.cd,fixed zerofill,precision=1},legend pos=south west]
	\addlegendimage{no markers, orange};
	\addlegendimage{only marks, mark=triangle*, red};
	\addplot [color = orange ,mark=no,thick,solid] table[x index=0, y index=1]{R_Sig_NGC5466.txt};
	\addplot[color = red, only marks, mark=triangle*, mark options={red}, error bars,y dir=both, y explicit] table [x index=0, y index=1, y error index=2]{Rd_Sigd_Errbar_Err_NGC5466.txt};
	\node[black] at (-0.2,-1.6) {(\normalsize{$r_\text{c}=1\textquotesingle.26, \, \delta=0.087$})} ; 
	\end{axis}
	\end{tikzpicture}\hspace{0.3cm}
	\begin{tikzpicture}
	\begin{axis}[width=8cm,height=6cm, grid=major,xmin=-1.5,xmax=1.5,ymin=-4.0,ymax=0.5,legend style={cells={align=left}},legend entries={\normalsize{ss-OAFP}\\\normalsize{$(\bar{c}=1.81)$}, \normalsize{NGC 6121(n)}},yticklabel style={/pgf/number format/.cd,fixed zerofill,precision=1},legend pos=south west]
	\addlegendimage{no markers, orange};
	\addlegendimage{only marks, mark=triangle*, red};
	\addplot [color = orange ,mark=no,thick,solid] table[x index=0, y index=1]{R_Sig_NGC6121.txt};
	\addplot[color = red, only marks, mark=triangle*, mark options={red}, error bars,y dir=both, y explicit] table [x index=0, y index=1, y error index=2]{Rd_Sigd_Errbar_Err_NGC6121.txt};
	\node[black] at (-0.5,-1.6) {(\normalsize{$r_\text{c}=1\textquotesingle.13, \, \delta=0.029$})} ; 
	\end{axis}
	\end{tikzpicture}
	
	\vspace{0.3cm}
	
	\begin{tikzpicture}
	\begin{axis}[width=8cm,height=4cm, grid=major,xmin=-1.0,xmax=1.5,ymin=-0.8,ymax=0.8,xlabel=\normalsize{$\log[R\text{(arcmin)}]$ },ylabel=\normalsize{$\Delta\log[\Sigma]$},legend style={cells={align=left}},legend entries={\normalsize{ NGC 5466(n)}},legend pos=south west]
	\addlegendimage{only marks, mark=triangle*, red};
	\addplot[color = red, only marks, mark=triangle*, mark options={red}, error bars,y dir=both, y explicit] table [x index=0, y index=3, y error index=2]{Rd_Sigd_Errbar_Err_NGC5466.txt};
	\draw[dashed] (-2.5,0)--(1.5,0);
	\end{axis}
	\end{tikzpicture}\hspace{0.3cm}
	\begin{tikzpicture}
	\begin{axis}[width=8cm,height=4cm, grid=major,xmin=-1.5,xmax=1.5,ymin=-0.8,ymax=0.8,xlabel=\normalsize{$\log[R\text{(arcmin)}]$ },legend style={cells={align=left}},legend entries={\normalsize{ NGC 6121(n)}},legend pos=south west]
	\addlegendimage{only marks, mark=triangle*, red};
	\addplot[color = red, only marks, mark=triangle*, mark options={red}, error bars,y dir=both, y explicit] table [x index=0, y index=3, y error index=2]{Rd_Sigd_Errbar_Err_NGC6121.txt};
	\draw[dashed] (-2.5,0)--(1.0,0);
	\end{axis}
	\end{tikzpicture}
	\caption{Fitting of the energy-truncated ss-OAFP model ($m=3.9$) to the projected density $\Sigma$ of NGC 288, NGC 1851, NGC 5466 and NGC 6121 reported in from \citep{Miocchi_2013}. The unit of $\Sigma$ is number per square of arcminutes and $\Sigma$ is normalized so that the density is unity at the smallest radius for data. In the legends, (n) means `normal' cluster that can be fitted by the King model as judged so in \citep{Djorgovski_1986}. $\Delta\log[\Sigma]$ is the corresponding deviation of $\Sigma$ from the model on log scale.}
	\label{fig:fitting_KM_Miocchi_1}
\end{figure}

\begin{figure}[H]
	\begin{tikzpicture}
	\begin{axis}[width=8cm,height=6cm, grid=major,xmin=-1.0,xmax=0.7,ymin=-5.0,ymax=0.5,ylabel=\normalsize{$\log[\Sigma]$},legend style={cells={align=left}},legend entries={\normalsize{ss-OAFP}\\\normalsize{$(\bar{c}=1.04)$}, \normalsize{Pal 14 (n)}},yticklabel style={/pgf/number format/.cd,fixed zerofill,precision=1},legend pos=south west]
	\addlegendimage{no markers, orange};
	\addlegendimage{only marks, mark=triangle*, red};
	\addplot [color = orange ,mark=no,thick,solid] table[x index=0, y index=1]{R_Sig_Pal14.txt};
	\addplot[color = red, only marks, mark=triangle*, mark options={red}, error bars,y dir=both, y explicit] table [x index=0, y index=1, y error index=2]{Rd_Sigd_Errbar_Err_Pal14.txt};
	\node[black] at (-0.3,-2.0) {(\normalsize{$r_\text{c}=0\textquotesingle.85, \, \delta=1.2$})} ; 
	\end{axis}
	\end{tikzpicture}\hspace{0.3cm}
	\begin{tikzpicture}
	\begin{axis}[width=8cm,height=6cm, grid=major,xmin=-1.7,xmax=0.7,ymin=-4.0,ymax=0.5,legend style={cells={align=left}},legend entries={\normalsize{ss-OAFP}\\\normalsize{$(\bar{c}=1.69)$}, \normalsize{Trz 5 (n)}},yticklabel style={/pgf/number format/.cd,fixed zerofill,precision=1},legend pos=south west]
	\addlegendimage{no markers, orange};
	\addlegendimage{only marks, mark=triangle*, red};
	\addplot [color = orange ,mark=no,thick,solid] table[x index=0, y index=1]{R_Sig_Trz5.txt};
	\addplot[color = red, only marks, mark=triangle*, mark options={red}, error bars,y dir=both, y explicit] table [x index=0, y index=1, y error index=2]{Rd_Sigd_Errbar_Err_Trz5.txt};
	\node[black] at (-0.9,-2.0) {(\normalsize{$r_\text{c}=0\textquotesingle.15, \, \delta=0.039$})} ; 
	\end{axis}
	\end{tikzpicture}
	
	\vspace{0.3cm}
	
	\begin{tikzpicture}
	\begin{axis}[width=8cm,height=4cm, grid=major,xmin=-1.0,xmax=0.7,ymin=-0.8,ymax=0.8,xlabel=\normalsize{$\log[R\text{(arcmin)}]$ },ylabel=\normalsize{$\Delta\log[\Sigma]$},legend style={cells={align=left}},legend entries={\normalsize{ Pal 14 (n)}},legend pos=south west]
	\addlegendimage{only marks, mark=triangle*, red};
	\addplot[color = red, only marks, mark=triangle*, mark options={red}, error bars,y dir=both, y explicit] table [x index=0, y index=3, y error index=2]{Rd_Sigd_Errbar_Err_Pal14.txt};
	\draw[dashed] (-2.5,0)--(1.0,0);
	\end{axis}
	\end{tikzpicture}\hspace{0.3cm}
	\begin{tikzpicture}
	\begin{axis}[width=8cm,height=4cm, grid=major,xmin=-1.7,xmax=0.7,ymin=-0.8,ymax=0.8,xlabel=\normalsize{$\log[R\text{(arcmin)}]$ },legend style={cells={align=left}},legend entries={\normalsize{ Trz 5 (n)}},legend pos=south west]
	\addlegendimage{only marks, mark=triangle*, red};
	\addplot[color = red, only marks, mark=triangle*, mark options={red}, error bars,y dir=both, y explicit] table [x index=0, y index=3, y error index=2]{Rd_Sigd_Errbar_Err_Trz5.txt};
	\draw[dashed] (-2.5,0)--(1.0,0);
	\end{axis}
	\end{tikzpicture}
	
	\vspace{0.6cm}
	
	\begin{tikzpicture}
	\begin{axis}[width=8cm,height=6cm, grid=major,xmin=-1.5,xmax=1.3,ymin=-5.0,ymax=0.5,ylabel=\normalsize{$\log[\Sigma]$},legend style={cells={align=left}},legend entries={\normalsize{ss-OAFP}\\\normalsize{$(\bar{c}=1.37)$}, \normalsize{6809 (n)}},yticklabel style={/pgf/number format/.cd,fixed zerofill,precision=1},legend pos=south west]
	\addlegendimage{no markers, orange};
	\addlegendimage{only marks, mark=triangle*, red};
	\addplot [color = orange ,mark=no,thick,solid] table[x index=0, y index=1]{R_Sig_NGC6809.txt};
	\addplot[color = red, only marks, mark=triangle*, mark options={red}, error bars,y dir=both, y explicit] table [x index=0, y index=1, y error index=2]{Rd_Sigd_Errbar_Err_NGC6809.txt};
	\node[black] at (-0.3,-2.0) {(\normalsize{$r_\text{c}=1\textquotesingle.63, \, \delta=0.1$})} ; 
	\end{axis}
	\end{tikzpicture}\hspace{0.3cm}
	
	\vspace{0.3cm}
	
	\begin{tikzpicture}
	\begin{axis}[width=8cm,height=4cm, grid=major,xmin=-1.5,xmax=1.3,ymin=-0.8,ymax=0.8,xlabel=\normalsize{$\log[R\text{(arcmin)}]$ },ylabel=\normalsize{$\Delta\log[\Sigma]$},legend style={cells={align=left}},legend entries={\normalsize{ NGC 6809 (n)}},legend pos=south west]
	\addlegendimage{only marks, mark=triangle*, red};
	\addplot[color = red, only marks, mark=triangle*, mark options={red}, error bars,y dir=both, y explicit] table [x index=0, y index=3, y error index=2]{Rd_Sigd_Errbar_Err_NGC6809.txt};
	\draw[dashed] (-2.5,0)--(1.5,0);
	\end{axis}
	\end{tikzpicture}\hspace{0.3cm}
	\caption{Fitting of the energy-truncated ss-OAFP model ($m=3.9$) to the projected density $\Sigma$ of Palomar 14, Terzan 5 and NGC 6809 reported in \citep{Miocchi_2013}. The unit of $\Sigma$ is number per square of arcminutes and $\Sigma$ is normalized so that the density is unity at the smallest radius for data. In the legends, (n) means `normal' cluster that can be fitted by the King model as judged so in \citep{Djorgovski_1986}. $\Delta\log[\Sigma]$ is the corresponding deviation of $\Sigma$ from the model on log scale.}
	\label{fig:fitting_KM_Miocchi_3}
\end{figure}
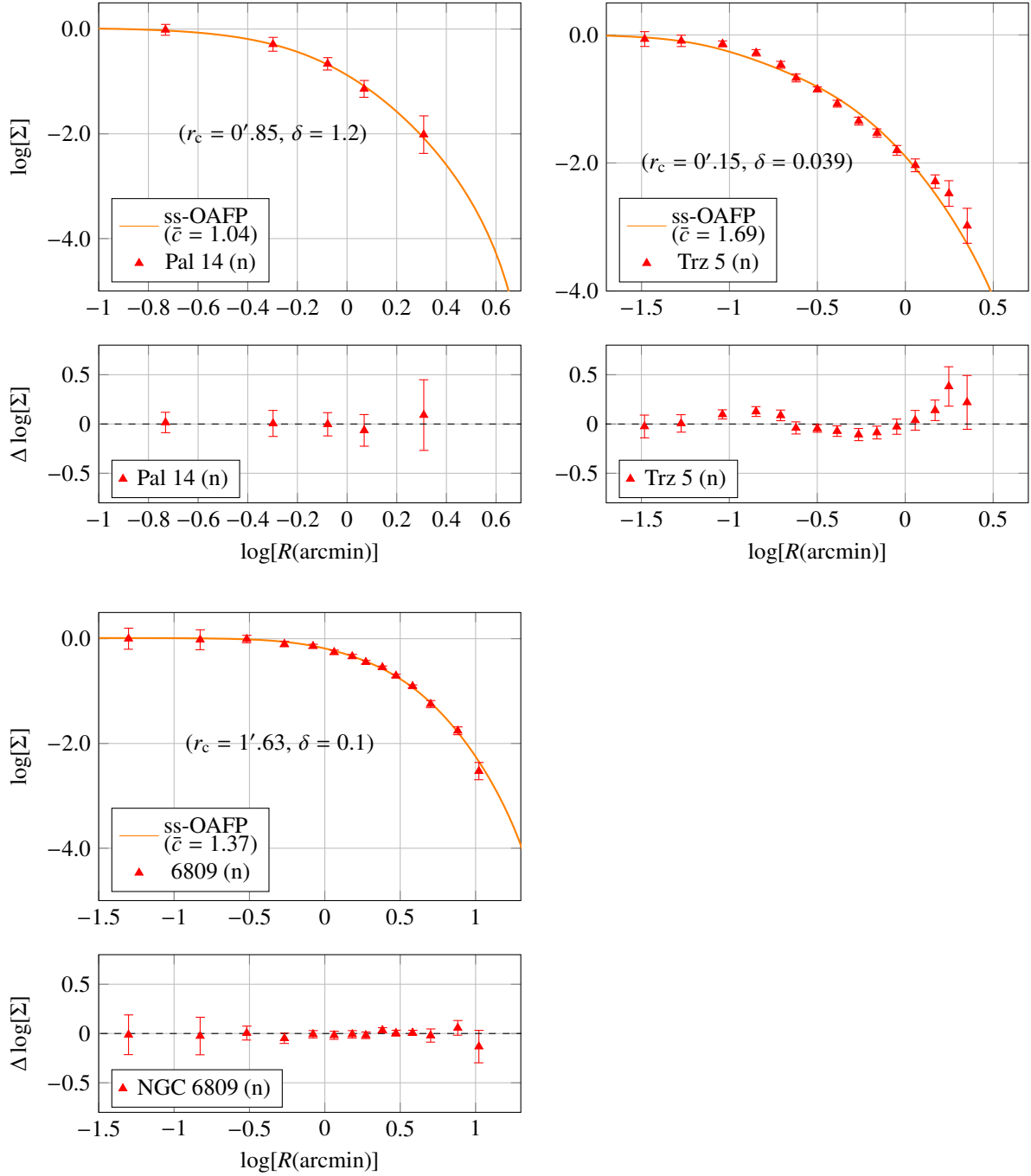

Figures \ref{fig:fitting_KM_Miocchi_1}, \ref{fig:fitting_KM_Miocchi_2} and \ref{fig:fitting_KM_Miocchi_3} depict the energy-truncated ss-OAFP model fitted to the projected density profiles of KM clusters reported in \citep{Miocchi_2013}. Table \ref{table:core_tidal_radius_moicchi} compares the structural parameters of our model, King model and Wilson model. Majority of the parameters obtained from our model is greater than those of the King model but less than those of the Wilson model. The energy-truncated ss-OAFP  model does not completely fit the structures of NGC 5466 and Terzan 5 while the King- and Wilson- models do fit them, which implies that they are less close to neither of the states of core collapse (or gravothermal instability phase) nor polytropic sphere (possibly collisionless phase).  

\begin{figure}[H]
	\begin{tikzpicture}
	\begin{axis}[width=8cm,height=6cm, grid=major,xmin=-1.5,xmax=1.2,ymin=-4.5,ymax=0.5,ylabel=\normalsize{$\log[\Sigma]$},legend style={cells={align=left}},legend entries={\normalsize{ss-OAFP}\\\normalsize{$(\bar{c}=1.64)$}, \normalsize{NGC 6254(n)}},yticklabel style={/pgf/number format/.cd,fixed zerofill,precision=1},legend pos=south west]
	\addlegendimage{no markers, orange};
	\addlegendimage{only marks, mark=triangle*, red};
	\addplot [color = orange ,mark=no,thick,solid] table[x index=0, y index=1]{R_Sig_NGC6254.txt};
	\addplot[color = red, only marks, mark=triangle*, mark options={red}, error bars,y dir=both, y explicit] table [x index=0, y index=1, y error index=2]{Rd_Sigd_Errbar_Err_NGC6254.txt};
	\node[black] at (-0.5,-1.6) {(\normalsize{$r_\text{c}=0\textquotesingle68, \, \delta=0.044$})} ; 
	\end{axis}
	\end{tikzpicture}\hspace{0.3cm}
	\begin{tikzpicture}
	\begin{axis}[width=8cm,height=6cm, grid=major,xmin=-2.0,xmax=1.0,ymin=-4.5,ymax=0.5,legend style={cells={align=left}},legend entries={\normalsize{ss-OAFP}\\\normalsize{$(\bar{c}=2.12)$}, \normalsize{NGC 6626(n)}},yticklabel style={/pgf/number format/.cd,fixed zerofill,precision=1},legend pos=south west]
	\addlegendimage{no markers, orange};
	\addlegendimage{only marks, mark=triangle*, red};
	\addplot [color = orange ,mark=no,thick,solid] table[x index=0, y index=1]{R_Sig_NGC6626.txt};
	\addplot[color = red, only marks, mark=triangle*, mark options={red}, error bars,y dir=both, y explicit] table [x index=0, y index=1, y error index=2]{Rd_Sigd_Errbar_Err_NGC6626.txt};
	\node[black] at (-1.0,-1.6) {(\normalsize{$r_\text{c}=0\textquotesingle.19, \, \delta=0.014$})} ; 
	\end{axis}
	\end{tikzpicture}
	
	\vspace{0.2cm}
	
	\begin{tikzpicture}
	\begin{axis}[width=8cm,height=4cm, grid=major,xmin=-1.5,xmax=1.2,ymin=-0.6,ymax=0.6,xlabel=\normalsize{$\log[R\text{(arcmin)}]$ },ylabel=\normalsize{$\Delta\log[\Sigma]$},legend style={cells={align=left}},legend entries={\normalsize{ NGC 6254(n)}},legend pos=south west]
	\addlegendimage{only marks, mark=triangle*, red};
	\addplot[color = red, only marks, mark=triangle*, mark options={red}, error bars,y dir=both, y explicit] table [x index=0, y index=3, y error index=2]{Rd_Sigd_Errbar_Err_NGC6254.txt};
	\draw[dashed] (-2.5,0)--(1.0,0);
	\end{axis}
	\end{tikzpicture}\hspace{0.3cm}
	\begin{tikzpicture}
	\begin{axis}[width=8cm,height=4cm, grid=major,xmin=-2.0,xmax=1.0,ymin=-0.6,ymax=0.6,xlabel=\normalsize{$\log[R\text{(arcmin)}]$ },legend style={cells={align=left}},legend entries={\normalsize{ NGC 6626(n)}},legend pos=south west]
	\addlegendimage{only marks, mark=triangle*, red};
	\addplot[color = red, only marks, mark=triangle*, mark options={red}, error bars,y dir=both, y explicit] table [x index=0, y index=3, y error index=2]{Rd_Sigd_Errbar_Err_NGC6626.txt};
	\draw[dashed] (-2.5,0)--(1.0,0);
	\end{axis}
	\end{tikzpicture}
	
	\vspace{0.7cm}
	
	\begin{tikzpicture}
	\begin{axis}[width=8cm,height=6cm, grid=major,xmin=-1.0,xmax=0.7,ymin=-5.0,ymax=0.5,ylabel=\normalsize{$\log[\Sigma]$},legend style={cells={align=left}},legend entries={\normalsize{ss-OAFP}\\\normalsize{$(\bar{c}=1.03)$}, \normalsize{Pal 3 (n)}},yticklabel style={/pgf/number format/.cd,fixed zerofill,precision=1},legend pos=south west]
	\addlegendimage{no markers, orange};
	\addlegendimage{only marks, mark=triangle*, red};
	\addplot [color = orange ,mark=no,thick,solid] table[x index=0, y index=1]{R_Sig_Pal3.txt};
	\addplot[color = red, only marks, mark=triangle*, mark options={red}, error bars,y dir=both, y explicit] table [x index=0, y index=1, y error index=2]{Rd_Sigd_Errbar_Err_Pal3.txt};
	\node[black] at (-0.3,-2.0) {(\normalsize{$r_\text{c}=0\textquotesingle.55, \, \delta=1.3$})} ; 
	\end{axis}
	\end{tikzpicture}\hspace{0.3cm}
	\begin{tikzpicture}
	\begin{axis}[width=8cm,height=6cm, grid=major,xmin=-1.5,xmax=0.7,ymin=-4.0,ymax=0.5,legend style={cells={align=left}},legend entries={\normalsize{ss-OAFP}\\\normalsize{$(\bar{c}=1.16)$}, \normalsize{Pal 4 (n)}},yticklabel style={/pgf/number format/.cd,fixed zerofill,precision=1},legend pos=south west]
	\addlegendimage{no markers, orange};
	\addlegendimage{only marks, mark=triangle*, red};
	\addplot [color = orange ,mark=no,thick,solid] table[x index=0, y index=1]{R_Sig_Pal4.txt};
	\addplot[color = red, only marks, mark=triangle*, mark options={red}, error bars,y dir=both, y explicit] table [x index=0, y index=1, y error index=2]{Rd_Sigd_Errbar_Err_Pal4.txt};
	\node[black] at (-0.7,-1.6) {(\normalsize{$r_\text{c}=0\textquotesingle.46, \, \delta=0.27$})} ; 
	\end{axis}
	\end{tikzpicture}
	
	\vspace{0.3cm}
	
	\begin{tikzpicture}
	\begin{axis}[width=8cm,height=4cm, grid=major,xmin=-1.0,xmax=0.7,ymin=-0.8,ymax=0.8,xlabel=\normalsize{$\log[R\text{(arcmin)}]$ },ylabel=\normalsize{$\Delta\log[\Sigma]$},legend style={cells={align=left}},legend entries={\normalsize{ Pal 3 (n)}},legend pos=south west]
	\addlegendimage{only marks, mark=triangle*, red};
	\addplot[color = red, only marks, mark=triangle*, mark options={red}, error bars,y dir=both, y explicit] table [x index=0, y index=3, y error index=2]{Rd_Sigd_Errbar_Err_Pal3.txt};
	\draw[dashed] (-2.5,0)--(1.0,0);
	\end{axis}
	\end{tikzpicture}\hspace{0.3cm}
	\begin{tikzpicture}
	\begin{axis}[width=8cm,height=4cm, grid=major,xmin=-1.5,xmax=0.7,ymin=-0.8,ymax=0.8,xlabel=\normalsize{$\log[R\text{(arcmin)}]$ },legend style={cells={align=left}},legend entries={\normalsize{ Pal 4 (n)}},legend pos=south west]
	\addlegendimage{only marks, mark=triangle*, red};
	\addplot[color = red, only marks, mark=triangle*, mark options={red}, error bars,y dir=both, y explicit] table [x index=0, y index=3, y error index=2]{Rd_Sigd_Errbar_Err_Pal4.txt};
	\draw[dashed] (-2.5,0)--(1.0,0);
	\end{axis}
	\end{tikzpicture}
	\caption{Fitting of the energy-truncated ss-OAFP model ($m=3.9$) to the projected density $\Sigma$ of NGC 6254, NGC 6626, Pal 3 and Pal 4 reported in \citep{Miocchi_2013}. The unit of $\Sigma$ is number per square of arcminutes and $\Sigma$ is normalized so that the density is unity at the smallest radius for data. In the legends, (n) means `normal' cluster that can be fitted by the King model as judged so in \citep{Djorgovski_1986}. $\Delta\log[\Sigma]$ is the corresponding deviation of $\Sigma$ from the model on log scale.}
	\label{fig:fitting_KM_Miocchi_2}
\end{figure}

\begin{table*}
	\ra{1.3}
	\begin{tabular}{@{}l|llr|llr|llr@{}}\toprule
		&\multicolumn{3}{l|}{NGC 288}    & \multicolumn{3}{l|}{NGC 1851} &\multicolumn{3}{l}{NGC 5466} \\ 
		\midrule
		&$c$ &$r_\text{c}$&$r_\text{tid}$&$c$ &$r_\text{c}$&$r_\text{tid}$&$c$ &$r_\text{c}$&$r_\text{tid}$\\
		\midrule
		\textbf{ss-OAFP model} &1.30 &1.43  &   28.9     &2.04 &  0.10    &  11.1      &1.41 &1.26 &32.4\\
		\midrule
		\textbf{King model}  &    &            &              &    &            &              &    &     &    \\
		\citep{Miocchi_2013} &1.20&1.17        &   21       &1.95&  0.09      &    8.3         &1.31&1.20&26.3\\	
		\midrule
		\textbf{Wilson model}&    &            &              &    &            &              &    &     &    \\
		\citep{Miocchi_2013} &1.10&1.53        &   25.8      &3.33&  0.09      &    204        &1.42&1.33&40\\
		\bottomrule
	\end{tabular}
	
	\vspace{0.4cm}
	
	\begin{tabular}{@{}l|llr|llr|llr@{}}\toprule
		&\multicolumn{3}{l|}{NGC 6121}    & \multicolumn{3}{l|}{NGC 6254} &\multicolumn{3}{l}{NGC 6626} \\ 
		\midrule
		&$c$ &$r_\text{c}$&$r_\text{tid}$&$c$ &$r_\text{c}$&$r_\text{tid}$&$c$ &$r_\text{c}$&$r_\text{tid}$\\
		\midrule
		\textbf{ss-OAFP model} &1.81 &1.13  &   73.6     &1.64 &  0.68    &  30.1      &2.12 &0.19 &25.4\\
		\midrule
		\textbf{King model}  &    &            &              &    &            &              &    &     &    \\
		\citep{Miocchi_2013} &1.68&1.07        &   53       &1.41&  0.68      &    19.0         &1.79&0.26&16\\	
		\midrule
		\textbf{Wilson model}&    &            &              &    &            &              &    &     &    \\
		\citep{Miocchi_2013} &2.08&1.08        &   1200      &1.80&  0.73      &   52        &3.1&0.26&380\\
		\bottomrule
	\end{tabular}
	
	\vspace{0.4cm}
	
	\begin{tabular}{@{}l|llr|llr|llr|llr@{}}\toprule
		&\multicolumn{3}{l|}{Pal 3}    & \multicolumn{3}{l|}{Pal 4} &\multicolumn{3}{l|}{Pal 14} &\multicolumn{3}{l}{Trz 5}\\ 
		\midrule
		&$c$ &$r_\text{c}$&$r_\text{tid}$&$c$ &$r_\text{c}$&$r_\text{tid}$&$c$ &$r_\text{c}$&$r_\text{tid}$&$c$ &$r_\text{c}$&$r_\text{tid}$\\
		\midrule
		\textbf{ss-OAFP model}&1.03 &0.55  &   5.91     &1.16 &  0.46    &  6.79      &1.04 &0.85 &9.23 &1.69 &0.15  &   7.25 \\
		\midrule
		\textbf{King model}  &    &            &              &    &            &              &    &     &  &    &            &   \\
		\citep{Miocchi_2013} &0.8&0.47        &   3.6       &1.1&  0.37      &    4.9         &0.9&0.68&6.4 &1.59&0.13        &   5.2 \\	
		\midrule
		\textbf{Wilson model}&    &            &              &    &            &              &    &     &  & & &   \\
		\citep{Miocchi_2013} &0.81&0.49        &   5.33      &1.3&  0.38      &   9        &1.0&0.70&10 &2.4&0.14        &   39\\
		\bottomrule
	\end{tabular}
	\caption{Core- and tidal- radii of finite ss-OAFP model applied to projected density profiles reported in \citep{Miocchi_2013}.}
	\label{table:core_tidal_radius_moicchi}
\end{table*}

\subsection{KM clusters \citep{Kron_1984}}\label{sec:KM_OAFP}

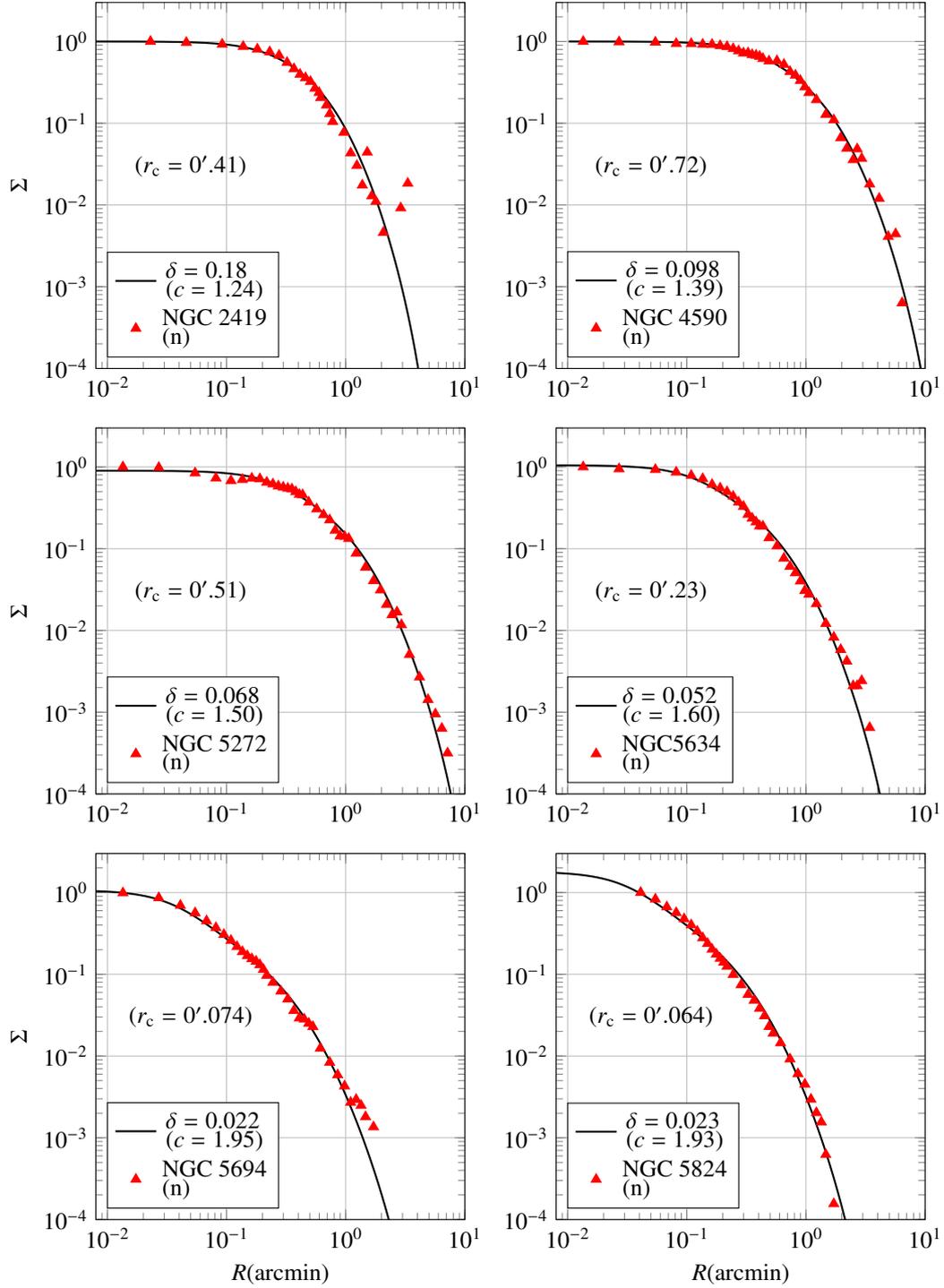
\begin{figure}[H]
	\begin{tikzpicture}
	\begin{loglogaxis}[width=7cm,height=7cm, grid=major,ylabel=\normalsize{$\Sigma$}, xmin=8e-3,xmax=10,ymin=1e-4,ymax=3,legend pos=south west, legend style={cells={align=left}}]
	\addplot [color = black ,mark=no,thick,solid ] table[x index=0, y index=1]{R_Sig_NGC2419.txt};
	\addlegendentry{\normalsize{$\delta=0.18$ }\\\normalsize{$(c=1.24)$}} 
	\addplot [only marks,color = red ,mark=triangle*,thick ] table[x index=0, y index=1]{Rd_Sigd_NGC2419.txt};
	\addlegendentry{\normalsize{NGC 2419}\\\normalsize{(n)}} 
\node[black] at (5e-2,3e-2) {(\normalsize{$r_\text{c}=0\textquotesingle.41$})} ; 
	\end{loglogaxis}
	\end{tikzpicture}
	\begin{tikzpicture}
	\begin{loglogaxis}[width=7cm,height=7cm, grid=major,xmin=8e-3,xmax=10,ymin=1e-4,ymax=3,legend pos=south west, legend style={cells={align=left}}]
	\addplot [color = black ,mark=no,thick,solid ] table[x index=0, y index=1]{R_Sig_NGC4590.txt};
	\addlegendentry{\normalsize{$\delta=0.098$ }\\\normalsize{$(c=1.39)$}} 
	\addplot [only marks,color = red ,mark=triangle*,thick ] table[x index=0, y index=1]{Rd_Sigd_NGC4590.txt};
	\addlegendentry{\normalsize{NGC 4590}\\ \normalsize{(n)}} 
\node[black] at (5e-2,3e-2) {(\normalsize{$r_\text{c}=0\textquotesingle.72$})} ; 
	\end{loglogaxis}
	\end{tikzpicture}
	
	\vspace{0.3cm}
	
	\begin{tikzpicture}
	\begin{loglogaxis}[width=7cm,height=7cm, grid=major, ylabel=\normalsize{$\Sigma$}, xmin=8e-3,xmax=10,ymin=1e-4,ymax=3,legend pos=south west, legend style={cells={align=left}}]
	\addplot [color = black ,mark=no,thick,solid ] table[x index=0, y index=1]{R_Sig_NGC5272.txt};
	\addlegendentry{\normalsize{$\delta=0.068$ }\\\normalsize{$(c=1.50)$}} 
	\addplot [only marks,color = red ,mark=triangle*,thick ] table[x index=0, y index=1]{Rd_Sigd_NGC5272.txt};
	\addlegendentry{\normalsize{NGC 5272}\\ \normalsize{(n)}} 
\node[black] at (5e-2,3e-2) {(\normalsize{$r_\text{c}=0\textquotesingle.51$})} ; 
	\end{loglogaxis}
	\end{tikzpicture}
	\begin{tikzpicture}
	\begin{loglogaxis}[width=7cm,height=7cm, grid=major,xmin=8e-3,xmax=10,ymin=1e-4,ymax=3,legend pos=south west, legend style={cells={align=left}}]
	\addplot [color = black ,mark=no,thick,solid ] table[x index=0, y index=1]{R_Sig_NGC5634.txt};
	\addlegendentry{\normalsize{$\delta=0.052$}\\ \normalsize{$(c=1.60)$}} 
	\addplot [only marks,color = red ,mark=triangle*,thick ] table[x index=0, y index=1]{Rd_Sigd_NGC5634.txt};
	\addlegendentry{\normalsize{NGC5634}\\\normalsize{(n)}} 
\node[black] at (5e-2,3e-2) {(\normalsize{$r_\text{c}=0\textquotesingle.23$})} ; 
	\end{loglogaxis}
	\end{tikzpicture}
	
	\vspace{0.3cm}
	
	\begin{tikzpicture}
	\begin{loglogaxis}[width=7cm,height=7cm, grid=major,xlabel=\normalsize{$R\text{(arcmin)}$}, ylabel=\normalsize{$\Sigma$}, xmin=8e-3,xmax=10,ymin=1e-4,ymax=3,legend pos=south west, legend style={cells={align=left}}]
	\addplot [color = black ,mark=no,thick,solid ] table[x index=0, y index=1]{R_Sig_NGC5694.txt};
	\addlegendentry{\normalsize{$\delta=0.022$ }\\\normalsize{$(c=1.95)$}} 
	\addplot [only marks,color = red ,mark=triangle*,thick ] table[x index=0, y index=1]{Rd_Sigd_NGC5694.txt};
	\addlegendentry{\normalsize{NGC 5694}\\ \normalsize{(n)}} 
\node[black] at (5e-2,3e-2) {(\normalsize{$r_\text{c}=0\textquotesingle.074$})} ; 
	\end{loglogaxis}
	\end{tikzpicture}
	\begin{tikzpicture}
	\begin{loglogaxis}[width=7cm,height=7cm, grid=major,xlabel=\normalsize{$R\text{(arcmin)}$}, xmin=8e-3,xmax=10,ymin=1e-4,ymax=3,legend pos=south west, legend style={cells={align=left}}]
	\addplot [color = black ,mark=no,thick,solid ] table[x index=0, y index=1]{R_Sig_NGC5824.txt};
	\addlegendentry{\normalsize{$\delta=0.023$ }\\ \normalsize{$(c=1.93)$}} 
	\addplot [only marks,color = red ,mark=triangle*,thick ] table[x index=0, y index=1]{Rd_Sigd_NGC5824.txt};
	\addlegendentry{\normalsize{NGC 5824}\\\normalsize{(n)}} 
\node[black] at (5e-2,3e-2) {(\normalsize{$r_\text{c}=0\textquotesingle.064$})} ; 
	\end{loglogaxis}
	\end{tikzpicture}
\caption{Fitting of the energy-truncated ss-OAFP model ($m=3.9$) to the projected density profiles NGC 2419, NGC 4590, NGC 5272, NGC 5634, NGC 5694 and NGC 5824 reported in \citep{Kron_1984}. The unit of the projected density $\Sigma$ is number per square of arcminutes. In the legends, (n) means normal cluster that can be fitted by the King model as judged so in \citep{Djorgovski_1986}.}
\label{fig:fitting_KM_1_6}
\end{figure}

Figures \ref{fig:fitting_KM_1_6}- \ref{fig:fitting_KM_19_24} show the projected density profiles reported in \citep{Kron_1984}, fitted by the energy-truncated ss-OAFP model. In \citep{Kron_1984}, the first several points are not included in fitting of King model due to uncertainty in data originating from too high brightness, yet the present work included them since almost all the data plots were well fitted by our model.

\begin{figure}[H]
	\centering
	\begin{tikzpicture}
	\begin{loglogaxis}[width=7cm,height=7cm, grid=major,ylabel=\normalsize{$\Sigma$}, xmin=8e-3,xmax=10,ymin=1e-4,ymax=3,legend pos=south west, legend style={cells={align=left}}]
	\addplot [color = black ,mark=no,thick,solid ] table[x index=0, y index=1]{R_Sig_NGC6093.txt};
	\addlegendentry{\normalsize{$\delta=0.036$ }\\\normalsize{$(c=1.74)$}} 
	\addplot [only marks,color = red ,mark=triangle*,thick ] table[x index=0, y index=1]{Rd_Sigd_NGC6093.txt};
	\addlegendentry{\normalsize{NGC 6093}\\\normalsize{(n)}} 
\node[black] at (5e-2,3e-2) {(\normalsize{$r_\text{c}=0\textquotesingle.20$})} ; 
	\end{loglogaxis}
	\end{tikzpicture}
	\begin{tikzpicture}
	\begin{loglogaxis}[width=7cm,height=7cm, grid=major,xmin=8e-3,xmax=10,ymin=1e-4,ymax=3,legend pos=south west, legend style={cells={align=left}}]
	\addplot [color = black ,mark=no,thick,solid ] table[x index=0, y index=1]{R_Sig_NGC6205.txt};
	\addlegendentry{\normalsize{$\delta=0.061$ }\\\normalsize{$(c=1.54)$}} 
	\addplot [only marks,color = red ,mark=triangle*,thick ] table[x index=0, y index=1]{Rd_Sigd_NGC6205.txt};
	\addlegendentry{\normalsize{NGC 6205}\\\normalsize{(n)}}
      \node[black] at (5e-2,3e-2) {(\normalsize{$r_\text{c}=0\textquotesingle.78$})} ;  
	\end{loglogaxis}
	\end{tikzpicture}
	
	\vspace{0.3cm}
	
	\begin{tikzpicture}
	\begin{loglogaxis}[width=7cm,height=7cm, grid=major,ylabel=\normalsize{$\Sigma$}, xmin=8e-3,xmax=10,ymin=1e-4,ymax=3,legend pos=south west, legend style={cells={align=left}}]
	\addplot [color = black ,mark=no,thick,solid ] table[x index=0, y index=1]{R_Sig_NGC6229.txt};
	\addlegendentry{\normalsize{$\delta=0.080$ }\\\normalsize{$(c=1.45)$}} 
	\addplot [only marks,color = red ,mark=triangle*,thick ] table[x index=0, y index=1]{Rd_Sigd_NGC6229.txt};
	\addlegendentry{\normalsize{NGC 6229}\\\normalsize{(n)}} 
\node[black] at (5e-2,3e-2) {(\normalsize{$r_\text{c}=0\textquotesingle.18$})} ; 
	\end{loglogaxis}
	\end{tikzpicture}
	\begin{tikzpicture}
	\begin{loglogaxis}[width=7cm,height=7cm, grid=major,xmin=8e-3,xmax=10,ymin=1e-4,ymax=3,legend pos=south west, legend style={cells={align=left}}]
	\addplot [color = black ,mark=no,thick,solid ] table[x index=0, y index=1]{R_Sig_NGC6273.txt};
	\addlegendentry{\normalsize{$\delta=0.045$ }\\\normalsize{$(c=1.65)$}} 
	\addplot [only marks,color = red ,mark=triangle*,thick ] table[x index=0, y index=1]{Rd_Sigd_NGC6273.txt};
	\addlegendentry{\normalsize{NGC 6273}\\\normalsize{(n)}} 
\node[black] at (5e-2,3e-2) {(\normalsize{$r_\text{c}=0\textquotesingle.49$})} ; 
	\end{loglogaxis}
	\end{tikzpicture}
	
	\vspace{0.3cm}
	
	\begin{tikzpicture}
	\begin{loglogaxis}[width=7cm,height=7cm, grid=major, xlabel=\normalsize{$R\text{(arcmin)}$}, ylabel=\normalsize{$\Sigma$}, xmin=8e-3,xmax=10,ymin=1e-4,ymax=3,legend pos=south west, legend style={cells={align=left}}]
	\addplot [color = black ,mark=no,thick,solid ] table[x index=0, y index=1]{R_Sig_NGC6304.txt};
	\addlegendentry{\normalsize{$\delta=0.044$ }\\\normalsize{$(c=1.66)$}} 
	\addplot [only marks,color = red ,mark=triangle*,thick ] table[x index=0, y index=1]{Rd_Sigd_NGC6304.txt};
	\addlegendentry{\normalsize{NGC 6304}\\\normalsize{(n$?$)}} 
\node[black] at (5e-2,3e-2) {(\normalsize{$r_\text{c}=0\textquotesingle.29$})} ; 
	\end{loglogaxis}
	\end{tikzpicture}
\begin{tikzpicture}
	\begin{loglogaxis}[width=7cm,height=7c m, grid=major, xlabel=\normalsize{$R\text{(arcmin)}$}, xmin=8e-3,xmax=10,ymin=1e-4,ymax=3,legend pos=south west, legend style={cells={align=left}}]
	\addplot [color = black ,mark=no,thick,solid ] table[x index=0, y index=1]{R_Sig_NGC6333.txt};
	\addlegendentry{\normalsize{$\delta=0.050$ }\\\normalsize{$(c=1.61)$}} 
	\addplot [only marks,color = red ,mark=triangle*,thick ] table[x index=0, y index=1]{Rd_Sigd_NGC6333.txt};
	\addlegendentry{\normalsize{NGC 6333}\\\normalsize{(n$?$)}} 
\node[black] at (5e-2,3e-2) {(\normalsize{$r_\text{c}=0\textquotesingle.37$})} ; 
	\end{loglogaxis}
	\end{tikzpicture}
	\caption{Fitting of the energy-truncated ss-OAFP model ($m=3.9$) to the projected density profiles of NGC 6093, NGC 6205, NGC 5229, NGC 6273, NGC 6304 and NGC 6333 reported in \citep{Kron_1984}. The unit of the projected density $\Sigma$ is number per square of arcminutes. In the legends, (n) means `normal' cluster that can be fitted by the King model and (n?) `probable normal' cluster  as judged so in \citep{Djorgovski_1986}.}
\label{fig:fitting_KM_7_12}
\end{figure}
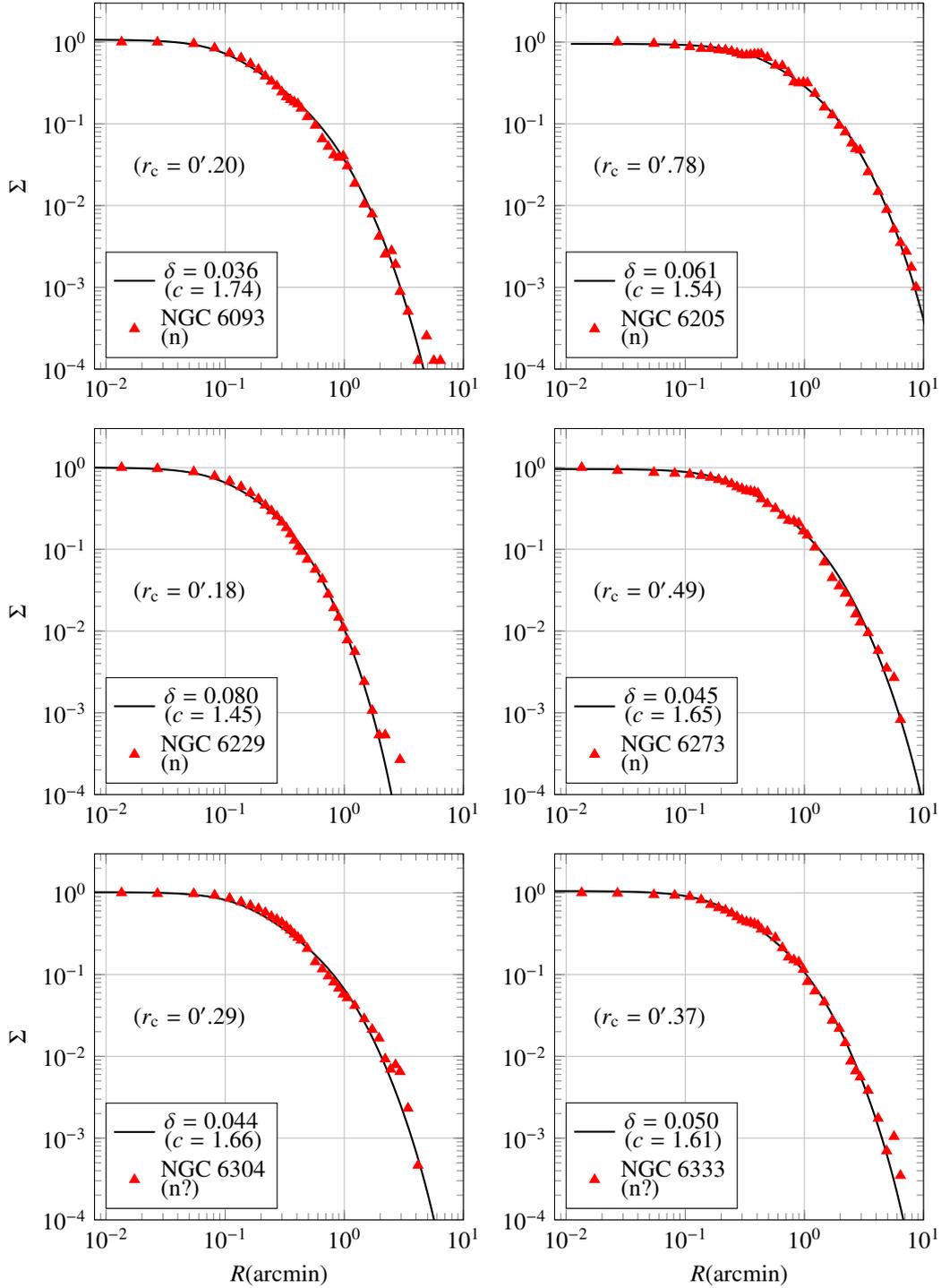

\begin{figure}[H]
	\begin{tikzpicture}
	\begin{loglogaxis}[width=7cm,height=7cm, grid=major,ylabel=\normalsize{$\Sigma$}, xmin=8e-3,xmax=10,ymin=1e-4,ymax=3,legend pos=south west, legend style={cells={align=left}}]
	\addplot [color = black ,mark=no,thick,solid ] table[x index=0, y index=1]{R_Sig_NGC6341.txt};
	\addlegendentry{\normalsize{$\delta=0.042$ }\\\normalsize{$(c=1.68)$}} 
	\addplot [only marks,color = red ,mark=triangle*,thick ] table[x index=0, y index=1]{Rd_Sigd_NGC6341.txt};
	\addlegendentry{\normalsize{NGC 6341}\\\normalsize{(n)}} 
\node[black] at (5e-2,3e-2) {(\normalsize{$r_\text{c}=0\textquotesingle.31$})} ; 
	\end{loglogaxis}
	\end{tikzpicture}
	\begin{tikzpicture}
	\begin{loglogaxis}[width=7cm,height=7cm, grid=major,xmin=8e-3,xmax=10,ymin=1e-4,ymax=3,legend pos=south west, legend style={cells={align=left}}]
	\addplot [color = black ,mark=no,thick,solid ] table[x index=0, y index=1]{R_Sig_NGC6356.txt};
	\addlegendentry{\normalsize{$\delta=0.042$ }\\\normalsize{$(c=1.68)$}} 
	\addplot [only marks,color = red ,mark=triangle*,thick ] table[x index=0, y index=1]{Rd_Sigd_NGC6356.txt};
	\addlegendentry{\normalsize{NGC 6356}\\\normalsize{(n)}} 
\node[black] at (5e-2,3e-2) {(\normalsize{$r_\text{c}=0\textquotesingle.27$})} ; 
	\end{loglogaxis}
	\end{tikzpicture}

\vspace{0.3cm}

	\begin{tikzpicture}
	\begin{loglogaxis}[width=7cm,height=7cm, grid=major,ylabel=\normalsize{$\Sigma$}, xmin=8e-3,xmax=10,ymin=1e-4,ymax=3,legend pos=south west, legend style={cells={align=left}}]
	\addplot [color = black ,mark=no,thick,solid ] table[x index=0, y index=1]{R_Sig_NGC6401.txt};
	\addlegendentry{\normalsize{$\delta=0.040$ }\\\normalsize{$(c=1.70)$}} 
	\addplot [only marks,color = red ,mark=triangle*,thick ] table[x index=0, y index=1]{Rd_Sigd_NGC6401.txt};
	\addlegendentry{\normalsize{NGC 6401}\\\normalsize{(n)}} 
\node[black] at (5e-2,3e-2) {(\normalsize{$r_\text{c}=0\textquotesingle.27$})} ; 
	\end{loglogaxis}
	\end{tikzpicture}
	\begin{tikzpicture}
	\begin{loglogaxis}[width=7cm,height=7cm, grid=major, xmin=8e-3,xmax=10,ymin=1e-4,ymax=3,legend pos=south west, legend style={cells={align=left}}]
	\addplot [color = black ,mark=no,thick,solid ] table[x index=0, y index=1]{R_Sig_NGC6440.txt};
	\addlegendentry{\normalsize{$\delta=0.038$ }\\\normalsize{$(c=1.72)$}} 
	\addplot [only marks,color = red ,mark=triangle*,thick ] table[x index=0, y index=1]{Rd_Sigd_NGC6440.txt};
	\addlegendentry{\normalsize{NGC 6440}\\\normalsize{(n)}} 
\node[black] at (5e-2,3e-2) {(\normalsize{$r_\text{c}=0\textquotesingle.16$})} ; 
	\end{loglogaxis}
	\end{tikzpicture}

\vspace{0.3cm}

	\begin{tikzpicture}
	\begin{loglogaxis}[width=7cm,height=7cm, grid=major, xlabel=\normalsize{$R\text{(arcmin)}$}, ylabel=\normalsize{$\Sigma$}, xmin=8e-3,xmax=10,ymin=1e-4,ymax=3,legend pos=south west, legend style={cells={align=left}}]
	\addplot [color = black ,mark=no,thick,solid ] table[x index=0, y index=1]{R_Sig_NGC6517.txt};
	\addlegendentry{\normalsize{$\delta=0.020$ }\\\normalsize{$(c=1.99)$}} 
	\addplot [only marks,color = red ,mark=triangle*,thick ] table[x index=0, y index=1]{Rd_Sigd_NGC6517.txt};
	\addlegendentry{\normalsize{NGC 6517}\\\normalsize{(n)}}
\node[black] at (5e-2,3e-2) {(\normalsize{$r_\text{c}=0\textquotesingle.16$})} ;  
	\end{loglogaxis}
	\end{tikzpicture}
\begin{tikzpicture}
	\begin{loglogaxis}[width=7cm,height=7cm, grid=major, xlabel=\normalsize{$R\text{(arcmin)}$}, xmin=8e-3,xmax=10,ymin=1e-4,ymax=3,legend pos=south west, legend style={cells={align=left}}]
	\addplot [color = black ,mark=no,thick,solid ] table[x index=0, y index=1]{R_Sig_NGC6553.txt};
	\addlegendentry{\normalsize{$\delta=0.108$ }\\\normalsize{$(c=1.36)$}} 
	\addplot [only marks,color = red ,mark=triangle*,thick ] table[x index=0, y index=1]{Rd_Sigd_NGC6553.txt};
	\addlegendentry{\normalsize{NGC 6553}\\\normalsize{(n)}} 
\node[black] at (5e-2,3e-2) {(\normalsize{$r_\text{c}=0\textquotesingle.55$})} ; 
	\end{loglogaxis}
	\end{tikzpicture}
	\caption{Fitting of the energy-truncated ss-OAFP model ($m=3.9$) to the projected density profiles NGC 6341, NGC 6356, NGC 6401, NGC 6440, NGC 6517 and NGC 6553 reported in \citep{Kron_1984}. The unit of the projected density $\Sigma$ is number per square of arcminutes. In the legends, (n) means normal cluster that can be fitted by the King model as judged so in \citep{Djorgovski_1986}.}
\label{fig:fitting_KM_13_18}
\end{figure}
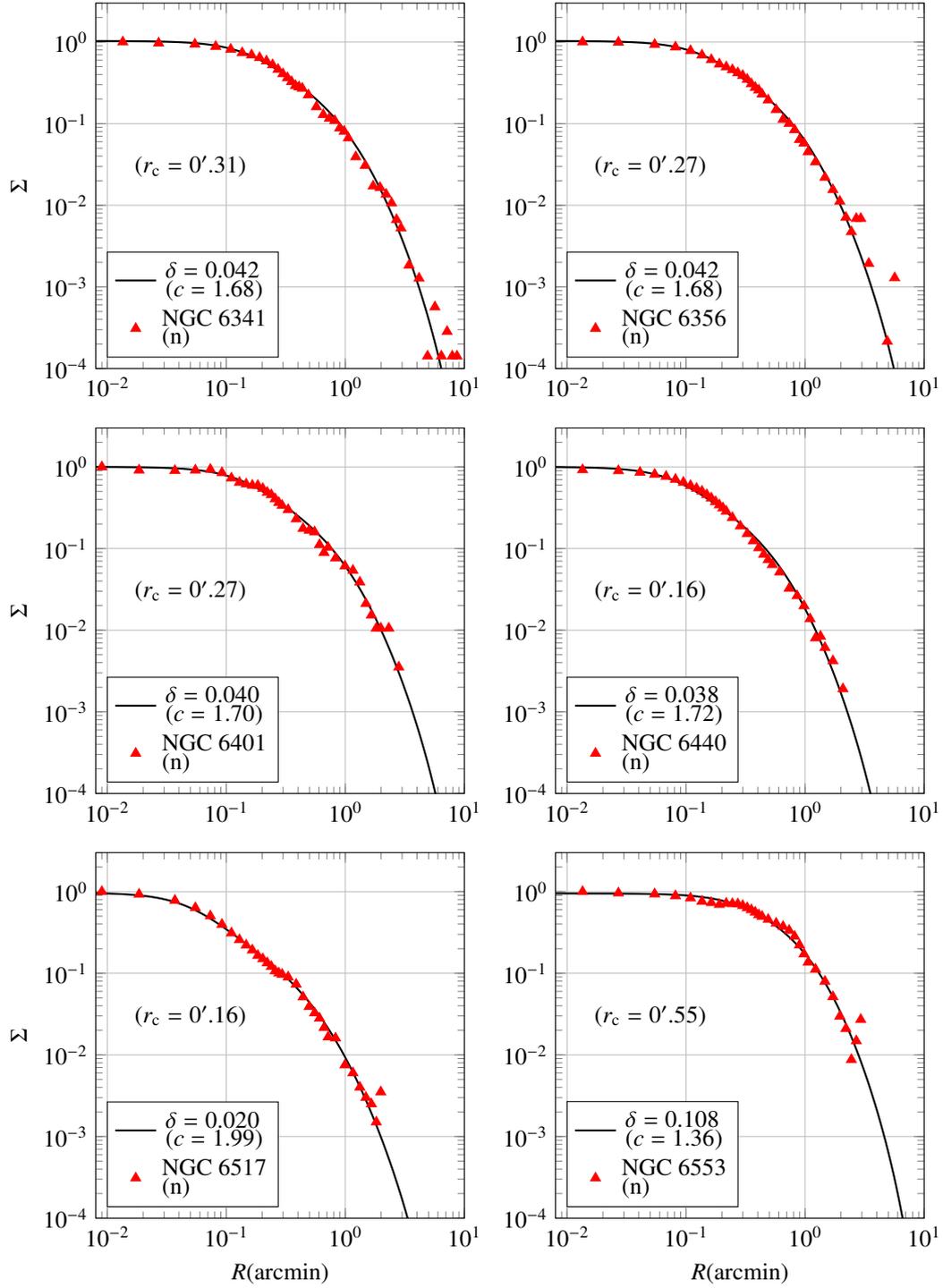

\begin{figure}[H]
	\begin{tikzpicture}
	\begin{loglogaxis}[width=7cm,height=7cm, grid=major,ylabel=\normalsize{$\Sigma$}, xmin=8e-3,xmax=10,ymin=1e-4,ymax=3,legend pos=south west, legend style={cells={align=left}}]
	\addplot [color = black ,mark=no,thick,solid ] table[x index=0, y index=1]{R_Sig_NGC6569.txt};
	\addlegendentry{\normalsize{$\delta=0.15$ }\\\normalsize{$(c=1.28)$}} 
	\addplot [only marks,color = red ,mark=triangle*,thick ] table[x index=0, y index=1]{Rd_Sigd_NGC6569.txt};
	\addlegendentry{\normalsize{NGC 6569}\\\normalsize{(n)}} 
\node[black] at (5e-2,3e-2) {(\normalsize{$r_\text{c}=0\textquotesingle.37$})} ; 
	\end{loglogaxis}
	\end{tikzpicture}
	\begin{tikzpicture}
	\begin{loglogaxis}[width=7cm,height=7cm, grid=major,xmin=8e-3,xmax=10,ymin=1e-4,ymax=3,legend pos=south west, legend style={cells={align=left}}]
	\addplot [color = black ,mark=no,thick,solid ] table[x index=0, y index=1]{R_Sig_NGC6638.txt};
	\addlegendentry{\normalsize{$\delta=0.105$ }\\\normalsize{$(c=1.37)$}} 
	\addplot [only marks,color = red ,mark=triangle*,thick ] table[x index=0, y index=1]{Rd_Sigd_NGC6638.txt};
	\addlegendentry{\normalsize{NGC 6638}\\\normalsize{(n)}} 
\node[black] at (5e-2,3e-2) {(\normalsize{$r_\text{c}=0\textquotesingle.25$})} ; 
	\end{loglogaxis}
	\end{tikzpicture}

\vspace{0.3cm}

	\begin{tikzpicture}
	\begin{loglogaxis}[width=7cm,height=7cm, grid=major,ylabel=\normalsize{$\Sigma$}, xmin=8e-3,xmax=10,ymin=1e-4,ymax=3,legend pos=south west, legend style={cells={align=left}}]
	\addplot [color = black ,mark=no,thick,solid ] table[x index=0, y index=1]{R_Sig_NGC6715.txt};
	\addlegendentry{\normalsize{$\delta=0.028$ }\\\normalsize{$(c=1.84)$}} 
	\addplot [only marks,color = red ,mark=triangle*,thick ] table[x index=0, y index=1]{Rd_Sigd_NGC6715.txt};
	\addlegendentry{\normalsize{NGC 6715}\\\normalsize{(n)}} 
\node[black] at (5e-2,3e-2) {(\normalsize{$r_\text{c}=0\textquotesingle.11$})} ; 
	\end{loglogaxis}
	\end{tikzpicture}
	\begin{tikzpicture}
	\begin{loglogaxis}[width=7cm,height=7cm, grid=major,xmin=8e-3,xmax=10,ymin=1e-4,ymax=3,legend pos=south west, legend style={cells={align=left}}]
	\addplot [color = black ,mark=no,thick,solid ] table[x index=0, y index=1]{R_Sig_NGC6864.txt};
	\addlegendentry{\normalsize{$\delta=0.029$ }\\\normalsize{$(c=1.83)$}} 
	\addplot [only marks,color = red ,mark=triangle*,thick ] table[x index=0, y index=1]{Rd_Sigd_NGC6864.txt};
	\addlegendentry{\normalsize{NGC 6864}\\\normalsize{(n)}} 
\node[black] at (5e-2,3e-2) {(\normalsize{$r_\text{c}=0\textquotesingle.12$})} ; 
	\end{loglogaxis}
	\end{tikzpicture}

	\vspace{0.3cm}
	
	\begin{tikzpicture}
	\begin{loglogaxis}[width=7cm,height=7cm, grid=major, xlabel=\normalsize{$R$ (arcmin)}, ylabel=\normalsize{$\Sigma$}, xmin=8e-3,xmax=10,ymin=1e-4,ymax=3,legend pos=south west, legend style={cells={align=left}}]
	\addplot [color = black ,mark=no,thick,solid ] table[x index=0, y index=1]{R_Sig_NGC6934.txt};
	\addlegendentry{\normalsize{$\delta=0.065$ }\\\normalsize{$(c=1.52)$}} 
	\addplot [only marks,color = red ,mark=triangle*,thick ] table[x index=0, y index=1]{Rd_Sigd_NGC6934.txt};
	\addlegendentry{\normalsize{NGC 6934}\\\normalsize{(n)}}
\node[black] at (5e-2,3e-2) {(\normalsize{$r_\text{c}=0\textquotesingle.24$})} ;  
	\end{loglogaxis}
	\end{tikzpicture}
\begin{tikzpicture}
	\begin{loglogaxis}[width=7cm,height=7cm, grid=major, xlabel=\normalsize{$R$ (arcmin)}, xmin=8e-3,xmax=10,ymin=1e-4,ymax=3,legend pos=south west, legend style={cells={align=left}}]
	\addplot [color = black ,mark=no,thick,solid ] table[x index=0, y index=1]{R_Sig_NGC7006.txt};
	\addlegendentry{\normalsize{$\delta=0.095$ }\\\normalsize{$(c=1.40)$}} 
	\addplot [only marks,color = red ,mark=triangle*,thick ] table[x index=0, y index=1]{Rd_Sigd_NGC7006.txt};
	\addlegendentry{\normalsize{NGC 7006}\\\normalsize{(n)}}
\node[black] at (5e-2,3e-2) {(\normalsize{$r_\text{c}=0\textquotesingle.18$})} ;  
	\end{loglogaxis}
	\end{tikzpicture}
	\caption{Fitting of the energy-truncated ss-OAFP model ($m=3.9$) to the projected density profiles of NGC 6569, NGC 6638, NGC 6715, NGC 6864, NGC 6934 and NGC 7006 reported in \citep{Kron_1984}. The unit of the projected density $\Sigma$ is number per square of arcminutes. In the legends, (n) means normal cluster that can be fitted by the King model as judged so in \citep{Djorgovski_1986}.}
\label{fig:fitting_KM_19_24}
\end{figure}
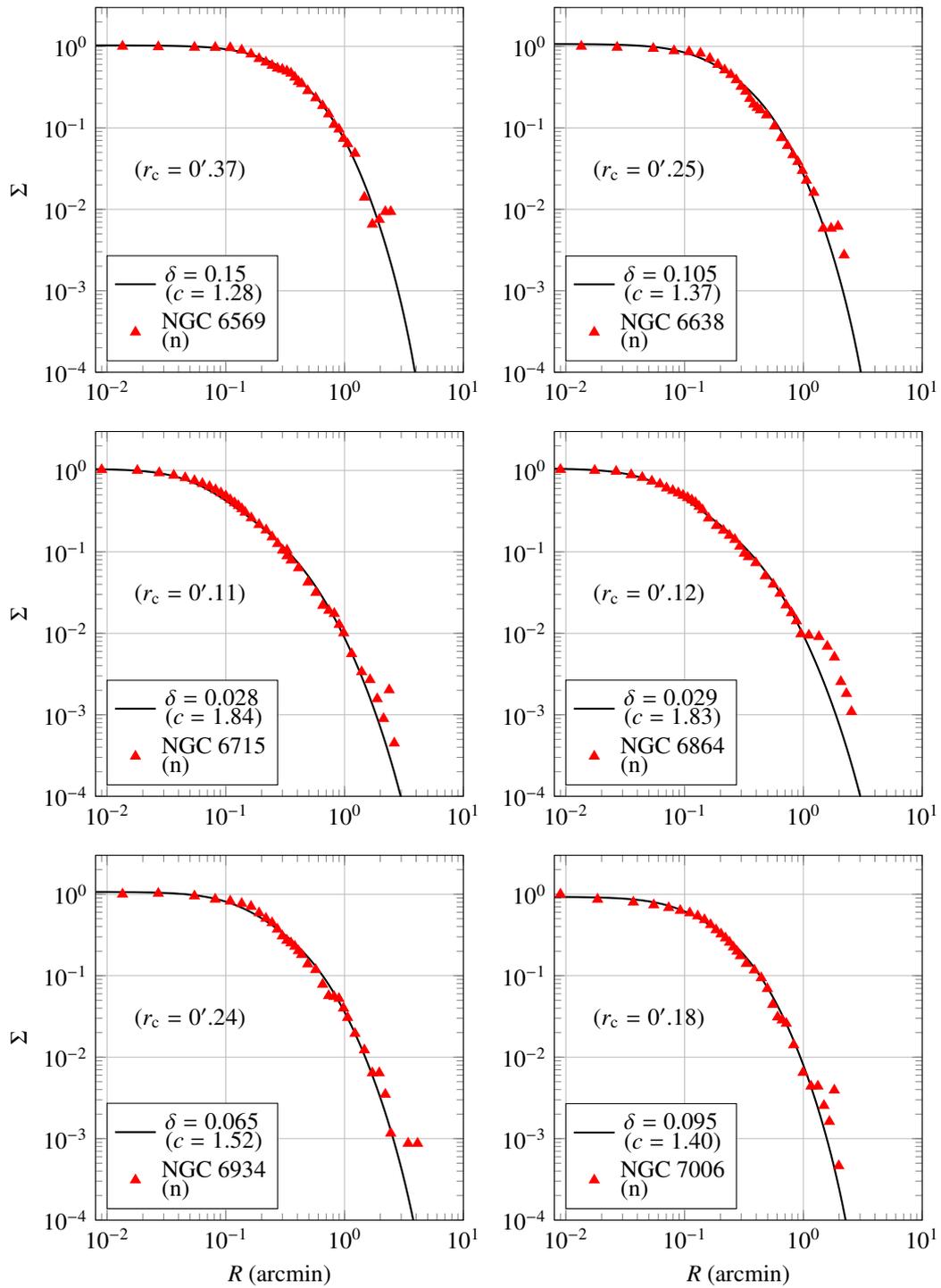

\begin{figure}[H]
	\begin{tikzpicture}
	\begin{loglogaxis}[width=7cm,height=7cm, grid=major, xlabel=\normalsize{$R$ (arcmin)}, ylabel=\normalsize{$\Sigma$}, xmin=1e-1,xmax=10,ymin=1e-4,ymax=3,legend pos=south west, legend style={cells={align=left}}]
	\addplot [color = black ,mark=no,thick,solid ] table[x index=0, y index=1]{R_Sig_NGC5053.txt};
	\addlegendentry{\normalsize{$\delta=5.5$ }\\\normalsize{$(c=1.00)$}} 
	\addplot [only marks,color = red ,mark=triangle*,thick ] table[x index=0, y index=1]{Rd_Sigd_NGC5053.txt};
	\addlegendentry{\normalsize{NGC 5053}\\\normalsize{(n)}}
\node[black] at (5e-1,3e-2) {(\normalsize{$r_\text{c}=2\textquotesingle.7$})} ;  
	\end{loglogaxis}
	\end{tikzpicture}
\begin{tikzpicture}
	\begin{loglogaxis}[width=7cm,height=7cm, grid=major, xlabel=\normalsize{$R$ (arcmin)}, xmin=1e-1,xmax=10,ymin=1e-4,ymax=3,legend pos=south west, legend style={cells={align=left}}]
	\addplot [color = black ,mark=no,thick,solid ] table[x index=0, y index=1]{R_Sig_NGC5897.txt};
	\addlegendentry{\normalsize{$\delta=0.78$ }\\\normalsize{$(c=1.06)$}} 
	\addplot [only marks,color = red ,mark=triangle*,thick ] table[x index=0, y index=1]{Rd_Sigd_NGC5897.txt};
	\addlegendentry{\normalsize{NGC 5897}\\\normalsize{(n)}} 
\node[black] at (5e-1,3e-2) {(\normalsize{$r_\text{c}=1\textquotesingle.64$})} ; 
	\end{loglogaxis}
	\end{tikzpicture}
	\caption{Fitting of the energy-truncated ss-OAFP model ($m=3.9$) to the projected density profiles NGC 5053  and NGC 5897 reported in \citep{Kron_1984}. The unit of the projected density $\Sigma$ is number per square of arcminutes. In the legends, (n) means normal cluster that can be fitted by the King model as judged so in \citep{Djorgovski_1986}. Following \citep{Kron_1984}, data at small radii are ignored due to the depletion of projected density profile.}
\label{fig:fitting_KM_25_26}
\end{figure}
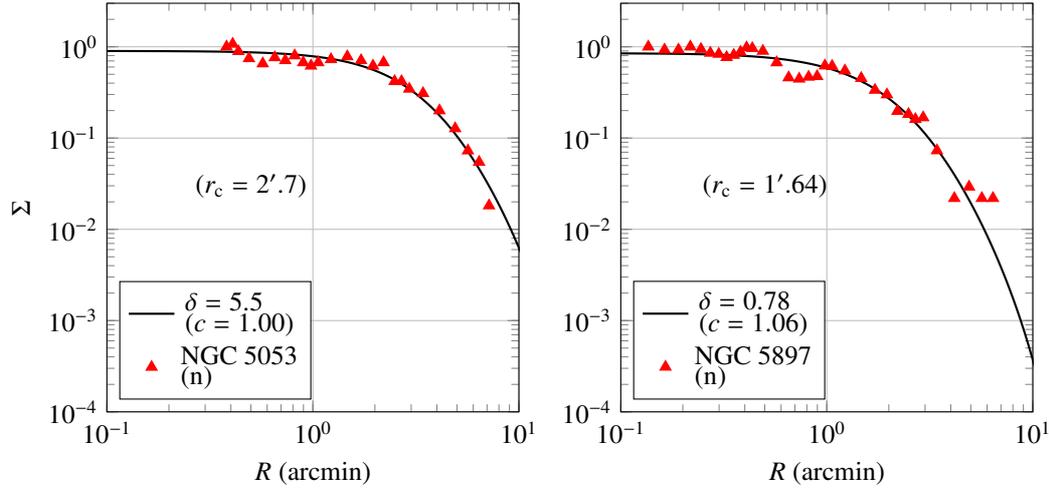

\subsection{KM clusters \citep{Noyola_2006}}\label{Fitting_KM_Noyola}

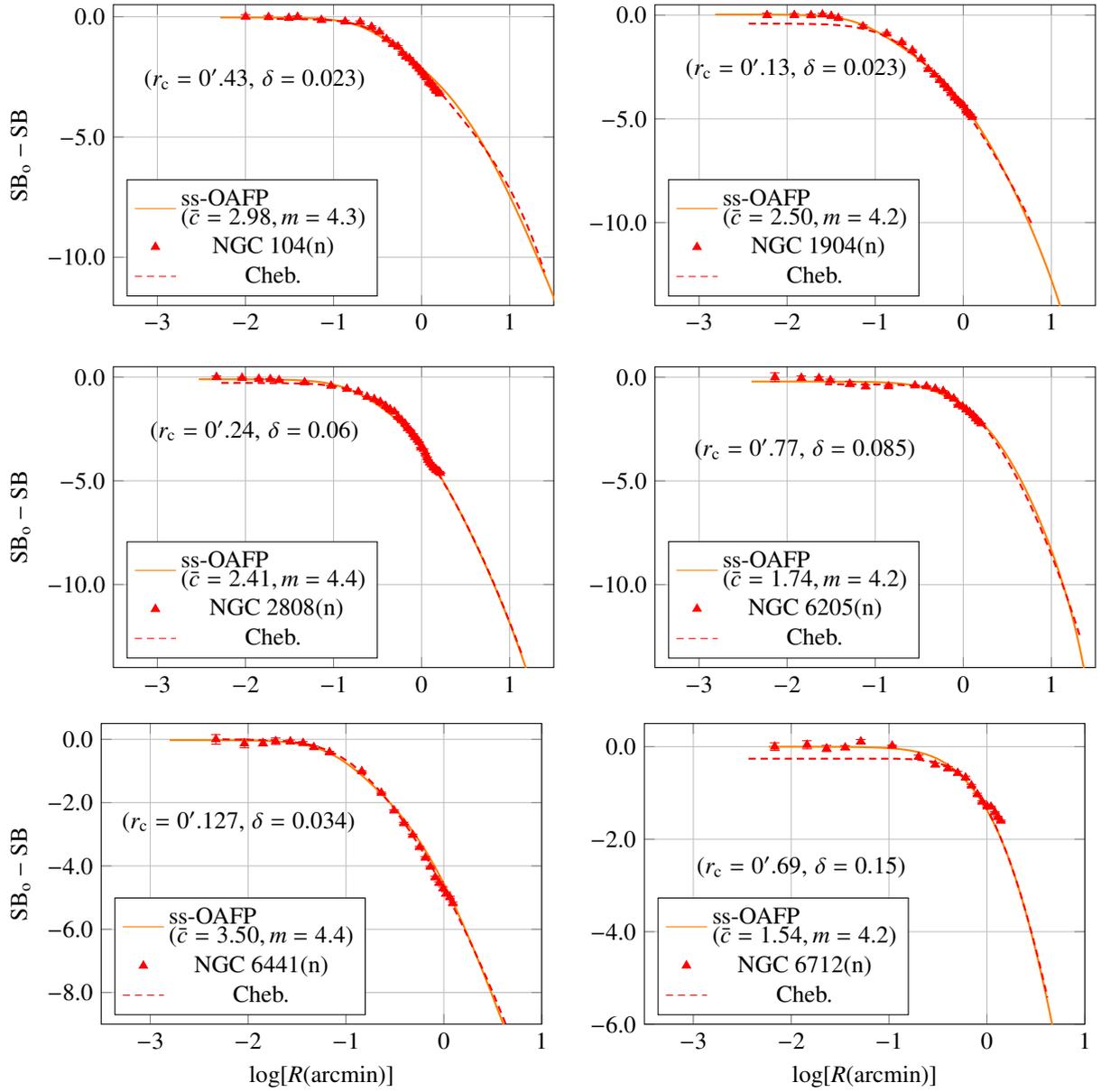
\begin{figure}[H]
	\begin{tikzpicture}
	\begin{axis}[width=8cm,height=6cm, grid=major,xmin=-3.5,xmax=1.5,ymin=-12,ymax=0.5,ylabel=\normalsize{$\text{SB}_\text{o}-\text{SB}$},legend style={cells={align=left}},legend entries={\normalsize{ss-OAFP}\\\normalsize{$(\bar{c}=2.98,m=4.3)$}, \normalsize{NGC 104(n)},\normalsize{Cheb.}},yticklabel style={/pgf/number format/.cd,fixed zerofill,precision=1},legend pos=south west]
	\addlegendimage{no markers, orange};
	\addlegendimage{only marks, mark=triangle*, red};
\addlegendimage{no markers, red, densely dashed};
	\addplot [color = orange ,mark=no,thick,solid] table[x index=0, y index=1]{R_Sig_NGC104_Noyora_whole.txt};
	\addplot[color = red, only marks, mark=triangle*, mark options={red}, error bars,y dir=both, y explicit] table [x index=0, y index=1, y error index=2]{Rd_Sigd_Errbar_Err_NGC104_Noyora_whole.txt};
\addplot [color = red ,mark=no,thick,densely dashed] table[x index=0, y index=1]{Rc_Sigc_NGC104_Trager_whole.txt};
	\node[black] at (-1.9,-2.6) {(\normalsize{$r_\text{c}=0\textquotesingle.43, \, \delta=0.023$})} ; 
	\end{axis}
	\end{tikzpicture}\hspace{0.3cm}
	\begin{tikzpicture}
	\begin{axis}[width=8cm,height=6cm, grid=major,xmin=-3.5,xmax=1.5,ymin=-14,ymax=0.5,legend style={cells={align=left}},legend entries={\normalsize{ss-OAFP}\\\normalsize{$(\bar{c}=2.50,m=4.2)$}, \normalsize{ NGC 1904(n)},\normalsize{Cheb.}},yticklabel style={/pgf/number format/.cd,fixed zerofill,precision=1},legend pos=south west]
	\addlegendimage{no markers, orange};
	\addlegendimage{only marks, mark=triangle*, red};
	\addlegendimage{no markers, red, densely dashed};
	\addplot [color = orange ,mark=no,thick,solid] table[x index=0, y index=1]{R_Sig_NGC1904_Noyora_whole.txt};
	\addplot[color = red, only marks, mark=triangle*, mark options={red}, error bars,y dir=both, y explicit] table [x index=0, y index=1, y error index=2]{Rd_Sigd_Errbar_Err_NGC1904_Noyora_whole.txt};
	\addplot [color = red ,mark=no,thick,densely dashed] table[x index=0, y index=1]{Rc_Sigc_NGC1904_Trager_whole.txt};
	\node[black] at (-1.9,-2.6) {(\normalsize{$r_\text{c}=0\textquotesingle.13, \, \delta=0.023$})} ; 
	\end{axis}
	\end{tikzpicture}

\vspace{0.3cm}

	\begin{tikzpicture}
	\begin{axis}[width=8cm,height=6cm, grid=major,xmin=-3.5,xmax=1.5,ymin=-14,ymax=0.5,ylabel=\normalsize{$\text{SB}_\text{o}-\text{SB}$},legend style={cells={align=left}},legend entries={\normalsize{ss-OAFP}\\\normalsize{$(\bar{c}=2.41,m=4.4)$}, \normalsize{ NGC 2808(n)},\normalsize{Cheb.}},yticklabel style={/pgf/number format/.cd,fixed zerofill,precision=1},legend pos=south west]
	\addlegendimage{no markers, orange};
	\addlegendimage{only marks, mark=triangle*, red};
	\addlegendimage{no markers, red, densely dashed};
	\addplot [color = orange ,mark=no,thick,solid] table[x index=0, y index=1]{R_Sig_NGC2808_Noyora_whole.txt};
	\addplot[color = red, only marks, mark=triangle*, mark options={red}, error bars,y dir=both, y explicit] table [x index=0, y index=1, y error index=2]{Rd_Sigd_Errbar_Err_NGC2808_Noyora_whole.txt};
	\addplot [color = red ,mark=no,thick,densely dashed] table[x index=0, y index=1]{Rc_Sigc_NGC2808_Trager_whole.txt};
	\node[black] at (-1.9,-2.6) {(\normalsize{$r_\text{c}=0\textquotesingle.24, \, \delta=0.06$})} ; 
	\end{axis}
	\end{tikzpicture}\hspace{0.3cm}
\begin{tikzpicture}
\begin{axis}[width=8cm,height=6cm, grid=major,xmin=-3.5,xmax=1.5,ymin=-14,ymax=0.5,legend style={cells={align=left}},legend entries={\normalsize{ss-OAFP}\\\normalsize{$(\bar{c}=1.74,m=4.2)$}, \normalsize{NGC 6205(n)}, \normalsize{Cheb.}},yticklabel style={/pgf/number format/.cd,fixed zerofill,precision=1},legend pos=south west]
\addlegendimage{no markers, orange};
\addlegendimage{only marks, mark=triangle*, red};
\addlegendimage{no markers, red, densely dashed};
\addplot [color = orange ,mark=no,thick,solid] table[x index=0, y index=1]{R_Sig_NGC6205_Noyora_whole.txt};
\addplot[color = red, only marks, mark=triangle*, mark options={red}, error bars,y dir=both, y explicit] table [x index=0, y index=1, y error index=2]{Rd_Sigd_Errbar_Err_NGC6205_Noyora_whole.txt};
\addplot [color = red ,mark=no,thick,densely dashed] table[x index=0, y index=1]{Rc_Sigc_NGC6205_Trager_whole.txt};
\node[black] at (-1.8,-3.5) {(\normalsize{$r_\text{c}=0\textquotesingle.77, \, \delta=0.085$})} ; 
\end{axis}
\end{tikzpicture}

\vspace{0.3cm}

	\begin{tikzpicture}
	\begin{axis}[width=8cm,height=6cm, grid=major,xmin=-3.5,xmax=1,ymin=-9,ymax=0.5,ylabel=\normalsize{$\text{SB}_\text{o}-\text{SB}$},xlabel=\normalsize{$\log[R(\text{arcmin})]$},legend style={cells={align=left}},legend entries={\normalsize{ss-OAFP}\\\normalsize{$(\bar{c}=3.50,m=4.4)$}, \normalsize{NGC 6441(n)},\normalsize{Cheb.}},yticklabel style={/pgf/number format/.cd,fixed zerofill,precision=1},legend pos=south west]
	\addlegendimage{no markers, orange};
	\addlegendimage{only marks, mark=triangle*, red};
	\addlegendimage{no markers, red, densely dashed};
	\addplot [color = orange ,mark=no,thick,solid] table[x index=0, y index=1]{R_Sig_NGC6441_Noyora_whole.txt};
	\addplot[color = red, only marks, mark=triangle*, mark options={red}, error bars,y dir=both, y explicit] table [x index=0, y index=1, y error index=2]{Rd_Sigd_Errbar_Err_NGC6441_Noyora_whole.txt};
	\addplot [color = red ,mark=no,thick,densely dashed] table[x index=0, y index=1]{Rc_Sigc_NGC6441_Trager_whole.txt};
	\node[black] at (-2.1,-2.6) {(\normalsize{$r_\text{c}=0\textquotesingle.127, \, \delta=0.034$})} ; 
	\end{axis}
	\end{tikzpicture}\hspace{0.3cm}
\begin{tikzpicture}
	\begin{axis}[width=8cm,height=6cm, grid=major,xmin=-3.5,xmax=1,ymin=-6,ymax=0.5,xlabel=\normalsize{$\log[R(\text{arcmin})]$},legend style={cells={align=left}},legend entries={\normalsize{ss-OAFP}\\\normalsize{$(\bar{c}=1.54,m=4.2)$}, \normalsize{NGC 6712(n)},\normalsize{Cheb.}},yticklabel style={/pgf/number format/.cd,fixed zerofill,precision=1},legend pos=south west]
	\addlegendimage{no markers, orange};
	\addlegendimage{only marks, mark=triangle*, red};
	\addlegendimage{no markers, red, densely dashed};
	\addplot [color = orange ,mark=no,thick,solid] table[x index=0, y index=1]{R_Sig_NGC6712_Noyora_whole.txt};
	\addplot[color = red, only marks, mark=triangle*, mark options={red}, error bars,y dir=both, y explicit] table [x index=0, y index=1, y error index=2]{Rd_Sigd_Errbar_Err_NGC6712_Noyora_whole.txt};
	\addplot [color = red ,mark=no,thick,densely dashed] table[x index=0, y index=1]{Rc_Sigc_NGC6712_Trager_whole.txt};
	\node[black] at (-1.9,-2.6) {(\normalsize{$r_\text{c}=0\textquotesingle.69, \, \delta=0.15$})} ; 
	\end{axis}
	\end{tikzpicture}
	\caption{Fitting of the energy-truncated ss-OAFP model to the surface brightness of NGC 104, NGC 1904, NGC 2808, NGC 6205, NGC 6441, and NGC 6712 reported in \citep{Noyola_2006}. The unit of the surface brightness (SB) is V magnitude per square of arcseconds. The brightness is normalized by the magnitude $\text{SB}_\text{o}$ at the smallest radius point. In the legends, `Cheb.' means the Chebyshev approximation of the surface brightness reported in \citep{Trager_1995} and `(n)' means KM cluster as judged so in \citep{Djorgovski_1986}.}
	\label{fig:fitting_KM_Noyora}
\end{figure}

\begin{figure}[H]
\begin{tikzpicture}
\begin{axis}[width=8cm,height=9cm, grid=major,xmin=-3.5,xmax=1.5,ymin=-14,ymax=0.5,ylabel=\normalsize{$\text{SB}_\text{o}-\text{SB}$},legend style={cells={align=left}},legend entries={\normalsize{ss-OAFP}\\ \normalsize{$(\bar{c}=2.59, m=4.2)$},\normalsize{approx. } \\ \normalsize{$(\bar{c}=2.72, m=4.8)$}, \normalsize{NGC 1851(n)}, \normalsize{Cheb.}},yticklabel style={/pgf/number format/.cd,fixed zerofill,precision=1},legend pos=south west]
\addlegendimage{no markers, orange, densely dotted};
\addlegendimage{no markers, black};
\addlegendimage{only marks, mark=triangle*, red};
\addlegendimage{no markers, red, densely dashed};
\addplot [color = orange ,mark=no,thick,densely dotted] table[x index=0, y index=1]{R_Sig_NGC1851_Noyora_whole.txt};
\addplot [color = black ,mark=no,thick,solid] table[x index=0, y index=1]{R_Sig_NGC1851_Noyora_whole_apprx.txt};
\addplot[color = red, only marks, mark=triangle*, mark options={red}, error bars,y dir=both, y explicit] table [x index=0, y index=1, y error index=2]{Rd_Sigd_Errbar_Err_NGC1851_Noyora_whole.txt};
\addplot [color = red ,mark=no,thick,densely dashed] table[x index=0, y index=1]{Rc_Sigc_NGC1851_Trager_whole.txt};
\node[black] at (-1.6,-7) {(\normalsize{$r_\text{c}=0\textquotesingle.0528, \, \delta=0.055$})} ; 
\end{axis}
\end{tikzpicture}\hspace{0.3cm}
\begin{tikzpicture}
\begin{axis}[width=8cm,height=9cm, grid=major,xmin=-3.5,xmax=1.0,ymin=-14,ymax=0.5,ylabel=\normalsize{$\text{SB}_\text{o}-\text{SB}$},xlabel=\normalsize{$\log[R(\text{arcmin})]$},legend style={cells={align=left}},legend entries={\normalsize{ss-OAFP}\\ \normalsize{$(\bar{c}=2.30,m=4.2)$},\normalsize{approx. } \\ \normalsize{$(\bar{c}=2.84, m=4.95)$}, \normalsize{NGC 5694(n)}, \normalsize{Cheb.}},yticklabel style={/pgf/number format/.cd,fixed zerofill,precision=1},legend pos=south west]
\addlegendimage{no markers, orange, densely dotted};
\addlegendimage{no markers, black};
\addlegendimage{only marks, mark=triangle*, red};
\addlegendimage{no markers, red, densely dashed};
\addplot [color = orange ,mark=no,thick,densely dotted] table[x index=0, y index=1]{R_Sig_NGC5694_Noyora_whole.txt};
\addplot [color = black ,mark=no,thick,solid] table[x index=0, y index=1]{R_Sig_NGC5694_Noyora_whole_apprx.txt};
\addplot[color = red, only marks, mark=triangle*, mark options={red}, error bars,y dir=both, y explicit] table [x index=0, y index=1, y error index=2]{Rd_Sigd_Errbar_Err_NGC5694_Noyora_whole.txt};
\addplot [color = red ,mark=no,thick,densely dashed] table[x index=0, y index=1]{Rc_Sigc_NGC5694_Trager_whole.txt};
\node[black] at (-1.9,-7) {(\normalsize{$r_\text{c}=0\textquotesingle.0388, \, \delta=0.10$})} ; 
\end{axis}
\end{tikzpicture}

\vspace{0.3cm}

\begin{tikzpicture}
\begin{axis}[width=8cm,height=9cm, grid=major,xmin=-3.5,xmax=1.5,ymin=-14,ymax=0.5,ylabel=\normalsize{$\text{SB}_\text{o}-\text{SB}$},xlabel=\normalsize{$\log[R(\text{arcmin})]$},legend style={cells={align=left}},legend entries={\normalsize{ss-OAFP}\\ \normalsize{$(\bar{c}=2.44,m=4.2)$},\normalsize{approx.} \\ \normalsize{ $(\bar{c}=2.83,m=4.95)$}, \normalsize{NGC 5824(n)}, \normalsize{Cheb.}},yticklabel style={/pgf/number format/.cd,fixed zerofill,precision=1},legend pos=south west]
\addlegendimage{no markers, orange, densely dotted};
\addlegendimage{no markers, black};
\addlegendimage{only marks, mark=triangle*, red};
\addlegendimage{no markers, red, densely dashed};
\addplot [color = orange ,mark=no,thick,densely dotted] table[x index=0, y index=1]{R_Sig_NGC5824_Noyora_whole.txt};
\addplot [color = black ,mark=no,thick,solid] table[x index=0, y index=1]{R_Sig_NGC5824_Noyora_whole_apprx.txt};
\addplot[color = red, only marks, mark=triangle*, mark options={red}, error bars,y dir=both, y explicit] table [x index=0, y index=1, y error index=2]{Rd_Sigd_Errbar_Err_NGC5824_Noyora_whole.txt};
\addplot [color = red ,mark=no,thick,densely dashed] table[x index=0, y index=1]{Rc_Sigc_NGC5824_Trager_whole.txt};
\node[black] at (-1.9,-7) {(\normalsize{$r_\text{c}=0\textquotesingle.0388, \, \delta=0.090$})} ; 
\end{axis}
\end{tikzpicture}
	\caption{Fitting of the energy-truncated ss-OAFP model and its approximated form to the surface brightness profiles of NGC 1851, NGC 5694 and NGC 5824 reported in \citep{Noyola_2006}.  The unit of the surface brightness (SB) is V magnitude per square of arcseconds. The brightness is normalized by the magnitude $\text{SB}_\text{o}$ at the smallest radius point. The values of $r_c$ and $\delta$ were obtained from the approximated form. In the legends, `Cheb.' means the Chebyshev approximation of the surface brightness reported in \citep{Trager_1995} and `(n)' means KM cluster as judged so in \citep{Djorgovski_1986}.}
\label{fig:fitting_PCC_Noyora_whole}
\end{figure}
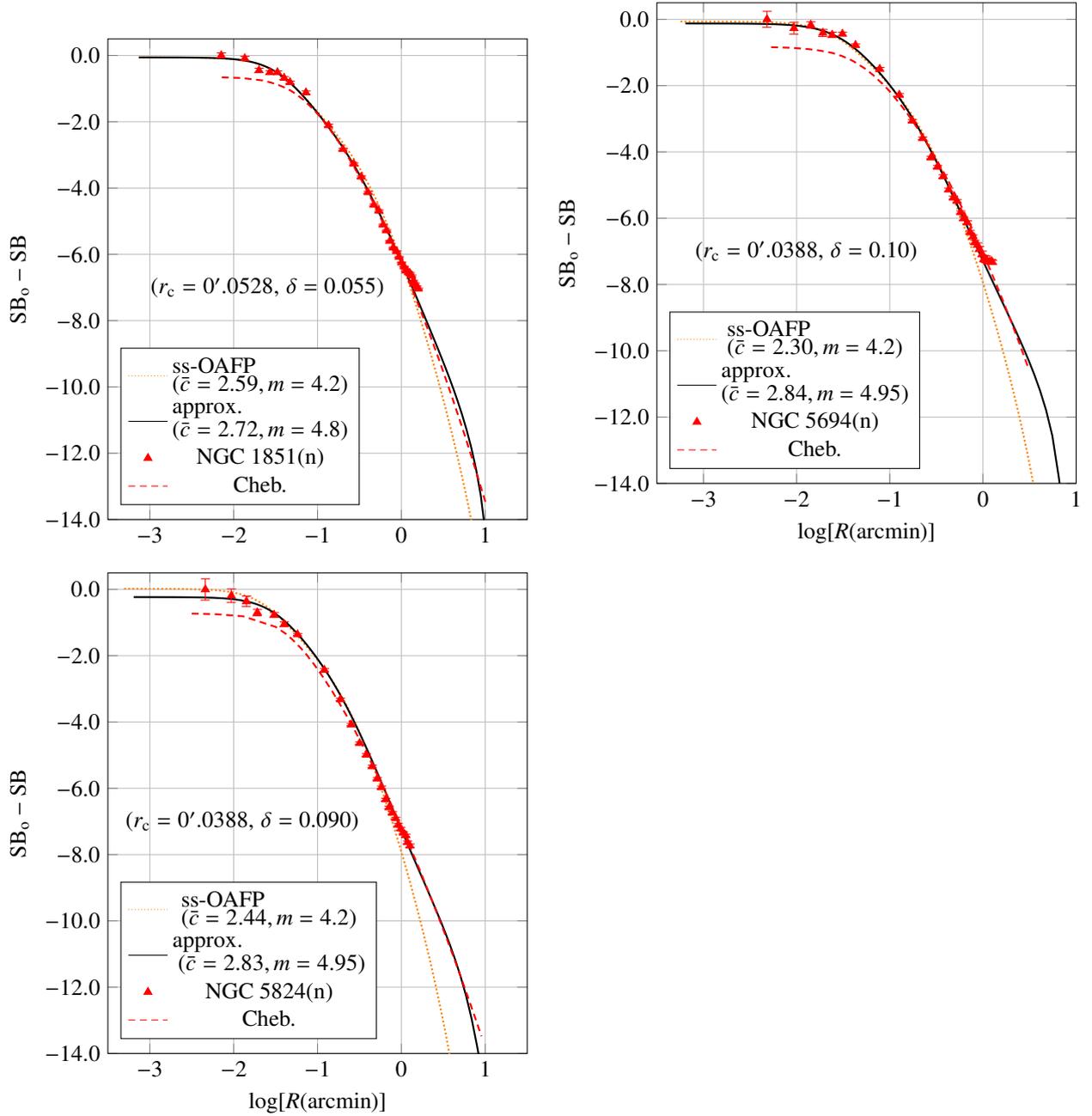

Figure \ref{fig:fitting_KM_Noyora} shows the fitting of the energy-truncated ss-OAFP model to the surface brightness ( $V$-band magnitude per arseconds squared) with flat cores reported in \citep{Noyola_2006}. The brightness is depicted together with the Chebyshev approximation of the brightness reported in \citep{Trager_1995}. To fit our model to the profiles, we employed polytropic indeces $m=4.2$ through $m=4.4$. Figure \ref{fig:fitting_PCC_Noyora_fail_1} depicts the surface brightness profile of globular clusters that our model could not fit due to the cusps in the cores. Figure \ref{fig:fitting_PCC_Noyora_whole} depicts the surface brightness profiles fitted by the energy-truncated ss-OAFP model and its approximated model. For approximated models, we needed to employ high polytropic indeces $m=4.8\sim4.95$. 

\begin{figure}[H]
\begin{tikzpicture}
\begin{axis}[width=8cm,height=6cm, grid=major,xmin=-3.5,xmax=1.5,ymin=-14,ymax=0.5,ylabel=\normalsize{$\text{SB}_\text{o}-\text{SB}$},legend style={cells={align=left}},legend entries={\normalsize{ss-OAFP}\\\normalsize{$(\bar{c}=2.08, m=4.20)$}, \normalsize{NGC 5286(n)}, \normalsize{Cheb.}},yticklabel style={/pgf/number format/.cd,fixed zerofill,precision=1},legend pos=south west]
\addlegendimage{no markers, orange};
\addlegendimage{only marks, mark=triangle*, red};
\addlegendimage{no markers, red, densely dashed};
\addplot [color = orange ,mark=no,thick,solid] table[x index=0, y index=1]{R_Sig_NGC5286_Noyora_whole.txt};
\addplot[color = red, only marks, mark=triangle*, mark options={red}, error bars,y dir=both, y explicit] table [x index=0, y index=1, y error index=2]{Rd_Sigd_Errbar_Err_NGC5286_Noyora_whole.txt};
\addplot [color = red ,mark=no,thick,densely dashed] table[x index=0, y index=1]{Rc_Sigc_NGC5286_Trager_whole.txt};
\node[black] at (-1.9,-5.5) {(\normalsize{$r_\text{c}=0\textquotesingle.230, \, \delta=0.044$})} ; 
\end{axis}
\end{tikzpicture}\hspace{0.3cm}
\begin{tikzpicture}
	\begin{axis}[width=8cm,height=6cm, grid=major,xmin=-3.5,xmax=1.0,ymin=-12,ymax=0.5,legend style={cells={align=left}},legend entries={\normalsize{ss-OAFP}\\\normalsize{$(\bar{c}=2.44,m=4.2)$}, \normalsize{NGC 6093(n)},\normalsize{Cheb.}},yticklabel style={/pgf/number format/.cd,fixed zerofill,precision=1},legend pos=south west]
	\addlegendimage{no markers, orange};
	\addlegendimage{only marks, mark=triangle*, red};
	\addlegendimage{no markers, red, densely dashed};
	\addplot [color = orange ,mark=no,thick,solid] table[x index=0, y index=1]{R_Sig_NGC6093_Noyora_whole.txt};
	\addplot[color = red, only marks, mark=triangle*, mark options={red}, error bars,y dir=both, y explicit] table [x index=0, y index=1, y error index=2]{Rd_Sigd_Errbar_Err_NGC6093_Noyora_whole.txt};
	\addplot [color = red ,mark=no,thick,densely dashed] table[x index=0, y index=1]{Rc_Sigc_NGC6093_Trager_whole.txt};
	\node[black] at (-1.9,-5.5)  {(\normalsize{$r_\text{c}=0\textquotesingle.121, \, \delta=0.025$})} ; 
	\end{axis}
	\end{tikzpicture}
	
\vspace{0.3cm}

\begin{tikzpicture}
	\begin{axis}[width=8cm,height=6cm, grid=major,xmin=-3.5,xmax=1.0,ymin=-12,ymax=0.5,ylabel=\normalsize{$\text{SB}_\text{o}-\text{SB}$},xlabel=\normalsize{$\log[R(\text{arcmin})]$}, legend style={cells={align=left}},legend entries={\normalsize{ss-OAFP}\\\normalsize{$(\bar{c}=2.03, m=4.2)$}, \normalsize{NGC 6535(n)},\normalsize{Cheb.}},yticklabel style={/pgf/number format/.cd,fixed zerofill,precision=1},legend pos=south west]
	\addlegendimage{no markers, orange};
	\addlegendimage{only marks, mark=triangle*, red};
	\addlegendimage{no markers, red, densely dashed};
	\addplot [color = orange ,mark=no,thick,solid] table[x index=0, y index=1]{R_Sig_NGC6535_Noyora_whole.txt};
	\addplot[color = red, only marks, mark=triangle*, mark options={red}, error bars,y dir=both, y explicit] table [x index=0, y index=1, y error index=2]{Rd_Sigd_Errbar_Err_NGC6535_Noyora_whole.txt};
	\addplot [color = red ,mark=no,thick,densely dashed] table[x index=0, y index=1]{Rc_Sigc_NGC6535_Trager_whole.txt};
	\node[black] at (-1.9,-5.5)  {(\normalsize{$r_\text{c}=0\textquotesingle.272, \, \delta=0.048$})} ; 
	\end{axis}
	\end{tikzpicture}
\begin{tikzpicture}
\begin{axis}[width=8cm,height=6cm, grid=major,xmin=-3.5,xmax=1.5,ymin=-14,ymax=0.5,xlabel=\normalsize{$\log[R(\text{arcmin})]$},legend style={cells={align=left}},legend entries={\normalsize{ss-OAFP}\\\normalsize{$(\bar{c}=2.59,m=4.2)$}, \normalsize{NGC 6541(n)}, \normalsize{Cheb.}},yticklabel style={/pgf/number format/.cd,fixed zerofill,precision=1},legend pos=south west]
\addlegendimage{no markers, orange};
\addlegendimage{only marks, mark=triangle*, red};
\addlegendimage{no markers, red, densely dashed};
\addplot [color = orange ,mark=no,thick,solid] table[x index=0, y index=1]{R_Sig_NGC6541_Noyora_whole.txt};
\addplot[color = red, only marks, mark=triangle*, mark options={red}, error bars,y dir=both, y explicit] table [x index=0, y index=1, y error index=2]{Rd_Sigd_Errbar_Err_NGC6541_Noyora_whole.txt};
\addplot [color = red ,mark=no,thick,densely dashed] table[x index=0, y index=1]{Rc_Sigc_NGC6541_Trager_whole.txt};
\node[black] at (-1.9,-5.5)  {(\normalsize{$r_\text{c}=0\textquotesingle.168, \, \delta=0.020$})} ; 
\end{axis}
\end{tikzpicture}\hspace{0.3cm}
	\caption{Failure of fitting of the energy-truncated ss-OAFP model to the surface brightness profiles of NGC 5286, NGC 6093, NGC 6535 and NGC 6541 reported in \citep{Noyola_2006}.  The unit of the surface brightness (SB) is V magnitude per square of arcseconds. The brightness is normalized by the magnitude $\text{SB}_\text{o}$ at the smallest radius point. In the legends, `Cheb.' means the Chebyshev approximation of the surface brightness reported in \citep{Trager_1995} and `(n)' means KM cluster as judged so in \citep{Djorgovski_1986}.}
\label{fig:fitting_PCC_Noyora_fail_1}
\end{figure}

	\section{Fitting of the polytropic sphere model to low-concentration clusters}\label{Appendix_poly_Fit}
	The present appendix shows the results of application of the energy-truncated ss-OAFP model to PCC clusters reported in \citep{Kron_1984,Lugger_1995,Noyola_2006}. For fitting of the model to \citep{Kron_1984,Lugger_1995}'s data, the infinite norm of the deviations of the model from the data is minimized while for fitting to \citep{Lugger_1995}, $\chi^{2}_\nu$-value is minimized. Sections \ref{sec:PCC?_OAFP}, \ref{sec:PCC_OAFP} and \ref{sec:PCC_OAFP_noyola} show the fitting to PCC clusters reported in \citep{Kron_1984}, \citep{Lugger_1995} and \citep{Noyola_2006}.

\subsection{PCC? clusters \citep{Kron_1984}}\label{sec:PCC?_OAFP}

\begin{figure}[H]
	\centering
	\begin{tikzpicture}
	\begin{loglogaxis}[width=7cm,height=7cm, grid=major,ylabel=\normalsize{$\Sigma$}, xmin=8e-3,xmax=10,ymin=1e-4,ymax=3,legend pos=south west, legend style={cells={align=left}}]
	\addplot [color = black ,mark=no,thick,solid ] table[x index=0, y index=1]{R_Sig_NGC1904.txt};
	\addlegendentry{\normalsize{$\delta=0.027$}\\ \normalsize{$(c=1.86)$}} 
	\addplot [only marks,color = blue, mark=o,thick ] table[x index=0, y index=1]{Rd_Sigd_NGC1904.txt};
	\addlegendentry{\normalsize{NGC 1904}\\ \normalsize{(c$?$)}} 
	\node[black] at (5e-2,3e-2) {(\normalsize{$r_\text{c}=0\textquotesingle.19$})} ; 
	\end{loglogaxis}
	\end{tikzpicture}
	\begin{tikzpicture}
	\begin{loglogaxis}[width=7cm,height=7cm, grid=major,xmin=8e-3,xmax=10,ymin=1e-4,ymax=3,legend pos=south west, legend style={cells={align=left}}]
	\addplot [color = black ,mark=no,thick,solid ] table[x index=0, y index=1]{R_Sig_NGC4147.txt};
	\addlegendentry{\normalsize{$\delta=0.025$} \\ \normalsize{ $(c=1.89)$}} 
	\addplot [only marks,color = blue, mark=o,,thick ] table[x index=0, y index=1]{Rd_Sigd_NGC4147.txt};
	\addlegendentry{\normalsize{NGC 4147} \\ \normalsize{(n$?$c$?$)}}
	\node[black] at (5e-2,3e-2) {(\normalsize{$r_\text{c}=0\textquotesingle.14$})} ;  
	\end{loglogaxis}
	\end{tikzpicture}

	\vspace{0.3cm}
	
	\begin{tikzpicture}
	\begin{loglogaxis}[width=7cm,height=7cm, grid=major,xmin=8e-3,xmax=10,ymin=1e-4,ymax=3,ylabel=\normalsize{$\Sigma$},xlabel=\normalsize{$R$(arcmin)},legend pos=south west, legend style={cells={align=left}}]
	\addplot [color = black ,mark=no,thick,solid ] table[x index=0, y index=1]{R_Sig_NGC6544.txt};
	\addlegendentry{\normalsize{$\delta=0.050$ }\\\normalsize{$(c=1.61)$}} 
	\addplot [only marks,color = blue, mark=o,thick ] table[x index=0, y index=1]{Rd_Sigd_NGC6544.txt};
	\addlegendentry{\normalsize{NGC 6544}\\\normalsize{(c$?$)}} 
	\node[black] at (5e-2,3e-2) {(\normalsize{$r_\text{c}=0\textquotesingle.50$})} ; 
	\end{loglogaxis}
	\end{tikzpicture}
	\begin{tikzpicture}
	\begin{loglogaxis}[width=7cm,height=7cm, grid=major,xlabel=\normalsize{$R$ (arcmin)},xmin=8e-3,xmax=10,ymin=1e-4,ymax=3,legend pos=south west, legend style={cells={align=left}}]
	\addplot [color = black ,mark=no,thick,solid ] table[x index=0, y index=1]{R_Sig_NGC6652.txt};
	\addlegendentry{\normalsize{$\delta=0.028$}\\\normalsize{$(c=1.84)$}} 
	\addplot [only marks,color = blue, mark=o,thick ] table[x index=0, y index=1]{Rd_Sigd_NGC6652.txt};
	\addlegendentry{\normalsize{NGC 6652}\\\normalsize{(n$?$c$?$)}} 
	\node[black] at (5e-2,3e-2) {(\normalsize{$r_\text{c}=0\textquotesingle.14$})} ; 
	\end{loglogaxis}
	\end{tikzpicture}
	\caption{ Fitting of the energy-truncated ss-OAFP model to the projected densities of NGC 1904, NGC 4147, NGC 6544 and NGC 6652 reported in \citep{Kron_1984}. The unit of the projected density $\Sigma$ is number per square of arcminutes. In the legends, (c) means PCC cluster, (c?) probable/possible PCC and (n?c?) weak indications of PCC as judged so in \citep{Djorgovski_1986}.}
	\label{fig:fitting_PCC_1_6}
\end{figure}
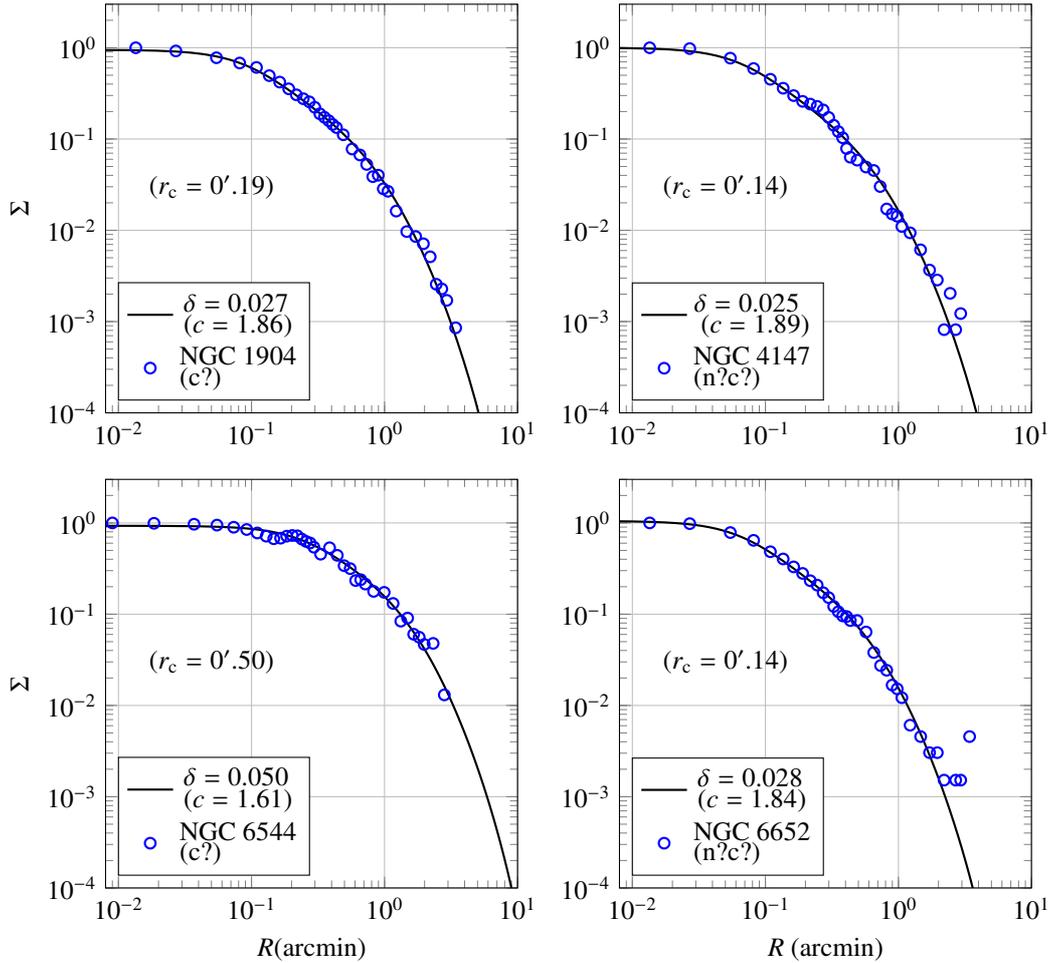

Figure \ref{fig:fitting_PCC_1_6} depicts the energy-truncated ss-OAFP model with $m=3.9$ fitted to the projected density profiles of possible PCC clusters reported in \citep{Kron_1984} in which NGC 1904,  NGC 4147, NGC 6544 and NGC 6652 are considered possibly PCC as described in \citep{Djorgovski_1986} by `probable/possible' PCC clusters or `the weak indications of PCC' clusters.

\subsection{PCC clusters \citep{Lugger_1995} and \citep{Djorgovski_1986}}\label{sec:PCC_OAFP}

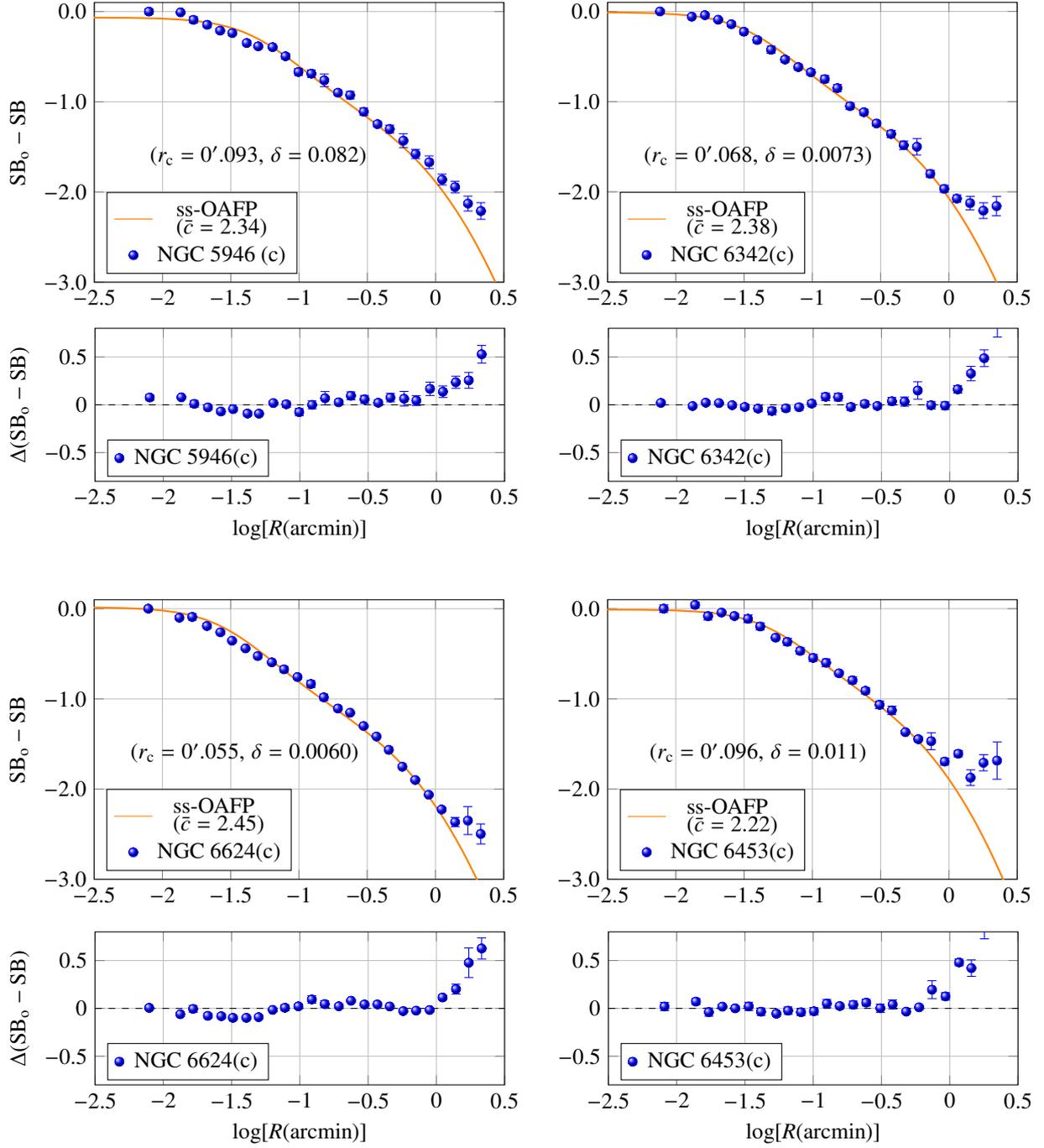
\begin{figure}[H]
	\begin{tikzpicture}
	\begin{axis}[width=8cm,height=6cm, grid=major,xmin=-2.5,xmax=0.5,ymin=-3.0,ymax=0.1,ylabel=\normalsize{$\text{SB}_\text{o}-\text{SB}$},legend style={cells={align=left}},legend entries={\normalsize{ss-OAFP}\\\normalsize{$(\bar{c}=2.34)$}, \normalsize{NGC 5946 (c)}},yticklabel style={/pgf/number format/.cd,fixed zerofill,precision=1},legend pos=south west]
	\addlegendimage{no markers, orange};
	\addlegendimage{only marks, mark=ball, blue};
	\addplot [color = orange ,mark=no,thick,solid] table[x index=0, y index=1]{R_Sig_NGC5946.txt};
	\addplot[color = blue, only marks, mark=ball, mark options={blue}, error bars,y dir=both, y explicit] table [x index=0, y index=1, y error index=2]{Rd_Sigd_Errbar_Err_NGC5946.txt};
	\node[black] at (-1.3,-1.6) {(\normalsize{$r_\text{c}=0\textquotesingle.093, \, \delta=0.082$})} ; 
	\end{axis}
	\end{tikzpicture}\hspace{0.3cm}
	\begin{tikzpicture}
	\begin{axis}[width=8cm,height=6cm, grid=major,xmin=-2.5,xmax=0.5,ymin=-3.0,ymax=0.1,legend style={cells={align=left}},legend entries={\normalsize{ss-OAFP}\\\normalsize{$(\bar{c}=2.38)$}, \normalsize{NGC 6342(c)}},yticklabel style={/pgf/number format/.cd,fixed zerofill,precision=1},legend pos=south west]
	\addlegendimage{no markers, orange};
	\addlegendimage{only marks, mark=ball, blue};
	\addplot [color = orange ,mark=no,thick,solid] table[x index=0, y index=1]{R_Sig_NGC6342.txt};
	\addplot[color = blue, only marks, mark=ball, mark options={blue}, error bars,y dir=both, y explicit] table [x index=0, y index=1, y error index=2]{Rd_Sigd_Errbar_Err_NGC6342.txt};
	\node[black] at (-1.4,-1.6) {(\normalsize{$r_\text{c}=0\textquotesingle.068, \, \delta=0.0073$})} ; 
	\end{axis}
	\end{tikzpicture}
	
	\vspace{0.2cm}
	
	\begin{tikzpicture}
	\begin{axis}[width=8cm,height=4cm, grid=major,xmin=-2.5,xmax=0.5,ymin=-0.8,ymax=0.8,xlabel=\normalsize{$\log[R(\text{arcmin})]$ },ylabel=\normalsize{$\Delta(\text{SB}_\text{o}-\text{SB})$},legend style={cells={align=left}},legend entries={\normalsize{ NGC 5946(c)}},legend pos=south west]
	\addlegendimage{only marks, mark=ball, blue};
	\addplot[color = blue, only marks, mark=ball, mark options={blue}, error bars,y dir=both, y explicit] table [x index=0, y index=3, y error index=2]{Rd_Sigd_Errbar_Err_NGC5946.txt};
	\draw[dashed] (-2.5,0)--(1.0,0);
	\end{axis}
	\end{tikzpicture}\hspace{0.3cm}
	\begin{tikzpicture}
	\begin{axis}[width=8cm,height=4cm, grid=major,xmin=-2.5,xmax=0.5,ymin=-0.8,ymax=0.8,xlabel=\normalsize{$\log[R(\text{arcmin})]$ },legend style={cells={align=left}},legend entries={\normalsize{ NGC 6342(c)}},legend pos=south west]
	\addlegendimage{only marks, mark=ball, blue};
	\addplot[color = blue, only marks, mark=ball, mark options={blue}, error bars,y dir=both, y explicit] table [x index=0, y index=3, y error index=2]{Rd_Sigd_Errbar_Err_NGC6342.txt};
	\draw[dashed] (-2.5,0)--(1.0,0);
	\end{axis}
	\end{tikzpicture}
	
	\vspace{0.7cm}
	
	\begin{tikzpicture}
	\begin{axis}[width=8cm,height=6cm, grid=major,xmin=-2.5,xmax=0.5,ymin=-3.0,ymax=0.1,ylabel=\normalsize{$\text{SB}_\text{o}-\text{SB}$},legend style={cells={align=left}},legend entries={\normalsize{ss-OAFP}\\\normalsize{$(\bar{c}=2.45)$}, \normalsize{NGC 6624(c)}},yticklabel style={/pgf/number format/.cd,fixed zerofill,precision=1},legend pos=south west]
	\addlegendimage{no markers, orange};
	\addlegendimage{only marks, mark=ball, blue};
	\addplot [color = orange ,mark=no,thick,solid] table[x index=0, y index=1]{R_Sig_NGC6624.txt};
	\addplot[color = blue, only marks, mark=ball, mark options={blue}, error bars,y dir=both, y explicit] table [x index=0, y index=1, y error index=2]{Rd_Sigd_Errbar_Err_NGC6624.txt};
	\node[black] at (-1.4,-1.6) {(\normalsize{$r_\text{c}=0\textquotesingle.055, \, \delta=0.0060$})} ; 
	\end{axis}
	\end{tikzpicture}\hspace{0.3cm}
	\begin{tikzpicture}
	\begin{axis}[width=8cm,height=6cm, grid=major,xmin=-2.5,xmax=0.5,ymin=-3.0,ymax=0.1,legend style={cells={align=left}},legend entries={\normalsize{ss-OAFP}\\\normalsize{$(\bar{c}=2.22)$}, \normalsize{NGC 6453(c)}},yticklabel style={/pgf/number format/.cd,fixed zerofill,precision=1},legend pos=south west]
	\addlegendimage{no markers, orange};
	\addlegendimage{only marks, mark=ball, blue};
	\addplot [color = orange ,mark=no,thick,solid] table[x index=0, y index=1]{R_Sig_NGC6453.txt};
	\addplot[color = blue, only marks, mark=ball, mark options={blue}, error bars,y dir=both, y explicit] table [x index=0, y index=1, y error index=2]{Rd_Sigd_Errbar_Err_NGC6453.txt};
	\node[black] at (-1.4,-1.6) {(\normalsize{$r_\text{c}=0\textquotesingle.096, \, \delta=0.011$})} ; 
	\end{axis}
	\end{tikzpicture}
	
	\vspace{0.3cm}
	
	\begin{tikzpicture}
	\begin{axis}[width=8cm,height=4cm, grid=major,xmin=-2.5,xmax=0.5,ymin=-0.8,ymax=0.8,xlabel=\normalsize{$\log[R(\text{arcmin})]$ },ylabel=\normalsize{$\Delta(\text{SB}_\text{o}-\text{SB})$},legend style={cells={align=left}},legend entries={\normalsize{ NGC 6624(c)}},legend pos=south west]
	\addlegendimage{only marks, mark=ball, blue};
	\addplot[color = blue, only marks, mark=ball, mark options={blue}, error bars,y dir=both, y explicit] table [x index=0, y index=3, y error index=2]{Rd_Sigd_Errbar_Err_NGC6624.txt};
	\draw[dashed] (-2.5,0)--(1.0,0);
	\end{axis}
	\end{tikzpicture}\hspace{0.3cm}
	\begin{tikzpicture}
	\begin{axis}[width=8cm,height=4cm, grid=major,xmin=-2.5,xmax=0.5,ymin=-0.8,ymax=0.8,xlabel=\normalsize{$\log[R(\text{arcmin})]$ },legend style={cells={align=left}},legend entries={\normalsize{ NGC 6453(c)}},legend pos=south west]
	\addlegendimage{only marks, mark=ball, blue};
	\addplot[color = blue, only marks, mark=ball, mark options={blue}, error bars,y dir=both, y explicit] table [x index=0, y index=3, y error index=2]{Rd_Sigd_Errbar_Err_NGC6453.txt};
	\draw[dashed] (-2.5,0)--(1.0,0);
	\end{axis}
	\end{tikzpicture}
	\caption{Fitting of the energy-truncated ss-OAFP model to the surface brightness profiles of NGC 5946, NGC 6342, NGC 6624 and NGC 6453 reported in \citep{Lugger_1995}. The unit of the surface brightness (SB) is U magnitude per square of arcseconds. The brightness is normalized by the magnitude $\text{SB}_\text{o}$ at the smallest radius point. In the legend, `(c)' means PCC cluster as judged so in \citep{Djorgovski_1986}. $\Delta(\text{SB}_\text{o}-\text{SB})$ is the corresponding deviation of $\text{SB}_\text{o}-\text{SB}$ from the model.}
	\label{fig:fitting_PCC_Lugger_1}
\end{figure}

Figures \ref{fig:fitting_PCC_Lugger_1} and \ref{fig:fitting_PCC_Lugger_3} show the results of fitting of energy-truncated ss-OAFP model with $m=3.9$ to the surface brightness profiles of PCC clusters reported in \citep{Lugger_1995} and Terzan 1 in \citep{Djorgovski_1986}. For clusters with resolved cores, our model well fits the core and halo at up to 1 arcminute. On one hand, as expected the fitting of our model to the clusters with unresolved core (NGC 5946 and NGC 6624) is not satisfactory while NGC 6342 is one of them but appears reasonably fitted by our model.

\begin{figure}[H]
	\begin{tikzpicture}
	\begin{axis}[width=8cm,height=6cm, grid=major,xmin=-2.5,xmax=0.5,ymin=-3.0,ymax=0.1,ylabel=\normalsize{$\text{SB}_\text{o}-\text{SB}$},legend style={cells={align=left}},legend entries={\normalsize{ss-OAFP}\\\normalsize{$(\bar{c}=2.29)$}, \normalsize{NGC 6522(c)}},yticklabel style={/pgf/number format/.cd,fixed zerofill,precision=1},legend pos=south west]
	\addlegendimage{no markers, orange};
	\addlegendimage{only marks, mark=ball, blue};
	\addplot [color = orange ,mark=no,thick,solid] table[x index=0, y index=1]{R_Sig_NGC6522.txt};
	\addplot[color = blue, only marks, mark=ball, mark options={blue}, error bars,y dir=both, y explicit] table [x index=0, y index=1, y error index=2]{Rd_Sigd_Errbar_Err_NGC6522.txt};
	\node[black] at (-1.3,-1.6) {(\normalsize{$r_\text{c}=0\textquotesingle.096, \, \delta=0.0094$})} ; 
	\end{axis}
	\end{tikzpicture}\hspace{0.3cm}
	\begin{tikzpicture}
	\begin{axis}[width=8cm,height=6cm, grid=major,xmin=-2.5,xmax=0.5,ymin=-3.0,ymax=0.1,legend style={cells={align=left}},legend entries={\normalsize{ss-OAFP}\\\normalsize{$(\bar{c}=2.51)$}, \normalsize{NGC 6558(c)}},yticklabel style={/pgf/number format/.cd,fixed zerofill,precision=1},legend pos=south west]
	\addlegendimage{no markers, orange};
	\addlegendimage{only marks, mark=ball, blue};
	\addplot [color = orange ,mark=no,thick,solid] table[x index=0, y index=1]{R_Sig_NGC6558.txt};
	\addplot[color = blue, only marks, mark=ball, mark options={blue}, error bars,y dir=both, y explicit] table [x index=0, y index=1, y error index=2]{Rd_Sigd_Errbar_Err_NGC6558.txt};
	\node[black] at (-1.3,-1.6) {(\normalsize{$r_\text{c}=0\textquotesingle.085, \, \delta=0.0052$})} ; 
	\end{axis}
	\end{tikzpicture}
	
	\vspace{0.3cm}
	
	\begin{tikzpicture}
	\begin{axis}[width=8cm,height=4cm, grid=major,xmin=-2.5,xmax=0.5,ymin=-0.8,ymax=0.8,xlabel=\normalsize{$\log[R(\text{arcmin})]$ },ylabel=\normalsize{$\Delta(\text{SB}_\text{o}-\text{SB})$},legend style={cells={align=left}},legend entries={\normalsize{ NGC 6522(c)}},legend pos=south west]
	\addlegendimage{only marks, mark=ball, blue};
	\addplot[color = blue, only marks, mark=ball, mark options={blue}, error bars,y dir=both, y explicit] table [x index=0, y index=3, y error index=2]{Rd_Sigd_Errbar_Err_NGC6522.txt};
	\draw[dashed] (-2.5,0)--(1.0,0);
	\end{axis}
	\end{tikzpicture}\hspace{0.3cm}
	\begin{tikzpicture}
	\begin{axis}[width=8cm,height=4cm, grid=major,xmin=-2.5,xmax=0.5,ymin=-0.8,ymax=0.8,xlabel=\normalsize{$\log[R(\text{arcmin})]$ },legend style={cells={align=left}},legend entries={\normalsize{ NGC 6558(c)}},legend pos=south west]
	\addlegendimage{only marks, mark=ball, blue};
	\addplot[color = blue, only marks, mark=ball, mark options={blue}, error bars,y dir=both, y explicit] table [x index=0, y index=3, y error index=2]{Rd_Sigd_Errbar_Err_NGC6558.txt};
	\draw[dashed] (-2.5,0)--(1.0,0);
	\end{axis}
	\end{tikzpicture}
	
	\vspace{0.7cm}
	
	\begin{tikzpicture}
	\begin{axis}[width=8cm,height=6cm, grid=major,xmin=-2.5,xmax=0.5,ymin=-3.0,ymax=0.1,ylabel=\normalsize{$\text{SB}_\text{o}-\text{SB}$},legend style={cells={align=left}},legend entries={\normalsize{ss-OAFP}\\\normalsize{$(\bar{c}=2.37)$}, \normalsize{NGC 7099(c)}},yticklabel style={/pgf/number format/.cd,fixed zerofill,precision=1},legend pos=south west]
	\addlegendimage{no markers, orange};
	\addlegendimage{only marks, mark=ball, blue};
	\addplot [color = orange ,mark=no,thick,solid] table[x index=0, y index=1]{R_Sig_NGC7099.txt};
	\addplot[color = blue, only marks, mark=ball, mark options={blue}, error bars,y dir=both, y explicit] table [x index=0, y index=1, y error index=2]{Rd_Sigd_Errbar_Err_NGC7099.txt};
	\node[black] at (-1.3,-1.6) {(\normalsize{$r_\text{c}=0\textquotesingle.086, \, \delta=0.0076$})} ; 
	\end{axis}
	\end{tikzpicture}\hspace{0.3cm}
	\begin{tikzpicture}
	\begin{axis}[width=8cm,height=6cm, grid=major,xmin=-2.5,xmax=0.5,ymin=-3.0,ymax=0.1,legend style={cells={align=left}},legend entries={\normalsize{ss-OAFP}\\\normalsize{$(\bar{c}=2.15)$}, \normalsize{ Trz 1(c)}},yticklabel style={/pgf/number format/.cd,fixed zerofill,precision=1},legend pos=south west]
	\addlegendimage{no markers, orange};
	\addlegendimage{only marks, mark=ball, blue};
	\addplot [color = orange ,mark=no,thick,solid] table[x index=0, y index=1]{R_Sig_Trz1.txt};
	\addplot[color = blue, only marks, mark=ball, mark options={blue}, error bars,y dir=both, y explicit] table [x index=0, y index=1, y error index=2]{Rd_Sigd_Errbar_Err_Trz1.txt};
	\node[black] at (-1.3,-1.6) {(\normalsize{$r_\text{c}=0\textquotesingle.149, \, \delta=0.0087$})} ; 
	\end{axis}
	\end{tikzpicture}
	
	\vspace{0.3cm}
	
	\begin{tikzpicture}
	\begin{axis}[width=8cm,height=4cm, grid=major,xmin=-2.5,xmax=0.5,ymin=-0.8,ymax=0.8,xlabel=\normalsize{$\log[R(\text{arcmin})]$ },ylabel=\normalsize{$\Delta(\text{SB}_\text{o}-\text{SB})$},legend style={cells={align=left}},legend entries={\normalsize{ NGC7099(c)}},legend pos=south west]
	\addlegendimage{only marks, mark=ball, blue};
	\addplot[color = blue, only marks, mark=ball, mark options={blue}, error bars,y dir=both, y explicit] table [x index=0, y index=3, y error index=2]{Rd_Sigd_Errbar_Err_NGC7099.txt};
	\draw[dashed] (-2.5,0)--(1.0,0);
	\end{axis}
	\end{tikzpicture}\hspace{0.3cm}
	\begin{tikzpicture}
	\begin{axis}[width=8cm,height=4cm, grid=major,xmin=-2.5,xmax=0.5,ymin=-0.8,ymax=0.8,xlabel=\normalsize{$\log[R(\text{arcmin})]$ },legend style={cells={align=left}},legend entries={\normalsize{ Trz 1 (c)}},legend pos=south west]
	\addlegendimage{only marks, mark=ball, blue};
	\addplot[color = blue, only marks, mark=ball, mark options={blue}, error bars,y dir=both, y explicit] table [x index=0, y index=3, y error index=2]{Rd_Sigd_Errbar_Err_Trz1.txt};
	\draw[dashed] (-2.5,0)--(1.0,0);
	\end{axis}
	\end{tikzpicture}
	\caption{Fitting of the energy-truncated ss-OAFP model to the surface brightness profiles of NGC 6522, NGC 6558 and  NGC 7099 reported in \citep{Lugger_1995} and Terzan 1 reported in \citep{Djorgovski_1986}.  The unit of the surface brightness (SB) is U magnitude per square of arcseconds except for Terzan 1 for which B band is used. The brightness is normalized by the magnitude $\text{SB}_\text{o}$ at the smallest radius point. In the legend, `(c)' means PCC cluster as judged so in \citep{Djorgovski_1986}. $\Delta(\text{SB}_\text{o}-\text{SB})$ is the corresponding deviation of $\text{SB}_\text{o}-\text{SB}$ from the model.}
	\label{fig:fitting_PCC_Lugger_3}
\end{figure}

\subsection{PCC clusters \citep{Noyola_2006}}\label{sec:PCC_OAFP_noyola}

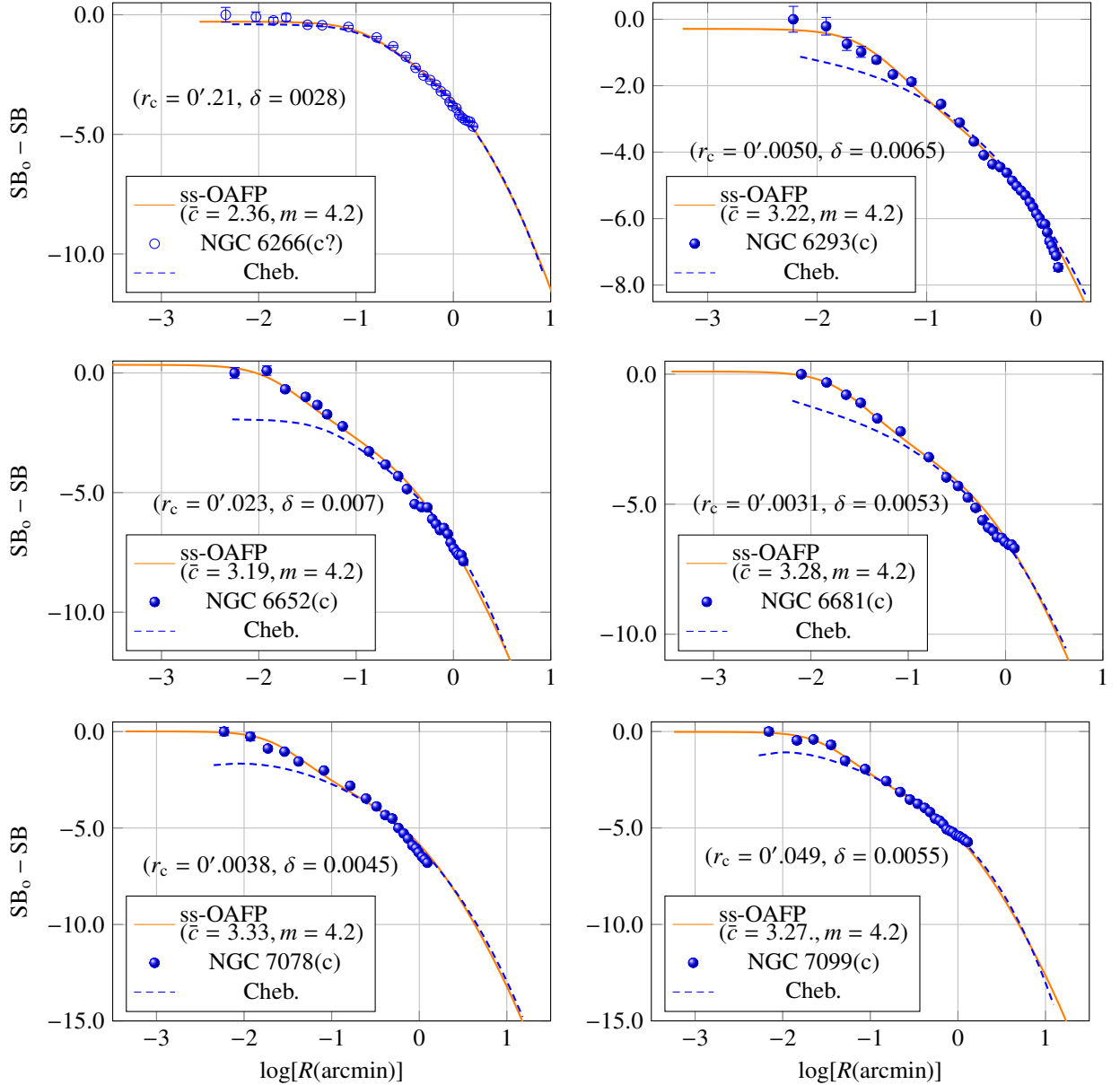
\begin{figure}[H]
	\begin{tikzpicture}
	\begin{axis}[width=8cm,height=6cm, grid=major,xmin=-3.5,xmax=1.0,ymin=-12,ymax=0.5,ylabel=\normalsize{$\text{SB}_\text{o}-\text{SB}$},legend style={cells={align=left}},legend entries={\normalsize{ss-OAFP}\\\normalsize{$(\bar{c}=2.36, m=4.2)$}, \normalsize{NGC 6266(c?)},\normalsize{Cheb.}},yticklabel style={/pgf/number format/.cd,fixed zerofill,precision=1},legend pos=south west]
	\addlegendimage{no markers, orange};
	\addlegendimage{only marks, mark=o, blue};
	\addlegendimage{no markers, blue, densely dashed};
	\addplot [color = orange ,mark=no,thick,solid] table[x index=0, y index=1]{R_Sig_NGC6266_Noyora_whole.txt};
	\addplot[color = blue, only marks, mark=o, mark options={blue}, error bars,y dir=both, y explicit] table [x index=0, y index=1, y error index=2]{Rd_Sigd_Errbar_Err_NGC6266_Noyora_whole.txt};
	\addplot [color = blue ,mark=no,thick,densely dashed] table[x index=0, y index=1]{Rc_Sigc_NGC6266_Trager_whole.txt};
	\node[black] at (-2.2,-3.5) {(\normalsize{$r_\text{c}=0\textquotesingle.21, \, \delta=0028$})} ; 
	\end{axis}
	\end{tikzpicture}\hspace{0.3cm}
	\begin{tikzpicture}
	\begin{axis}[width=8cm,height=6cm, grid=major,xmin=-3.5,xmax=0.5,ymin=-8.5,ymax=0.5,legend style={cells={align=left}},legend entries={\normalsize{ss-OAFP}\\\normalsize{$(\bar{c}=3.22,m=4.2)$}, \normalsize{NGC 6293(c)},\normalsize{Cheb.}},yticklabel style={/pgf/number format/.cd,fixed zerofill,precision=1},legend pos=south west]
	\addlegendimage{no markers, orange};
	\addlegendimage{only marks, mark=ball, blue};
	\addlegendimage{no markers, blue, densely dashed};
	\addplot [color = orange ,mark=no,thick,solid] table[x index=0, y index=1]{R_Sig_NGC6293_Noyora_whole.txt};
	\addplot[color = blue, only marks, mark=ball, mark options={blue}, error bars,y dir=both, y explicit] table [x index=0, y index=1, y error index=2]{Rd_Sigd_Errbar_Err_NGC6293_Noyora_whole.txt};
	\addplot [color = blue ,mark=no,thick,densely dashed] table[x index=0, y index=1]{Rc_Sigc_NGC6293_Trager_whole.txt};
	\node[black] at (-2.0,-4) {(\normalsize{$r_\text{c}=0\textquotesingle.0050, \, \delta=0.0065$})} ; 
	\end{axis}
	\end{tikzpicture}
	
	\vspace{0.3cm}
	
	\begin{tikzpicture}
	\begin{axis}[width=8cm,height=6cm, grid=major,xmin=-3.5,xmax=1.0,ymin=-12,ymax=0.5,ylabel=\normalsize{$\text{SB}_\text{o}-\text{SB}$},legend style={cells={align=left}},legend entries={\normalsize{ss-OAFP}\\\normalsize{$(\bar{c}=3.19, m=4.2)$}, \normalsize{NGC 6652(c)},\normalsize{Cheb.}},yticklabel style={/pgf/number format/.cd,fixed zerofill,precision=1},legend pos=south west]
	\addlegendimage{no markers, orange};
	\addlegendimage{only marks, mark=ball, blue};
	\addlegendimage{no markers, blue, densely dashed};
	\addplot [color = orange ,mark=no,thick,solid] table[x index=0, y index=1]{R_Sig_NGC6652_Noyora_whole.txt};
	\addplot[color = blue, only marks, mark=ball, mark options={blue}, error bars,y dir=both, y explicit] table [x index=0, y index=1, y error index=2]{Rd_Sigd_Errbar_Err_NGC6652_Noyora_whole.txt};
	\addplot [color = blue ,mark=no,thick,densely dashed] table[x index=0, y index=1]{Rc_Sigc_NGC6652_Trager_whole.txt};
	\node[black] at (-1.9,-5.5) {(\normalsize{$r_\text{c}=0\textquotesingle.023, \, \delta=0.007$})} ; 
	\end{axis}
	\end{tikzpicture}\hspace{0.3cm}
	\begin{tikzpicture}
	\begin{axis}[width=8cm,height=6cm, grid=major,xmin=-3.5,xmax=1.0,ymin=-11,ymax=0.5,legend style={cells={align=left}},legend entries={\normalsize{ss-OAFP}\\ \normalsize{$(\bar{c}=3.28,m=4.2)$}, \normalsize{ NGC 6681(c)},\normalsize{Cheb.}},yticklabel style={/pgf/number format/.cd,fixed zerofill,precision=1},legend pos=south west]
	\addlegendimage{no markers, orange};
	\addlegendimage{only marks, mark=ball, blue};
	\addlegendimage{no markers, blue, densely dashed};
	\addplot [color = orange ,mark=no,thick,solid] table[x index=0, y index=1]{R_Sig_NGC6681_Noyora_whole.txt};
	\addplot[color = blue, only marks, mark=ball, mark options={blue}, error bars,y dir=both, y explicit] table [x index=0, y index=1, y error index=2]{Rd_Sigd_Errbar_Err_NGC6681_Noyora_whole.txt};
	\addplot [color = blue ,mark=no,thick,densely dashed] table[x index=0, y index=1]{Rc_Sigc_NGC6681_Trager_whole.txt};
	\node[black] at (-1.9,-5) {(\normalsize{$r_\text{c}=0\textquotesingle.0031, \, \delta=0.0053$})} ; 
	\end{axis}
	\end{tikzpicture}
	
	\vspace{0.3cm}
	
	\begin{tikzpicture}
	\begin{axis}[width=8cm,height=6cm, grid=major,xmin=-3.5,xmax=1.5,ymin=-15,ymax=0.5,ylabel=\normalsize{$\text{SB}_\text{o}-\text{SB}$},xlabel=\normalsize{$\log[R(\text{arcmin})]$},legend style={cells={align=left}},legend entries={\normalsize{ss-OAFP}\\\normalsize{$(\bar{c}=3.33, m=4.2)$}, \normalsize{ NGC 7078(c)},\normalsize{Cheb.}},yticklabel style={/pgf/number format/.cd,fixed zerofill,precision=1},legend pos=south west]
	\addlegendimage{no markers, orange};
	\addlegendimage{only marks, mark=ball, blue};
	\addlegendimage{no markers, blue, densely dashed};
	\addplot [color = orange ,mark=no,thick,solid] table[x index=0, y index=1]{R_Sig_NGC7078_Noyora_whole.txt};
	\addplot[color = blue, only marks, mark=ball, mark options={blue}, error bars,y dir=both, y explicit] table [x index=0, y index=1, y error index=2]{Rd_Sigd_Errbar_Err_NGC7078_Noyora_whole.txt};
	\addplot [color = blue ,mark=no,thick,densely dashed] table[x index=0, y index=1]{Rc_Sigc_NGC7078_Trager_whole.txt};
	\node[black] at (-1.7,-7) {(\normalsize{$r_\text{c}=0\textquotesingle.0038, \, \delta=0.0045$})} ; 
	\end{axis}
	\end{tikzpicture}\hspace{0.3cm}
	\begin{tikzpicture}
	\begin{axis}[width=8cm,height=6cm, grid=major,xmin=-3.5,xmax=1.5,ymin=-15,ymax=0.5,xlabel=\normalsize{$\log[R(\text{arcmin})]$},legend style={cells={align=left}},legend entries={\normalsize{ss-OAFP}\\\normalsize{$(\bar{c}=3.27., m=4.2)$}, \normalsize{NGC 7099(c)},\normalsize{Cheb.}},yticklabel style={/pgf/number format/.cd,fixed zerofill,precision=1},legend pos=south west]
	\addlegendimage{no markers, orange};
	\addlegendimage{only marks, mark=ball, blue};
	\addlegendimage{no markers, blue, densely dashed};
	\addplot [color = orange ,mark=no,thick,solid] table[x index=0, y index=1]{R_Sig_NGC7099_Noyora_whole.txt};
	\addplot[color = blue, only marks, mark=ball, mark options={blue}, error bars,y dir=both, y explicit] table [x index=0, y index=1, y error index=2]{Rd_Sigd_Errbar_Err_NGC7099_Noyora_whole.txt};
	\addplot [color = blue ,mark=no,thick,densely dashed] table[x index=0, y index=1]{Rc_Sigc_NGC7099_Trager_whole.txt};
	\node[black] at (-1.5,-6.5) {(\normalsize{$r_\text{c}=0\textquotesingle.049, \, \delta=0.0055$})} ; 
	\end{axis}
	\end{tikzpicture}
	\caption{Fitting of the energy-truncated ss-OAFP model to the surface brightness profiles of NGC 6266, NGC 6293,  NGC 6652, NGC 6681, NGC 7078 and NGC 7099 reported in \citep{Noyola_2006}.   The unit of the surface brightness (SB) is V magnitude per square of arcseconds. The brightness is normalized by the magnitude $\text{SB}_\text{o}$ at the smallest radius point. In the legend, `Cheb.' means the Chebyshev approximation of the surface brightness profiles reported in \citep{Trager_1995} and `(c)' means PCC cluster as judged so in \citep{Djorgovski_1986}. $\Delta(\text{SB}_\text{o}-\text{SB})$ is the corresponding deviation of $\text{SB}_\text{o}-\text{SB}$ from the model}
	\label{fig:fitting_PCC_Noyora}
\end{figure}

\begin{figure}[H]
	\begin{tikzpicture}
	\begin{axis}[width=8cm,height=6cm, grid=major,xmin=-3.5,xmax=1.5,ymin=-14,ymax=0.5,ylabel=\normalsize{$\text{SB}_\text{o}-\text{SB}$},xlabel=\normalsize{$\log[R(\text{arcmin})]$}, legend style={cells={align=left}},legend entries={\normalsize{ss-OAFP}\\\normalsize{$(\bar{c}=2.59, m=4.2)$}, \normalsize{NGC 6541(c?)}, \normalsize{Cheb.}},yticklabel style={/pgf/number format/.cd,fixed zerofill,precision=1},legend pos=south west]
	\addlegendimage{no markers, orange};
	\addlegendimage{only marks, mark=o, blue};
	\addlegendimage{no markers, blue, densely dashed};
	\addplot [color = orange ,mark=no,thick,solid] table[x index=0, y index=1]{R_Sig_NGC6541_Noyora_whole.txt};
	\addplot[color = blue, only marks, mark=o, mark options={blue}, error bars,y dir=both, y explicit] table [x index=0, y index=1, y error index=2]{Rd_Sigd_Errbar_Err_NGC6541_Noyora_whole.txt};
	\addplot [color = blue ,mark=no,thick,densely dashed] table[x index=0, y index=1]{Rc_Sigc_NGC6541_Trager_whole.txt};
	\node[black] at (-1.8,-5.5) {(\normalsize{$r_\text{c}=0\textquotesingle.168, \, \delta=0.02$})} ; 
	\end{axis}
	\end{tikzpicture}\hspace{0.3cm}
	\begin{tikzpicture}
	\begin{axis}[width=8cm,height=6cm, grid=major,xmin=-3.5,xmax=1.0,ymin=-12,ymax=0.5,xlabel=\normalsize{$\log[R(\text{arcmin})]$},legend style={cells={align=left}},legend entries={\normalsize{ss-OAFP}\\\normalsize{$(\bar{c}=3.21, m=4.2)$}, \normalsize{NGC 6624(c)},\normalsize{Cheb.}},yticklabel style={/pgf/number format/.cd,fixed zerofill,precision=1},legend pos=south west]
	\addlegendimage{no markers, orange};
	\addlegendimage{only marks, mark=ball, blue};
	\addlegendimage{no markers, blue, densely dashed};
	\addplot [color = orange ,mark=no,thick,solid] table[x index=0, y index=1]{R_Sig_NGC6624_Noyora_whole.txt};
	\addplot[color = blue, only marks, mark=ball, mark options={blue}, error bars,y dir=both, y explicit] table [x index=0, y index=1, y error index=2]{Rd_Sigd_Errbar_Err_NGC6624_Noyora_whole.txt};
	\addplot [color = blue ,mark=no,thick,densely dashed] table[x index=0, y index=1]{Rc_Sigc_NGC6624_Trager_whole.txt};
	\node[black] at (-1.8,-5.5) {(\normalsize{$r_\text{c}=0\textquotesingle.0461, \, \delta=0.0066$})} ; 
	\end{axis}
	\end{tikzpicture}
	\caption{Failure of fitting of the energy-truncated ss-OAFP model to the surface brightness profiles of NGC 6541 and NGC 6624 reported in \citep{Noyola_2006}.   The unit of the surface brightness (SB) is V magnitude per square of arcseconds. The brightness is normalized by the magnitude $\text{SB}_\text{o}$ at the smallest radius point. In the legend, `Cheb.' means the Chebyshev approximation of the surface brightness profiles reported in \citep{Trager_1995} and `(c)' means PCC cluster and `(c?)' possibly PCC cluster as judged so in \citep{Djorgovski_1986}. $\Delta(\text{SB}_\text{o}-\text{SB})$ is the corresponding deviation of $\text{SB}_\text{o}-\text{SB}$ from the model}
	\label{fig:fitting_PCC_Noyora_fail3}
\end{figure}
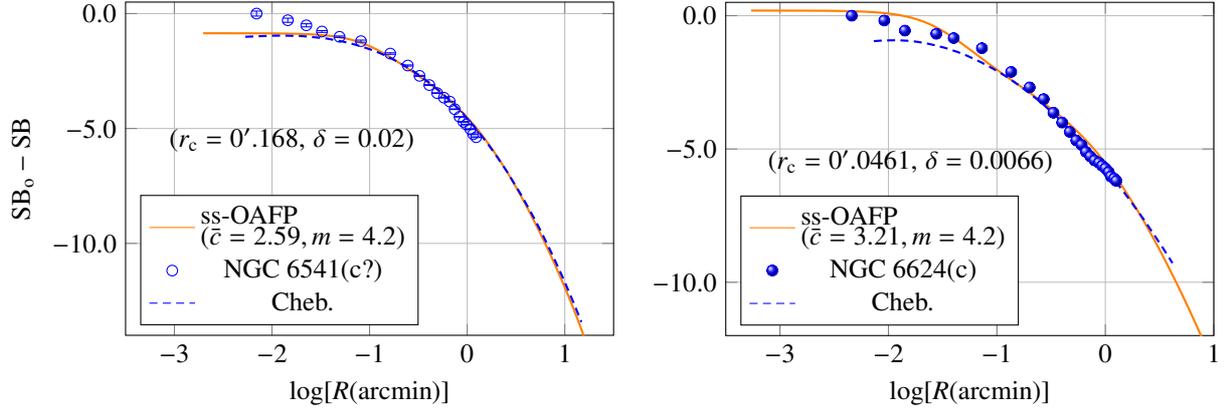

The energy-truncated ss-OAFP model with $m=4.2$ well fits the surface brightness profiles of PCC clusters reported in \citep{Noyola_2006} as shown in Figure \ref{fig:fitting_PCC_Noyora}. However, NGC 6624 is an example of failure of our model as shown in Figure \ref{fig:fitting_PCC_Noyora_fail3}. This failure also occurred to the same cluster but reported in \citep{Lugger_1995} (Appendix \ref{sec:PCC_OAFP}). It appears that NGC 6624 needs more realistic effects, such as the binary heating and mass function. Also, NGC 6541 has also a cusp in the core that our model can not fit.

	\section{Fitting of the finite ss-OAFP model to Post-collapsed-core clusters}\label{Appendix_OAFP_Fit_PCC}
	The present Appendix shows the fitting of polytropic sphere of index $m$ to the projected density profiles and surface brightness of low-concentration globular clusters reported in \citep{Miocchi_2013,Trager_1995,Kron_1984}. Fitting of the polytrope is done for \citep{Miocchi_2013}'s data by minimizing $\chi^{2}_{\nu}$ and for \citep{Kron_1984,Trager_1995}'s data by minimizing the infinite norm of residues for the differences between the fitted curve and the data.

\subsection{Polytropic cluster \citep{Miocchi_2013} and \citep{Kron_1984}} 

\begin{figure}[H]	\centering
	\begin{tikzpicture}
	\begin{loglogaxis}[width=7cm,height=6cm, grid=major, xlabel=\normalsize{$R$ (arcmin)}, ylabel=\normalsize{$\Sigma$}, xmin=1e-1,xmax=10,ymin=1e-4,ymax=3,legend pos=south west, legend style={cells={align=left}}]
	\addplot [color = black ,mark=no,thick,solid ] table[x index=0, y index=1]{R_Sig_NGC4590_poly.txt};
	\addlegendentry{\normalsize{polytrope$(m=4.99)$}} 
	\addplot [only marks,color = red ,mark=triangle*,thick ] table[x index=0, y index=1]{Rd_Sigd_NGC4590_poly.txt};
	\addlegendentry{\normalsize{NGC 4590}\normalsize{(n)}} 
	\end{loglogaxis}
	\end{tikzpicture}
	\caption{Partial success of fitting of the polytropic sphere of index $m$ to the projected density of NGC 4590 reported in \citep{Kron_1984}. The unit of the projected density $\Sigma$ is number per square of arcminutes. In the legends, (n) means normal or KM cluster as judged so in \citep{Djorgovski_1986}. Following \citep{Kron_1984}, data at small radii are ignored due to the depletion of projected density profile.}
	\label{fig:fitting_poly_fail}
\end{figure}
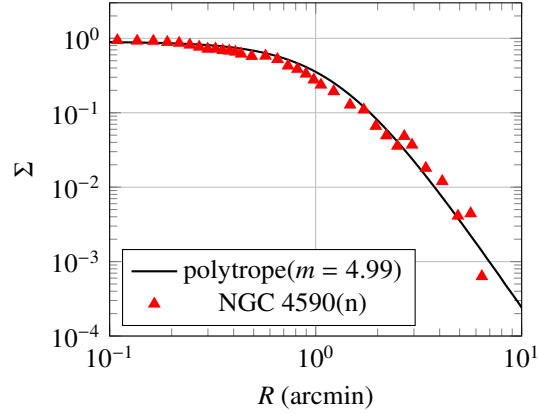

Figures \ref{fig:fitting_poly_1} and \ref{fig:fitting_poly_2} show successful application of the polytrope model to the projected density profiles of NGC 5466, NGC 6809, Palomar 3, Palomar 4 and Palomar 14 reported in \citep{Miocchi_2013}.  In the figures, the polytropes of $3.0<m<5$ reasonably fit to the projected surface densities of the globular clusters whose concentrations range $1<\bar{c}<1.4$. Figure \ref{fig:fitting_poly_fail} shows the projected density of NGC 4590 fitted with a polytrope. NGC 4590 is one of the examples that could fall in either of polytropic and non-polytropic .  

\begin{figure}[H]
	\begin{tikzpicture}
	\begin{axis}[width=8cm,height=6cm, grid=major,xmin=-1.5,xmax=1.5,ymin=-4.5,ymax=0.5,ylabel=\normalsize{$\log[\Sigma]$},legend style={cells={align=left}},legend entries={\normalsize{polytrope}\\\normalsize{$(m=4.97)$}, \normalsize{NGC 5466(n)}},yticklabel style={/pgf/number format/.cd,fixed zerofill,precision=1},legend pos=south west]
	\addlegendimage{no markers, orange};
	\addlegendimage{only marks, mark=triangle*, red};
	\addplot [color = orange ,mark=no,thick,solid] table[x index=0, y index=1]{R_Sig_NGC5466_poly.txt};
	\addplot[color = red, only marks, mark=triangle*, mark options={red}, error bars,y dir=both, y explicit] table [x index=0, y index=1, y error index=2]{Rd_Sigd_Errbar_Err_NGC5466_poly.txt};
	\node[black] at (-1.0,-1.6) {(\normalsize{$\chi^{2}_\nu=0.37$})} ; 
	\end{axis}
	\end{tikzpicture}\hspace{0.3cm}
	\begin{tikzpicture}
	\begin{axis}[width=8cm,height=6cm, grid=major,xmin=-1.0,xmax=1.5,ymin=-5.0,ymax=0.5,legend style={cells={align=left}},legend entries={\normalsize{polytrope}\\\normalsize{$(m=4.6)$}, \normalsize{NGC 6809(n)}},yticklabel style={/pgf/number format/.cd,fixed zerofill,precision=1},legend pos=south west]
	\addlegendimage{no markers, orange};
	\addlegendimage{only marks, mark=triangle*, red};
	\addplot [color = orange ,mark=no,thick,solid] table[x index=0, y index=1]{R_Sig_NGC6809_poly.txt};
	\addplot[color = red, only marks, mark=triangle*, mark options={red}, error bars,y dir=both, y explicit] table [x index=0, y index=1, y error index=2]{Rd_Sigd_Errbar_Err_NGC6809_poly.txt};
	\node[black] at (-0.2,-1.6) {(\normalsize{$\chi^{2}_\nu=1.17$})} ; 
	\end{axis}
	\end{tikzpicture}
	
	\vspace{0.2cm}
	
	\begin{tikzpicture}
	\begin{axis}[width=8cm,height=4cm, grid=major,xmin=-1.5,xmax=1.5,ymin=-0.6,ymax=0.6,xlabel=\normalsize{$\log[R(\text{arcmin})]$ },ylabel=\normalsize{$\Delta\log[\Sigma]$},legend style={cells={align=left}},legend entries={\normalsize{ NGC 5466(n)}},legend pos=south west]
	\addlegendimage{only marks, mark=triangle*, red};
	\addplot[color = red, only marks, mark=triangle*, mark options={red}, error bars,y dir=both, y explicit] table [x index=0, y index=3, y error index=2]{Rd_Sigd_Errbar_Err_NGC5466_poly.txt};
	\draw[dashed] (-2.5,0)--(1.5,0);
	\end{axis}
	\end{tikzpicture}\hspace{0.3cm}
	\begin{tikzpicture}
	\begin{axis}[width=8cm,height=4cm, grid=major,xmin=-1.0,xmax=1.5,ymin=-0.8,ymax=0.8,xlabel=\normalsize{$\log[R(\text{arcmin})]$ },legend style={cells={align=left}},legend entries={\normalsize{ NGC 6809 (n)}},legend pos=south west]
	\addlegendimage{only marks, mark=triangle*, red};
	\addplot[color = red, only marks, mark=triangle*, mark options={red}, error bars,y dir=both, y explicit] table [x index=0, y index=3, y error index=2]{Rd_Sigd_Errbar_Err_NGC6809_poly.txt};
	\draw[dashed] (-2.5,0)--(1.5,0);
	\end{axis}
	\end{tikzpicture}
	
	\vspace{0.7cm}
	
	\begin{tikzpicture}
	\begin{axis}[width=8cm,height=6cm, grid=major,xmin=-1.5,xmax=1,ymin=-4.0,ymax=0.5,ylabel=\normalsize{$\log[\Sigma]$},legend style={cells={align=left}},legend entries={\normalsize{polytrope}\\\normalsize{$(m=4.2)$}, \normalsize{Pal 3(n)}},yticklabel style={/pgf/number format/.cd,fixed zerofill,precision=1},legend pos=south west]
	\addlegendimage{no markers, orange};
	\addlegendimage{only marks, mark=triangle*, red};
	\addplot [color = orange ,mark=no,thick,solid] table[x index=0, y index=1]{R_Sig_Pal3_poly.txt};
	\addplot[color = red, only marks, mark=triangle*, mark options={red}, error bars,y dir=both, y explicit] table [x index=0, y index=1, y error index=2]{Rd_Sigd_Errbar_Err_Pal3_poly.txt};
	\node[black] at (-0.5,-1.6) {(\normalsize{$\chi^{2}_\nu=0.004$})} ; 
	\end{axis}
	\end{tikzpicture}\hspace{0.3cm}
	\begin{tikzpicture}
	\begin{axis}[width=8cm,height=6cm, grid=major,xmin=-1.5,xmax=1.0,ymin=-4.5,ymax=0.5,legend style={cells={align=left}},legend entries={\normalsize{polytrope}\\\normalsize{$(m=4.80)$}, \normalsize{Pal 4 (n)}},yticklabel style={/pgf/number format/.cd,fixed zerofill,precision=1},legend pos=south west]
	\addlegendimage{no markers, orange};
	\addlegendimage{only marks, mark=triangle*, red};
	\addplot [color = orange ,mark=no,thick,solid] table[x index=0, y index=1]{R_Sig_Pal4_poly.txt};
	\addplot[color = red, only marks, mark=triangle*, mark options={red}, error bars,y dir=both, y explicit] table [x index=0, y index=1, y error index=2]{Rd_Sigd_Errbar_Err_Pal4_poly.txt};
	\node[black] at (-0.5,-1.6) {(\normalsize{$\chi^{2}_\nu=0.32$})} ; 
	\end{axis}
	\end{tikzpicture}
	
	\vspace{0.3cm}
	\begin{tikzpicture}
	\begin{axis}[width=8cm,height=4cm, grid=major,xmin=-1.5,xmax=1.0,ymin=-0.8,ymax=0.8,xlabel=\normalsize{$\log[R(\text{arcmin})]$ },ylabel=\normalsize{$\Delta\log[\Sigma]$},legend style={cells={align=left}},legend entries={\normalsize{ Pal 3(n)}},legend pos=south west]
	\addlegendimage{only marks, mark=triangle*, red};
	\addplot[color = red, only marks, mark=triangle*, mark options={red}, error bars,y dir=both, y explicit] table [x index=0, y index=3, y error index=2]{Rd_Sigd_Errbar_Err_Pal3_poly.txt};
	\draw[dashed] (-2.5,0)--(1.0,0);
	\end{axis}
	\end{tikzpicture}\hspace{0.3cm}
	\begin{tikzpicture}
	\begin{axis}[width=8cm,height=4cm, grid=major,xmin=-1.5,xmax=1.0,ymin=-0.6,ymax=0.6,xlabel=\normalsize{$\log[R(\text{arcmin})]$ },legend style={cells={align=left}},legend entries={\normalsize{ Pal 4(n)}},legend pos=south west]
	\addlegendimage{only marks, mark=triangle*, red};
	\addplot[color = red, only marks, mark=triangle*, mark options={red}, error bars,y dir=both, y explicit] table [x index=0, y index=3, y error index=2]{Rd_Sigd_Errbar_Err_Pal4_poly.txt};
	\draw[dashed] (-2.5,0)--(1.0,0);
	\end{axis}
	\end{tikzpicture}
	\caption{Fitting of the polytropic sphere of index $m$ to the projected density $\Sigma$ of NGC 5466, NGC 6809, Palomar 3 and Palomar 4 reported in \citep{Miocchi_2013}. The unit of $\Sigma$ is number per square of arcminutes and $\Sigma$ is normalized so that the density is unity at smallest radius for data. In the legends, (n) means normal or KM cluster as judged so in \citep{Djorgovski_1986}. $\Delta\log[\Sigma]$ is the corresponding deviation of $\Sigma$ from the model on log scale.}
	\label{fig:fitting_poly_1}
\end{figure}

\begin{figure}[H]
	\hspace{0.3cm}
	\begin{tikzpicture}
	\begin{axis}[width=8cm,height=6cm, grid=major,xmin=-1.5,xmax=1.0,ymin=-4.5,ymax=0.5,xlabel=\normalsize{$\log[R]$}, ylabel=\normalsize{$\log[\Sigma]$},legend style={cells={align=left}},legend entries={\normalsize{polytrope}\\\normalsize{$(m=4.6)$}, \normalsize{Pal14(n)}},yticklabel style={/pgf/number format/.cd,fixed zerofill,precision=1},legend pos=south west]
	\addlegendimage{no markers, orange};
	\addlegendimage{only marks, mark=triangle*, red};
	\addplot [color = orange ,mark=no,thick,solid] table[x index=0, y index=1]{R_Sig_Pal14_poly.txt};
	\addplot[color = red, only marks, mark=triangle*, mark options={red}, error bars,y dir=both, y explicit] table [x index=0, y index=1, y error index=2]{Rd_Sigd_Errbar_Err_Pal14_poly.txt};
	\node[black] at (-1.0,-1.6) {(\normalsize{$\chi^{2}_\nu=0.15$})} ; 
	\end{axis}
	\end{tikzpicture}\hspace{0.2cm}
	\begin{tikzpicture}
	\begin{axis}[width=8cm,height=4cm, grid=major,xmin=-1.5,xmax=1.0,ymin=-0.6,ymax=0.6,xlabel=\normalsize{$\log[R(\text{arcmin})]$ },ylabel=\normalsize{$\Delta\log[\Sigma]$},legend style={cells={align=left}},legend entries={\normalsize{ Pal 14 (n)}},legend pos=south west]
	\addlegendimage{only marks, mark=triangle*, red};
	\addplot[color = red, only marks, mark=triangle*, mark options={red}, error bars,y dir=both, y explicit] table [x index=0, y index=3, y error index=2]{Rd_Sigd_Errbar_Err_Pal14_poly.txt};
	\draw[dashed] (-2.5,0)--(1.0,0);
	\end{axis}
	\end{tikzpicture}
	\caption{Fitting of the polytropic sphere of index $m$ to the projected density of Palomar 14 reported in \citep{Miocchi_2013}.  The unit of the projected density $\Sigma$ is number per square of arcminutes. In the legends, (n) means normal or KM cluster as judged so in \citep{Djorgovski_1986}.}
	\label{fig:fitting_poly_2}
\end{figure}
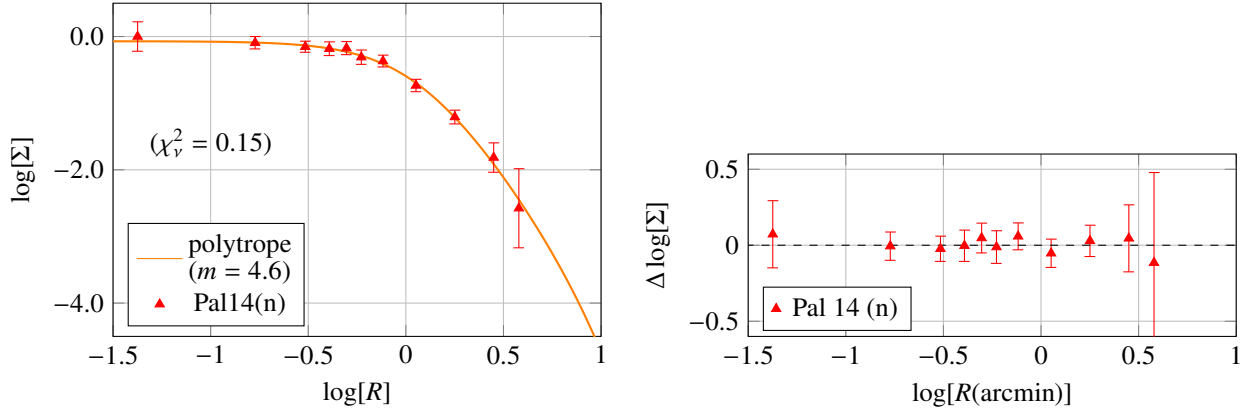

\subsection{Polytropic cluster \citep{Trager_1995}}

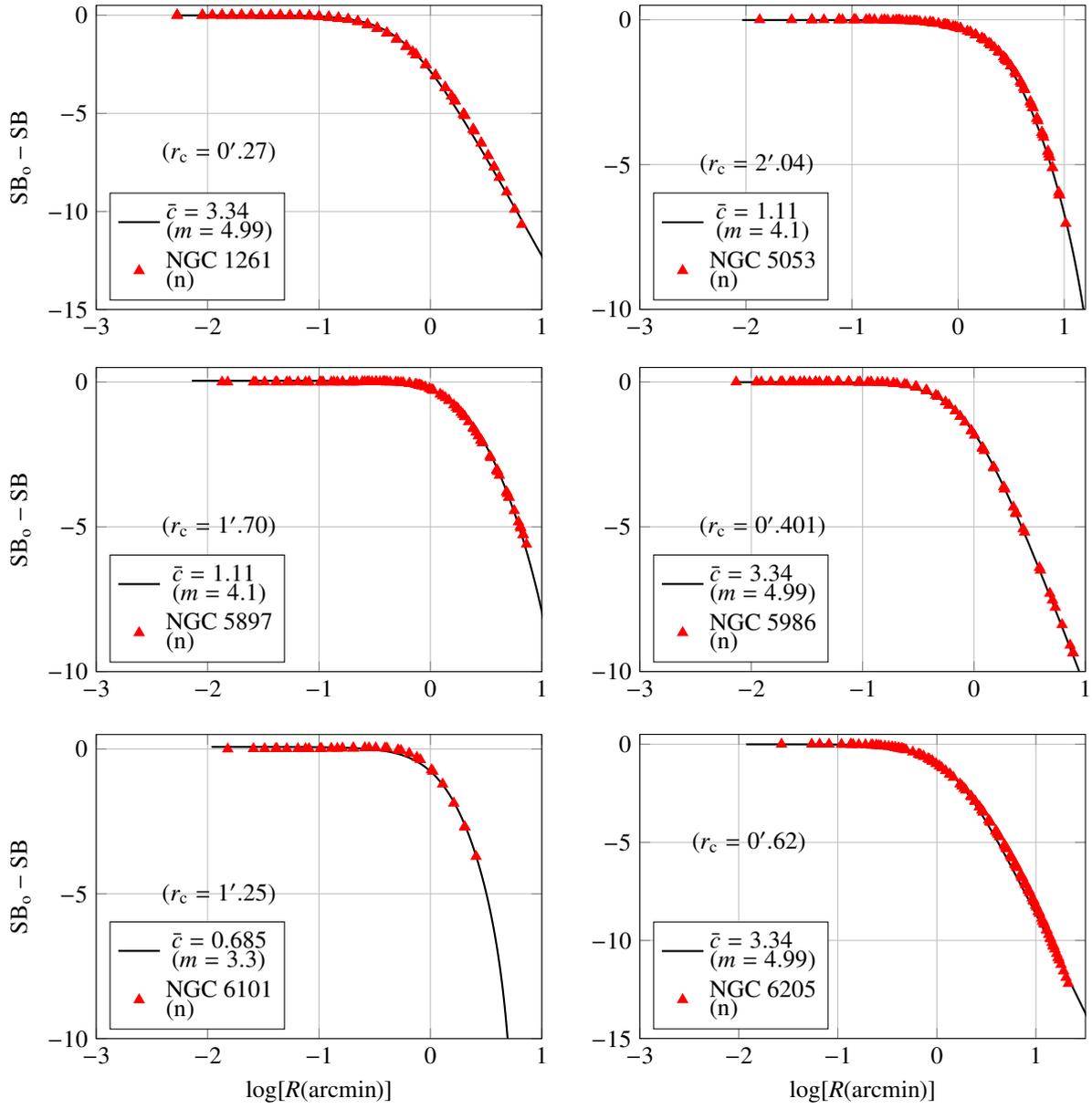
\begin{figure}[H]
	\begin{tikzpicture}
	\begin{axis}[width=8cm,height=6cm, grid=major, ylabel=\normalsize{$\text{SB}_\text{o}-\text{SB}$}, xmin=-3.0,xmax=1.0,ymin=-15,ymax=0.5,legend pos=south west, legend style={cells={align=left}}]
	\addplot [color = black ,mark=no,thick,solid ] table[x index=0, y index=1]{R_Sig_poly_NGC1261.txt};
	\addlegendentry{\normalsize{$\bar{c}=3.34$ }\\ \normalsize{$(m=4.99)$}} 
	\addplot [only marks,color = red ,mark=triangle*,thick ] table[x index=0, y index=1]{Rd_Sigd_poly_NGC1261.txt};
	\addlegendentry{\normalsize{NGC 1261}\\ \normalsize{(n)}} 
	\node[black] at (-1.9,-7) {(\normalsize{$r_\text{c}=0\textquotesingle.27$})} ; 
	\end{axis}
	\end{tikzpicture}\hspace{0.3cm}
	\begin{tikzpicture}
	\begin{axis}[width=8cm,height=6cm, grid=major, xmin=-3.0,xmax=1.2,ymin=-10,ymax=0.5,legend pos=south west, legend style={cells={align=left}}]
	\addplot [color = black ,mark=no,thick,solid ] table[x index=0, y index=1]{R_Sig_poly_NGC5053.txt};
	\addlegendentry{\normalsize{$\bar{c}=1.11$}\\\normalsize{$(m=4.1)$}} 
	\addplot [only marks,color = red ,mark=triangle*,thick ] table[x index=0, y index=1]{Rd_Sigd_poly_NGC5053.txt};
	\addlegendentry{\normalsize{NGC 5053}\\\normalsize{(n)}} 
	\node[black] at (-1.9,-5) {(\normalsize{$r_\text{c}=2\textquotesingle.04$})} ; 
	\end{axis}
	\end{tikzpicture}
	
	\vspace{0.3cm}
	
	\begin{tikzpicture}
	\begin{axis}[width=8cm,height=6cm, grid=major, ylabel=\normalsize{$\text{SB}_\text{o}-\text{SB}$}, xmin=-3.0,xmax=1.0,ymin=-10,ymax=0.5,legend pos=south west, legend style={cells={align=left}}]
	\addplot [color = black ,mark=no,thick,solid ] table[x index=0, y index=1]{R_Sig_poly_NGC5897.txt};
	\addlegendentry{\normalsize{$\bar{c}=1.11$ }\\\normalsize{$(m=4.1)$}} 
	\addplot [only marks,color = red ,mark=triangle*,thick ] table[x index=0, y index=1]{Rd_Sigd_poly_NGC5897.txt};
	\addlegendentry{\normalsize{NGC 5897}\\\normalsize{(n)}} 
	\node[black] at (-1.9,-5) {(\normalsize{$r_\text{c}=1\textquotesingle.70$})} ; 
	\end{axis}
	\end{tikzpicture}\hspace{0.3cm}
	\begin{tikzpicture}
	\begin{axis}[width=8cm,height=6cm, grid=major, xmin=-3.0,xmax=1.0,ymin=-10,ymax=0.5,legend pos=south west, legend style={cells={align=left}}]
	\addplot [color = black ,mark=no,thick,solid ] table[x index=0, y index=1]{R_Sig_poly_NGC5986.txt};
	\addlegendentry{\normalsize{$\bar{c}=3.34$ }\\\normalsize{$(m=4.99)$}} 
	\addplot [only marks,color = red ,mark=triangle*,thick ] table[x index=0, y index=1]{Rd_Sigd_poly_NGC5986.txt};
	\addlegendentry{\normalsize{NGC 5986}\\\normalsize{(n)}} 
	\node[black] at (-1.9,-5) {(\normalsize{$r_\text{c}=0\textquotesingle.401$})} ; 
	\end{axis}
	\end{tikzpicture}
	
	\vspace{0.3cm}
	
	\begin{tikzpicture}
	\begin{axis}[width=8cm,height=6cm, grid=major, xlabel=\normalsize{$\log [R(\text{arcmin})]$}, ylabel=\normalsize{$\text{SB}_\text{o}-\text{SB}$}, xmin=-3.0,xmax=1.0, ymin=-10,ymax=0.5,legend pos=south west, legend style={cells={align=left}}]
	\addplot [color = black ,mark=no,thick,solid ] table[x index=0, y index=1]{R_Sig_poly_NGC6101.txt};
	\addlegendentry{\normalsize{$\bar{c}=0.685$ }\\\normalsize{$(m=3.3)$}} 
	\addplot [only marks,color = red ,mark=triangle*,thick ] table[x index=0, y index=1]{Rd_Sigd_poly_NGC6101.txt};
	\addlegendentry{\normalsize{NGC 6101}\\\normalsize{(n)}} 
	\node[black] at (-1.9,-5) {(\normalsize{$r_\text{c}=1\textquotesingle.25$})} ; 
	\end{axis}
	\end{tikzpicture}\hspace{0.3cm}
	\begin{tikzpicture}
	\begin{axis}[width=8cm,height=6cm, grid=major, xlabel=\normalsize{$\log [R(\text{arcmin})]$}, xmin=-3.0,xmax=1.5,ymin=-15,ymax=0.5,legend pos=south west, legend style={cells={align=left}}]
	\addplot [color = black ,mark=no,thick,solid ] table[x index=0, y index=1]{R_Sig_poly_NGC6205.txt};
	\addlegendentry{\normalsize{$\bar{c}=3.34$ }\\\normalsize{$(m=4.99)$}} 
	\addplot [only marks,color = red ,mark=triangle*,thick ] table[x index=0, y index=1]{Rd_Sigd_poly_NGC6205.txt};
	\addlegendentry{\normalsize{NGC 6205}\\\normalsize{(n)}} 
	\node[black] at (-1.9,-5) {(\normalsize{$r_\text{c}=0\textquotesingle.62$})} ; 
	\end{axis}
	\end{tikzpicture}
	\caption{Fitting of the polytropic sphere of index $m$ to the surface brightness profiles of NGC 1261, NGC 5053, NGC 5897, NGC 5986, NGC 6101 and NGC 6205 reported in \cite{Trager_1995}. The unit of the surface brightness (SB) is V magnitude per square of arcseconds. The brightness is normalized by the magnitude $\text{SB}_\text{o}$ at the smallest radius point. In the legends, `(n)' means KM cluster as judged so in \citep{Djorgovski_1986}.}
	\label{fig:fitting_poly_Trager1}
\end{figure}

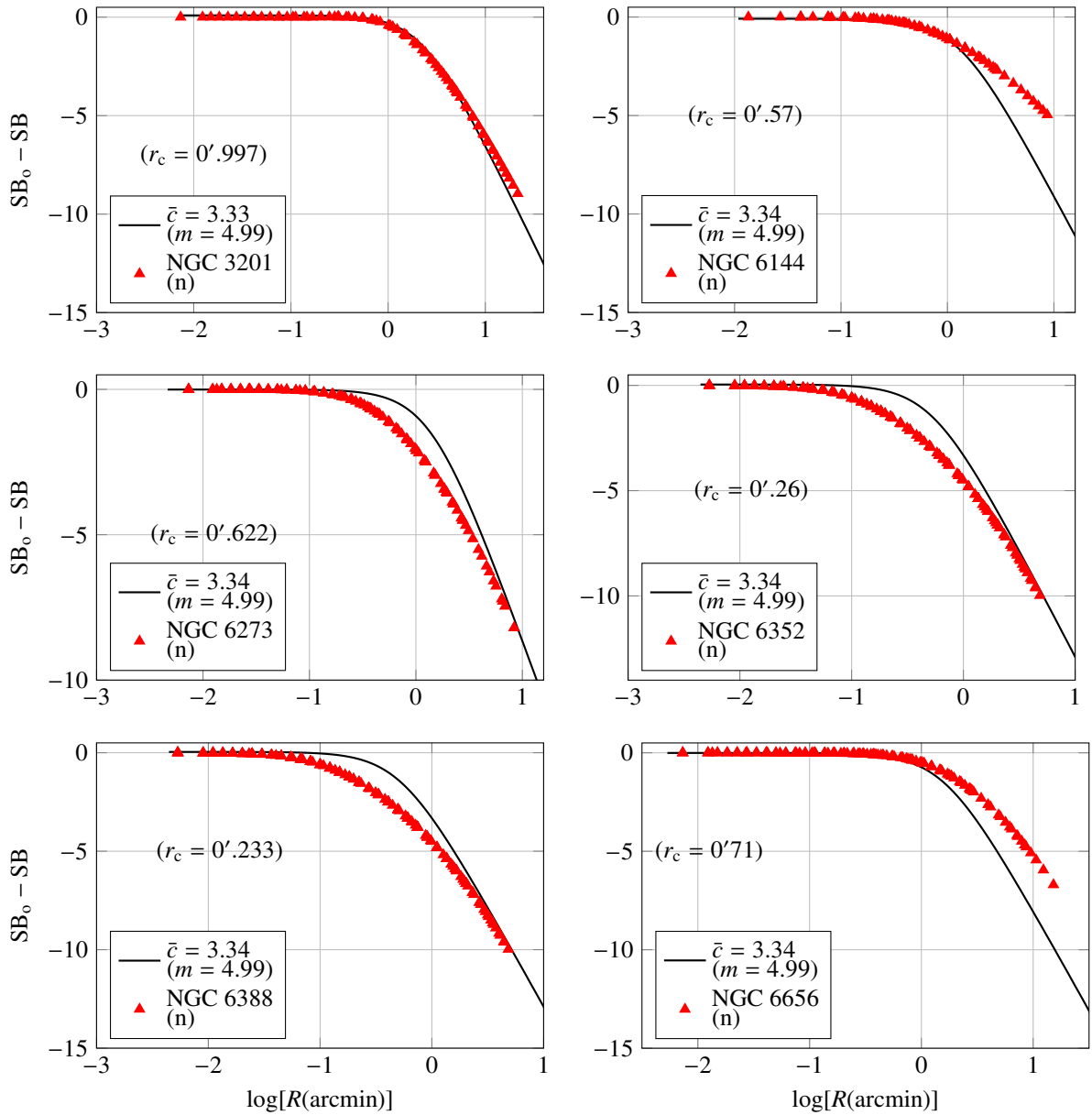
\begin{figure}[H]
	\begin{tikzpicture}
	\begin{axis}[width=8cm,height=6cm, grid=major, ylabel=\normalsize{$\text{SB}_\text{o}-\text{SB}$}, xmin=-3.0,xmax=1.6,ymin=-15,ymax=0.5,legend pos=south west, legend style={cells={align=left}}]
	\addplot [color = black ,mark=no,thick,solid ] table[x index=0, y index=1]{R_Sig_poly_NGC3201.txt};
	\addlegendentry{\normalsize{$\bar{c}=3.33$ }\\\normalsize{$(m=4.99)$}} 
	\addplot [only marks,color = red ,mark=triangle*,thick ] table[x index=0, y index=1]{Rd_Sigd_poly_NGC3201.txt};
	\addlegendentry{\normalsize{NGC 3201}\\\normalsize{(n)}} 
	\node[black] at (-1.9,-7) {(\normalsize{$r_\text{c}=0\textquotesingle.997$})} ; 
	\end{axis}
	\end{tikzpicture}\hspace{0.3cm}
	\begin{tikzpicture}
	\begin{axis}[width=8cm,height=6cm, grid=major, xmin=-3.0,xmax=1.2,ymin=-15,ymax=0.5,legend pos=south west, legend style={cells={align=left}}]
	\addplot [color = black ,mark=no,thick,solid ] table[x index=0, y index=1]{R_Sig_poly_NGC6144.txt};
	\addlegendentry{\normalsize{$\bar{c}=3.34$ }\\ \normalsize{$(m=4.99)$}} 
	\addplot [only marks,color = red ,mark=triangle*,thick ] table[x index=0, y index=1]{Rd_Sigd_poly_NGC6144.txt};
	\addlegendentry{\normalsize{NGC 6144}\\ \normalsize{(n)}} 
	\node[black] at (-1.9,-5) {(\normalsize{$r_\text{c}=0\textquotesingle.57$})} ; 
	\end{axis}
	\end{tikzpicture}
	
	\vspace{0.3cm}
	
	\begin{tikzpicture}
	\begin{axis}[width=8cm,height=6cm, grid=major,ylabel=\normalsize{$\text{SB}_\text{o}-\text{SB}$}, xmin=-3.0,xmax=1.2,ymin=-10,ymax=0.5,legend pos=south west, legend style={cells={align=left}}]
	\addplot [color = black ,mark=no,thick,solid ] table[x index=0, y index=1]{R_Sig_poly_NGC6273.txt};
	\addlegendentry{\normalsize{$\bar{c}=3.34$ }\\\normalsize{$(m=4.99)$}} 
	\addplot [only marks,color = red ,mark=triangle*,thick ] table[x index=0, y index=1]{Rd_Sigd_poly_NGC6273.txt};
	\addlegendentry{\normalsize{NGC 6273}\\\normalsize{(n)}} 
	\node[black] at (-1.9,-5) {(\normalsize{$r_\text{c}=0\textquotesingle.622$})} ; 
	\end{axis}
	\end{tikzpicture}\hspace{0.3cm}
	\begin{tikzpicture}
	\begin{axis}[width=8cm,height=6cm, grid=major, xmin=-3.0,xmax=1.0,ymin=-14,ymax=0.5,legend pos=south west, legend style={cells={align=left}}]
	\addplot [color = black ,mark=no,thick,solid ] table[x index=0, y index=1]{R_Sig_poly_NGC6352.txt};
	\addlegendentry{\normalsize{$\bar{c}=3.34$ }\\\normalsize{$(m=4.99)$}} 
	\addplot [only marks,color = red ,mark=triangle*,thick ] table[x index=0, y index=1]{Rd_Sigd_poly_NGC6352.txt};
	\addlegendentry{\normalsize{NGC 6352}\\\normalsize{(n)}} 
	\node[black] at (-1.9,-5) {(\normalsize{$r_\text{c}=0\textquotesingle.26$})} ; 
	\end{axis}
	\end{tikzpicture}
	
	\vspace{0.3cm}
	
	\begin{tikzpicture}
	\begin{axis}[width=8cm,height=6cm, grid=major, xlabel=\normalsize{$\log [R(\text{arcmin})]$}, ylabel=\normalsize{$\text{SB}_\text{o}-\text{SB}$}, xmin=-3.0,xmax=1.0,ymin=-15,ymax=0.5,legend pos=south west, legend style={cells={align=left}}]
	\addplot [color = black ,mark=no,thick,solid ] table[x index=0, y index=1]{R_Sig_poly_NGC6388.txt};
	\addlegendentry{\normalsize{$\bar{c}=3.34$}\\\normalsize{$(m=4.99)$}} 
	\addplot [only marks,color = red ,mark=triangle*,thick ] table[x index=0, y index=1]{Rd_Sigd_poly_NGC6388.txt};
	\addlegendentry{\normalsize{NGC 6388}\\\normalsize{(n)}} 
	\node[black] at (-1.9,-5) {(\normalsize{$r_\text{c}=0\textquotesingle.233$})} ; 
	\end{axis}
	\end{tikzpicture}\hspace{0.3cm}
	\begin{tikzpicture}
	\begin{axis}[width=8cm,height=6cm, grid=major, xlabel=\normalsize{$\log [R(\text{arcmin})]$},  xmin=-2.5,xmax=1.5,ymin=-15,ymax=0.5,legend pos=south west, legend style={cells={align=left}}]
	\addplot [color = black ,mark=no,thick,solid ] table[x index=0, y index=1]{R_Sig_poly_NGC6656.txt};
	\addlegendentry{\normalsize{$\bar{c}=3.34$}\\\normalsize{$(m=4.99)$}} 
	\addplot [only marks,color = red ,mark=triangle*,thick ] table[x index=0, y index=1]{Rd_Sigd_poly_NGC6656.txt};
	\addlegendentry{\normalsize{NGC 6656}\\\normalsize{(n)}} 
	\node[black] at (-1.9,-5) {(\normalsize{$r_\text{c}=0\textquotesingle71$})} ; 
	\end{axis}
	\end{tikzpicture}
	\caption{Failure of fitting of the polytropic sphere of index $m$ to the surface brightness profiles of  NGC 3201, NGC 6144, NGC 6273, NGC 6352, NGC 6388 and NGC 6656 reported in \citep{Trager_1995}. The unit of the surface brightness (SB) is V magnitude per square of arcseconds. The brightness is normalized by the magnitude $\text{SB}_\text{o}$ at the smallest radius point. In the legends, `(n)' means KM cluster as judged so in \citep{Djorgovski_1986}.}
	\label{fig:fitting_poly_Trager_fail}
\end{figure}

Figures \ref{fig:fitting_poly_Trager1} and \ref{fig:fitting_poly_Trager3} depict the fitting of the polytropic sphere model to the Chebyshev approximation to the surface brightness profiles reported in \citep{Trager_1995}. The polytropic indeces $m=3.3\sim4.99$ are useful to fit the polytrope model to the low-concentration clusters whose core relaxation time are order of $1$ Gyr (from \cite[][(2010 edition)]{Harris_1996}'s catalog). Since the polytrope itself does not rapidly decay near its limiting radius, the corresponding concentrations of the polytrope are relatively high such as $\bar{c}=3.34$ for $m=4.99$. On one hand, Figure \ref{fig:fitting_poly_Trager_fail} shows the clusters whose surface brightness profiles are not close to the polytrope. Such clusters have shorter core relaxation times $<0.5$ Gyr and relatively high concentrations $c>1.5$ (based on the King model) as shown in Table \ref{table:tcr_age_poly}.

\begin{figure}[H]
	\begin{tikzpicture}
	\begin{axis}[width=8cm,height=6cm, grid=major, ylabel=\normalsize{$\text{SB}_\text{o}-\text{SB}$}, xmin=-3.0,xmax=1.0,ymin=-10,ymax=0.5,legend pos=south west, legend style={cells={align=left}}]
	\addplot [color = black ,mark=no,thick,solid ] table[x index=0, y index=1]{R_Sig_poly_NGC6402.txt};
	\addlegendentry{\normalsize{$\bar{c}=2.24$}\\\normalsize{$(m=4.9)$}} 
	\addplot [only marks,color = red ,mark=triangle*,thick ] table[x index=0, y index=1]{Rd_Sigd_poly_NGC6402.txt};
	\addlegendentry{\normalsize{NGC 6402}\\\normalsize{(n)}} 
	\node[black] at (-1.9,-5) {(\normalsize{$r_\text{c}=0\textquotesingle.73$})} ; 
	\end{axis}
	\end{tikzpicture}\hspace{0.3cm}
	\begin{tikzpicture}
	\begin{axis}[width=8cm,height=6cm, grid=major, xmin=-3.0,xmax=1.0,ymin=-15,ymax=0.5,legend pos=south west, legend style={cells={align=left}}]
	\addplot [color = black ,mark=no,thick,solid ] table[x index=0, y index=1]{R_Sig_poly_NGC6496.txt};
	\addlegendentry{\normalsize{$\bar{c}=2.24$}\\ \normalsize{$(m=3.3)$}} 
	\addplot [only marks,color = red ,mark=triangle*,thick ] table[x index=0, y index=1]{Rd_Sigd_poly_NGC6496.txt};
	\addlegendentry{\normalsize{NGC 6496}\\ \normalsize{(n)}} 
	\node[black] at (-1.9,-5) {(\normalsize{$r_\text{c}=0\textquotesingle.73$})} ; 
	\end{axis}
	\end{tikzpicture}

	\vspace{0.3cm}
	
	\begin{tikzpicture}
	\begin{axis}[width=8cm,height=6cm, grid=major, ylabel=\normalsize{$\text{SB}_\text{o}-\text{SB}$}, xmin=-3.0,xmax=1.0,ymin=-10,ymax=0.5,legend pos=south west, legend style={cells={align=left}}]
	\addplot [color = black ,mark=no,thick,solid ] table[x index=0, y index=1]{R_Sig_poly_NGC6712.txt};
	\addlegendentry{\normalsize{$\bar{c}=3.34$}\\\normalsize{$(m=4.99)$}} 
	\addplot [only marks,color = red ,mark=triangle*,thick ] table[x index=0, y index=1]{Rd_Sigd_poly_NGC6712.txt};
	\addlegendentry{\normalsize{NGC 6712}\\\normalsize{(n)}} 
	\node[black] at (-1.9,-5) {(\normalsize{$r_\text{c}=0\textquotesingle.58$})} ; 
	\end{axis}
	\end{tikzpicture}\hspace{0.3cm}
	\begin{tikzpicture}
	\begin{axis}[width=8cm,height=6cm, grid=major,  xmin=-3.0,xmax=1.0,ymin=-10,ymax=0.5, xlabel=\normalsize{$\log [R(\text{arcmin})]$},legend pos=south west, legend style={cells={align=left}}]
	\addplot [color = black ,mark=no,thick,solid ] table[x index=0, y index=1]{R_Sig_poly_NGC6723.txt};
	\addlegendentry{\normalsize{$\bar{c}=3.34$}\\\normalsize{$(m=4.99)$}} 
	\addplot [only marks,color = red ,mark=triangle*,thick ] table[x index=0, y index=1]{Rd_Sigd_poly_NGC6723.txt};
	\addlegendentry{\normalsize{NGC 6723}\\\normalsize{(n)}} 
	\node[black] at (-1.9,-5) {(\normalsize{$r_\text{c}=0\textquotesingle.634$})} ; 
	\end{axis}
	\end{tikzpicture}
	
	\vspace{0.3cm}
	
	\begin{tikzpicture}
	\begin{axis}[width=8cm,height=6cm, grid=major, xlabel=\normalsize{$\log [R(\text{arcmin})]$}, ylabel=\normalsize{$\text{SB}_\text{o}-\text{SB}$}, xmin=-3.0,xmax=1.0,ymin=-10,ymax=0.5,legend pos=south west, legend style={cells={align=left}}]
	\addplot [color = black ,mark=no,thick,solid ] table[x index=0, y index=1]{R_Sig_poly_NGC6981.txt};
	\addlegendentry{\normalsize{$\bar{c}=3.34$}\\\normalsize{$(m=4.99)$}} 
	\addplot [only marks,color = red ,mark=triangle*,thick ] table[x index=0, y index=1]{Rd_Sigd_poly_NGC6981.txt};
	\addlegendentry{\normalsize{NGC 6981}\\\normalsize{(n)}} 
	\node[black] at (-1.9,-5) {(\normalsize{$r_\text{c}=0\textquotesingle.36$})} ; 
	\end{axis}
	\end{tikzpicture}
	\caption{Fitting of the polytropic sphere of index $m$ to the surface brightness profiles of NGC 6402,  NGC 6496, NGC 6712, NGC 6723 and NGC 6981 reported in \citep{Trager_1995}.  The unit of the surface brightness (SB) is V magnitude per square of arcseconds. The brightness is normalized by the magnitude $\text{SB}_\text{o}$ at the smallest radius point. In the legends, `(n)' means KM cluster as judged so in \citep{Djorgovski_1986}.}
	\label{fig:fitting_poly_Trager3}
\end{figure}

\end{appendices}
\bibliographystyle{elsarticle-harv} 
\bibliography{science}


\end{document}